\documentclass[fleqn,usenatbib]{mnras}

\usepackage{newtxtext,newtxmath}
\usepackage{multirow}
\usepackage{siunitx}

\usepackage[T1]{fontenc}

\usepackage{supertabular}
\usepackage{listings}

\DeclareRobustCommand{\VAN}[3]{#2}
\let\VANthebibliography\thebibliography
\def\thebibliography{\DeclareRobustCommand{\VAN}[3]{##3}\VANthebibliography}

\usepackage{graphicx}	
\usepackage{amsmath}	
\usepackage{amssymb}	
\usepackage{multirow}

\usepackage[dvipsnames]{xcolor}
\definecolor{lgreen}{HTML}{079d90}

\title[Inferring galaxy dark halo properties from visible matter with Machine Learning]{Inferring galaxy dark halo properties from visible matter with Machine Learning}

\author[Rodrigo von Marttens et al.]
{Rodrigo~von~Marttens$^{1}$,
Luciano~Casarini$^{2}$,
Nicola~R.~Napolitano$^{3,4}$\thanks{E-mail: napolitano@mail.sysu.edu.cn},
Sirui~Wu$^{3,4}$,
Valeria~Amaro$^{3}$,
\newauthor
Rui~Li$^{3}$,
Crescenzo Tortora$^{5}$,
Askery Canabarro$^{6}$
and Yang Wang$^{3,4}$
\\
$^{1}$Observat\'orio Nacional, 20921-400, Rio de Janeiro, RJ, Brasil\\
$^{2}$Department of Physics, Federal University of Sergipe, Avenida Marechal Rondon s/n, Jardim Rosa Elze, São Cristovão, SE, 49100-000, Brazil\\
$^{3}$School of Physics and Astronomy, Sun Yat-sen University, Zhuhai Campus, 2 Daxue Road, Xiangzhou District, Zhuhai, P. R. China\\
$^{4}$CSST Science Center for Guangdong-Hong Kong-Macau Great Bay Area, Zhuhai, China, 519082\\
$^{5}$INAF -- Osservatorio Astronomico di Capodimonte, Salita Moiariello 16, 80131 - Napoli, Italy\\ 
$^{6}$Grupo de F\'isica da Mat\'eria Condensada, N\'ucleo de Ci\^encias Exatas - NCEx, Federal University of Alagoas, 57309-005, Arapiraca, AL, Brazil}

\date{Accepted XXX. Received YYY; in original form ZZZ}

\pubyear{2015}

\begin{document}
\label{firstpage}
\pagerange{\pageref{firstpage}--\pageref{lastpage}}
\maketitle

\begin{abstract}
Next-generation surveys will provide photometric and spectroscopic data of millions to billions of galaxies with unprecedented precision. This offers a unique chance to improve our understanding of the galaxy evolution and the unresolved nature of dark matter (DM). At galaxy scales, the density distribution of DM is strongly affected by the astrophysical feedback processes, which are difficult to fully account for in classical techniques to derive mass models. 
In this work, we explore the capability of supervised learning algorithms to predict the DM content of galaxies from ``luminous'' observational-like parameters, using the public catalog of the TNG100 simulation. In particular, we use, \textit{Photometric} (magnitudes in different bands), \textit{Structural} (the stellar half-mass radius and three different baryonic masses) and \textit{Kinematic} (1D velocity dispersion of all particles and the maximum rotation velocity) parameters to predict the total DM mass, DM half-mass radius, DM mass inside one and two stellar half-mass radii. We adopt the coefficient of determination, $R^2$, as a reference metric to evaluate the accuracy of these predictions, whose values are between 0 (worst fit possible) and 1 (perfect match).
We find that the Photometric features alone are able to predict the total DM mass with fair accuracy ($R^2 \lesssim 0.86$), while Structural and Photometric features together are more effective to determine the DM inside the stellar half mass radius ($0.88 \lesssim R^2 \lesssim 0.94$), and the DM within twice the stellar half mass radius ($0.90 \lesssim R^2 \lesssim 0.92$). However, using all observational quantities together (Photometry, Structural and Kinematics) incredibly improves the overall accuracy for all DM quantities (up to $R^2 \sim 0.98$). This first test shows that Machine Learning tools are promising approaches to derive predictions of the DM in real galaxies. The next steps will be to improve observational realism of the training sets, by closely select samples which accurately reproduce the typical observed ``luminous'' scaling relations. The so-trained pipelines will be suitable 
for real galaxy data collected from the next-generation surveys like Rubin/LSST, Euclid, CSST, 4MOST, DESI, to derive, e.g., the properties of their central DM fractions.
\end{abstract}




\section{Introduction}
\label{sec:intro}
Machine learning (ML, hereafter) is a branch of artificial intelligence that allows to classify or fit data. ML algorithms automatically learn and improve their performances from their experience, without being explicitly programmed. In particular, ML tools are applied to complex systems where an analytical solution is difficult or not feasible to find. For this reason, ML has been widely used for astrophysical applications. These include the photometric classification of supernovas (\citealt{lochner}, \citealt{charnock}), the star/galaxy classification \citep{Baqui:2020sfd}, the analyses of gravitational waves data (\citealt{biswas}, \citealt{Carrillo}), the estimation of photometric redshift (\citealt{cavuoti2018photometric}, \citealt{amaro+19}), the search of quasars (\citealt{nakoneczny2019catalog}), the galaxy morphology (\citealt{gauci}, \citealt{ball}, \citealt{banerji}), the search of strong lensing events (\citealt{Petrillo+17_CNN,Petrillo+19_CNN,Petrillo+19_LinKS}, \citealt{schaefer}, \citealt{lanusse}, \citealt{hartley}, \citealt{pourrahmani}, \citealt{2019MNRAS.482..313L,Li+20_KiDS}), the prediction of cosmological parameters (\citealt{auld}, \citealt{auldfast}, \citealt{shimabukuro}, \citealt{zarrouk}), and the model-independent search for deviations from the $\Lambda$CDM model~\citep{vonMarttens:2018bvz,vonMarttens:2020apn}. 
In particular, ML tools are particularly suitable for application to large galaxy surveys, where the large amount of data are hard to handle with standard analysis tools (\citealt{kelleher2020fundamentals}).

In this respect, ML techniques will play a major role in the exploitation of datasets coming from future photometric sky survey like Vera Rubin/LSST (\citealt{LSST-cit}), Euclid mission (\citealt{Euclid-cit}), the Chinese Space Station Telescope (CSST, \citealt{CSST-cit}) and spectroscopic surveys, like 4MOST (\citealt{4MOST-cit}), WEAVE (\citealt{WEAVE-cit}) and DESI (\citealt{DESI-cit}). 

Imaging surveys will provide us with accurate total photometry that can be used to infer the stellar masses (\citealt{Maraston+13_BOSS}; \citealt{KV450}) and constrain other stellar population parameters (like age and metallicity, see e.g. \citealt{Labarbera2010}).
For the brightest galaxies with higher signal-to-noise ratios, we will be able to collect detailed surface photometry to measure structure parameters (i.e., the effective radius, total luminosity, ellipticity, S\'ersic index and B/D, see e.g., \citealt{Grogin:2011ua}, \citealt{Baldry2012, GAMA_Lange2016_size_mass}, \citealt{Roy2018}) for billions of galaxies up to redshift $\sim$1.5 (see e.g., \citealt{LSSTGSC-road}).

Spectroscopic surveys will provide redshifts and central velocity dispersion measurements for millions of galaxies (\citealt{4MOST-cit}), which can be used for galaxy scaling relations (e.g. \citealt{Bernardi2005_scaling},  \citealt{Nigoche-FJ}, \citealt{Califa_FJ_TF}, \citealt{Napolitano2020_lamost}) or for constraining the star formation of galaxies (\citealt{Bernardi2006_spec_lines_sfr}, \citealt{Califa_sfr}). 

Overall, the wealth of data expected from the future imaging and spectroscopic surveys will offer us a unique opportunity to have a comprehensive understanding of the dark matter content of galaxies. In this context, there have been a few attempts to use simple observational measurements, like galaxy structural parameters and central velocity dispersion from extensive surveys, to derive information on the galaxy central dark matter fractions (e.g., \citealt{HB09_FP}; \citealt{Tortora+09,TRN13_SPIDER_IMF,SPIDER-VI,Tortora+14_DMevol,Tortora+18_KiDS_DMevol}; \citealt{NRT10}; \citealt{Beifiori+14}; \citealt{Nigoche-Netro+16,Nigoche-Netro+19}). Most of these studies have been so far based on standard Jeans analyses, which needs strong assumptions on the dynamics (e.g., the overall equilibrium of galaxies and the internal anisotropy) and the stellar populations (e.g., the initial mass function, setting the galaxy total stellar masses). 
Moreover, it is well known that DM density profiles can be modified by the gravitational effects of collapsing (e.g. adiabatic contraction, \citealt{Gnedin+04}) or expanding (e.g. adiabatic expansion, see e.g. \citealt{2014MNRAS.437..415D}) baryons, overall changing their central slope. These effects imply strong degeneracies among the parameters in the galaxy models (see e.g. \citealt{NRT10}). 
Furthermore, the integration of the Jeans equations it is often a tedious process that needs computational approximations to quickly converge. For this reason, these techniques have been lately applied to relatively small samples \citep[e.g.][]{Cappellari+06,Cappellari2013_fDM}. 

In this respect, ML techniques can provide a wholly new paradigm for the study of dark matter (DM, hereafter) at different scales. For instance, \cite{ntampaka2018} used Convolutional Neural Networks (CNNs) to estimate the mass of galaxy clusters from an X-ray mock catalog constructed from the TNG100 simulation \citep{nelson2019illustristng}. Similarly, \cite{yan2020galaxy} have obtained predictions for the masses of clusters using images from \textit{BAHAMAS} hydrodynamical simulations.
These analyses provided new and interesting approaches to obtain the estimates of the DM content of hot gravitational systems, are based on previous knowledge of their baryonic component. 

In this paper, we test the ability of ML algorithms to make predictions about the DM halo properties of galaxies, starting from observed quantities derived by the luminous matter, in particular optical and near infrared imaging and spectroscopy.

We make use of catalogs of simulated galaxies from the Illustris simulations public data release (\citealt{Illustris2019_pub_rel}). In possession of this data, we train and test a variety of ML tools and optimize them to obtain the best accuracy in the DM properties of the same simulated galaxies. In particular, we will test what is the minimal set of parameters needed to produce reasonable accuracy in some relevant DM quantities, like the total dark mass and the dark mass within the half mass radius.

This first ``feasibility'' study is meant to set the stage for future analyses, where we plan to train the ML tools on different simulations (EAGLE: \citealt{EAGLE_res_effect}, DIANOGA: \citealt{Dianoga_Bassini2020}, Magneticum: \citealt{Magneticum_sim}) and apply the trained ML tools to real data to make specific predictions on the dark matter fractions of galaxies to be compared to literature analyses (e.g. \citealt{Cappellari2013_fDM,Alabi2017_fDMsluggs,Tortora+18_KiDS_DMevol}).

On a longer term, we aim at applying ML techniques to large sky survey observations to put constraints on the baryonic and the dark matter assembly in galaxies, in a unique ``universal'' scenario. 
Indeed, different cosmologies, combined with different galaxy formation recipes, are expected to provide different observational predictions (see e.g., \citealt{Camels2021}) that can be tested against real data to finally constrain the best cosmological + evolutionary scenario. In this case, ML tools can be used to effectively address this regression problem, which involves a multi-dimensional parameter space that includes standard cosmological parameters (e.g., the critical mass density parameter, $\Omega_m$, and the scatter of the scale of the 8 Mpc fluctuations, $\sigma_8$), plus different parameters regulating the star formation, stellar winds, and AGN power (see \citealt{Camels2021} for a discussion), possibly in different DM flavors (e.g. $\Lambda$CDM: \citealt{LCDM-cit_WMAP,LCDM-cit_planck}, Warm Dark Matter: \citealt{WDM-Kang_Maccio`}, or Self-interacting Dark Matter: \citealt{SIDM_spergel}). 
By training ``supervised'' ML tools over different simulations, where a given set of observational parameters is defined as the ones measured in the aforementioned large sky surveys, we can expect to derive the best ``universe'' interpreting all data.

This work is organized as follows: in \S\ref{sec:data} we present the simulation data from IllustrisTNG-100 used in this work, and the main physical quantities of the simulated galaxies that we will use throughout the paper. In \S\ref{sec:method} we describe the approach adopted to obtain the best pipeline to predict the galaxy DM parameters and the performance metrics used to evaluate the accuracy of these predictions. In \S\ref{sec:analyses} we discuss the results obtained in our analyses. In particular, we assess the accuracy and precision of the predicted quantities and perform a feature importance analysis. Finally, in \S\ref{sec:conclusions} we draw the conclusions and discuss future perspectives.

\section{Synthetic data}
\label{sec:data}

As mentioned in \S\ref{sec:intro}, in this work we use synthetic data from IllustrisTNG-100 (TNG100, hereafter) simulation for which both baryonic quantities (see also \S\ref{ssec:features}) and dark matter halos properties (see also \S\ref{ssec:targets}) are provided. 

In our approach the baryonic quantities are the ``features'', i.e., the input entries of the ML tools, while the dark halo properties are the ``targets'', i.e., the quantities that the ML tools need to predict for each galaxy corresponding to a given entry.

{Below, we present} details of the simulation and the catalog of `features'' and ``targets'' of galaxies extracted from the TNG100.

\subsection{Simulation and galaxy catalog}
\label{ssec:catalog}
The TNG100 simulation is a cosmological, magneto-hydrodynamic simulation using a comoving volume of $106.5 \, {\rm Mpc}$ by side.
The cosmological parameters are set accordingly to Planck 2015~\citep{Planck:2015fie}, i.e. $\Omega_{\Lambda}=0.6911$, $\Omega_{m}=0.3089$, $\Omega_b=0.0486$ and $H_0=67.74\rm\,{km\ s^{-1} Mpc^{-1}}$.

The simulation contains $1820^3$ dark matter particles, $1820^3$ hydrodynamic cells, and $1820^3$ Monte Carlo tracer particles, with a softening length of $0.74\ \rm{kpc}$ for both dark matter and stellar particles and $1.85\ \rm{kpc}$ for gas cells. The mass resolution of dark matter particles is $7.5\times10^6M_{\odot}$, while the mean mass resolution of baryon particles is $1.4\times10^6M_{\odot}$. 
The hydrodynamical part of the simulation includes updated recipes for star formation and evolution, chemical enrichment, cooling and feedbacks \citep{ Weinberger2017,Pillepich2018,Nelson2018}. It also accounts for AGN feedback \citep{Weinberger2017} and galactic winds model \citep{Pillepich2018}, mimicking supernovae feedback.

The simulation evolves from initial conditions, starting at redshift $z=127$, until $z=0$ and $136$ output snapshots are available.
For this work, we use only the $z=0$ snapshot, where the object catalog includes $1,048,574$ substructures.
Among these, for our analysis we select all substructures corresponding to friends-of-friends (FoF) selected SubHalos, having ``galaxies'' with DM mass and DM sizes larger than zero and stellar masses $M_*(z=0)>10^8M_{\odot}/h$. The  number of selected SubHalos in our catalog is $43,379$, corresponding to as many SubHalos, in the TNG. Hereafter we will refer to these objects as ``galaxies''\footnote{Further details on the TNG100 data can be found on \href{https://www.tng-project.org/data/docs/specifications}{https://www.tng-project.org/data/docs/specifications}.}.

\subsection{Targets}
\label{ssec:targets}

We start by defining the targets, i.e., the quantities that we want to predict and that represent the output of the ML algorithm. As clarified before, in this work we are interested in DM properties of galaxies. The TNG100 output delivers separately information on different matter species. In the full catalog, these different matter species are divided in groups denoted by {\tt Type\_i}, where {\tt i}=$0,...,5$ correspond respectively to: gas, DM, tracers, stars + wind particles and black holes\footnote{The {\tt Type\_2} matter species is not used.}. The TNG100 catalog provides us the following DM ({\tt Type\_1}) quantities:

\begin{itemize}
    \item {\bf T1:} Total DM matter. In the TNG100 catalog, this is the parameter {\tt SubhaloMass}. We denote this target as $M_{\rm DM}$, and it is given in units of $10^{10} M_{\sun}/h$;
    \item {\bf T2:} Comoving radius containing half of the DM mass. This corresponds to the parameter {\tt SubhaloHalfmassRad}. This target is denoted as $R_{\rm DM/2}$, and it is given in units of $c$kpc$/h$;
    \item {\bf T3:} DM mass within the stellar half mass radius (i.e. the radius which contains half of the total stellar mass, $R_{\rm */2}$, see also \S\ref{ssec:features}). This corresponds to the parameter {\tt SubhaloMassInHalfRad}. This target is denoted as $M_{\rm DM}(R_{\rm*/2})$, and it is given in units of $10^{10} M_{\sun} /h$;
    \item {\bf T4:} DM mass within twice the stellar half mass radius. This corresponds to the parameter {\tt SubhaloMassInRad}. This target is denoted as $M_{\rm DM}(2R_{\rm*/2})$, and it is given in units of $10^{10} M_{\sun} /h$;
\end{itemize}

\subsection{Features}
\label{ssec:features}

Features are the quantities the ML algorithms utilizes to predict the aforementioned targets. As we are interested in investigating the predictive power of the data collected by imaging and spectroscopic surveys in optical and near infrared, in this work we consider a total of 15 features related to the visible matter within the halos provided within the TNG100 catalog. We divide these features into three groups: {\it Photometric}, {\it Structural} and {\it Kinematic}. The features in each group are the following:

~\\
{\it Photometric:} These are the rest-frame absolute magnitudes in the following eight photometric bands:
\begin{itemize}
    \item {\bf P1:} Johnson-Bessel $U$ band ($\lambda=0.360\ \mu m$);
    \item {\bf P2:} Johnson-Bessel $B$ band ($\lambda=0.435\ \mu m$);
    \item {\bf P3:} Johnson-Bessel $V$ band ($\lambda=0.550\ \mu m$);
    \item {\bf P4:} Johnson-Bessel $K$ band ($\lambda=2.22\ \mu m$);
    \item {\bf P5:} SDSS $g$ band ($\lambda=0.469\ \mu m$);
    \item {\bf P6:} SDSS $r$ band ($\lambda=0.617\ \mu m$);
    \item {\bf P7:} SDSS $i$ band ($\lambda=0.748\ \mu m$);
    \item {\bf P8:} SDSS $z$ band ($\lambda=0.893\ \mu m$);
\end{itemize}
These features are all collected under the parameter {\tt SubhaloStellarPhotometrics\_i}, where {\tt i}=$1,...,8$, according to the list above. We have kept all filters, despite there is some redundancy among the different photometric systems because the ML will consider all of them independently. As one of the aims of this analysis is to establish the significance of the features in making predictions for any given target, this will allow us to consider the significance of each of the observational bands as above at once, with no preferred filter system. We also stress that NIR is only marginally covered by two ``extreme'' filters, SDSS $z$ and Johnson-Bessel $K$.

~\\
{\it Structural:} These features are related to some structural parameters from visible matter mass in the halos. Note that these features are the same quantities as the targets, introduced in \S\ref{ssec:targets}, but defined here for stellar mass, which corresponds to the {\tt Type\_4} 
particle.
\begin{itemize}
    \item {\bf S1:} Total stellar matter. This corresponds to the paramater {\tt SubhaloMass}. This feature is denoted as $M_{*}$, and it is given in units of $10^{10} M_{\sun}/h$;
    \item {\bf S2:} Comoving radius containing half of the stellar mass. This corresponds to the paramater {\tt SubhaloHalfmassRad\_4}. This feature is denoted as $R_{\rm */2}$, and it is given in units of $c$kpc$/h$;
    \item {\bf S3:} Stellar mass within the stellar half mass radius. This corresponds to the paramater {\tt SubhaloMassInHalfRadType\_4}. This feature is denoted as $M_{*}(R_{\rm */2})$, and it is given in units of $10^{10} M_{\sun}/h$;
    \item {\bf S4:} Stellar mass within twice the stellar half mass radius. This corresponds to the paramater {\tt SubhaloMassInRadType\_4}. This feature is denoted as $M_{*}(2R_{\rm */2})$, and it is given in units of $10^{10} M_{\sun}/h$;
\end{itemize}

~\\
{\it Kinematic:} Features related to the line-of-sight velocity distribution:
\begin{itemize}
    \item {\bf K1:} 
    One-dimensional velocity dispersion of all the member particles/cells. This corresponds to the parameter {\tt SubhaloVelDisp}. This feature is denoted by $\sigma_{V}$, and it is given in units of kms$^{-1}$;
    \item {\bf K2:} Maximum value of the spherically averaged rotation curve. This corresponds to the parameter {\tt SubhaloVmax}. This feature is denoted by $\langle V\rangle_{max}$, and it is given in units of kms$^{-1}$;
\end{itemize}

For the {\it Structural} and {\it Kinematic} groups, we take the logarithm (with base 10, ``Log'' hereafter) of all features. 
This is particularly convenient in our analysis because typical scaling relations involving correlations among structural parameters and kinematics with photometry are power laws. Hence, the logarithmic quantities produce generally linear correlations between the features and the targets, which improve the performance of the ML predictions.
The distribution of all features are shown in Appendix~\ref{sec:dist}. We remark here that magnitudes are, by definition, Log quantities, and hence they do not need to be transformed.

\subsubsection{Considerations about the correspondence of features and observations}
\label{sssec:considerations}

When defining the features above, we have not discussed the accuracy of these parameters derived in the Illustris simulations as compared to the ones measured in real galaxies. There are other analyses dedicated to this specific aspects, where simulated properties of galaxies like galaxy light profiles, profile of colors (see e.g.,~\citealt{2020A&A...641A..60P}), star formation rates (e.g.,~\cite{Pillepich:2017jle}), and even kinematics (see e.g., \citealt{SAMI_3D_simulations_vdeSande19}) have been compared to observations. As discussed in some of these analyses, there are yet unresolved problems in the match between observations and simulations: ($i$) simulations are still limited in resolution for large-enough cosmological volumes (\citealt{EAGLE_res_effect}); and ($ii$) realistic observational-like quantities from simulations are still unsatisfactory (see e.g. \citealt{Obs_sim_compar_guidi15,SAMI_3D_simulations_vdeSande19}). However, these are problems that will be possibly overcome in a near future\footnote{See e.g. the FIRE project: https://fire.northwestern.edu/}. Here, we need to remark that the present analysis is not meant to make predictions of DM in real galaxies, hence the ``observational realism'' of the quantities used as features is not an issue.
Training ML tools in more realistic simulations, will be a matter of future analyses. 

In this paper, we want to address the following questions:
\begin{itemize}
    \item[$i$)] Is it possible to make predictions of the dark component of galaxies, starting from simple observational parameters?
    \item[$ii$)] If so, what are the best DM parameters we can infer with these data?
\end{itemize}

To do that we are motivated to use observational-like parameters provided by hydrodynamical simulations that carry the physical information contained in real observations, even if not fully observational compliant. In particular, the {\it Photometric} features described above are similar to basic parameters provided by imaging surveys (e.g., SDSS:~\citealt{SDSS:2000hjo}; KiDS:~\citealt{deJong+15_KiDS_paperI}) and will be a fundamental part of the galaxy parameters provided by future large sky surveys (e.g. Rubin/LSST: \citealt{LSST-cit}; CSST: \citealt{CSST-cit}; Euclid: \citealt{Euclid-cit}).

The {\it Kinematical} features, like the central velocity dispersion of galaxies, are standard parameters provided by current (e.g., SDSS:~\citealt{2012The}; LAMOST:~\citealt{Napolitano2020_lamost}) and future (e.g., WEAVE: \citealt{Costantin+19_steps}; 4MOST: \citealt{deJongR+11_4MOST}; DESI: \citealt{2016arXiv161100036D}) spectroscopic surveys. In this respect the 1D velocity dispersion provided by TNG100, is a fair approximation, as it contains the information on the random motions of galaxies despite it is not limited to some central aperture\footnote{Note that in the plausible scenario of almost constant velocity dispersion profiles, the aperture has a minimal effect on the overall 1D estimate, with respect to other effects like the gravitational softening that might strongly impact stellar orbits around the galaxy centers \citep{10.1093/mnras/stz309}.}. 
Similar arguments can be used for $\langle V\rangle_{max}$, which is also a crude approximation of the observed galaxy rotation, although, as for $\sigma_V$ they show correlations with Photometry parameters that mirror the observed scaling relations (see \S\ref{ssec:corrmatrix}).

{\it Structural} features like the half-mass radius, are also derived from photometric surveys (e.g., SDSS: \citealt{Shen+03}; \citealt{Baldry2012}. \citealt{2003The}; KiDS:~\citealt{Roy2018}) and will be provided in future imaging surveys, while stellar masses are obtained either from multi-band photometry (e.g., \citealt{Maraston+13_BOSS}; \citealt{KV450}) or from spectroscopy (e.g., \citealt{Kauffmann+03}; \citealt{ThomasJ+11}). In this respect the stellar masses provided by the TNG100, like for the Photometry paraneters above, provide a fair representation of the quantities derived from observations.    

With our ML approach, we can test whether the use of a combination of all these data can give some accurate predictions of the dark matter content of galaxies. However, this will also give the chance to test whether there is a minimal set of parameters (e.g., photometry only or spectroscopy only or a combination of these parameters with structural parameters), which can also give accurate enough predictions. This will be a crucial step to optimize the number of observable parameters to collect for the predictive samples, i.e., real galaxies from specific surveys, to be used for making DM estimates using the ML tools we want to test in the following sections.

\subsection{Correlation Matrix}
\label{ssec:corrmatrix}

When we are faced with a statistical learning problem, either interpolation or classification, it is first of all appropriate to see the sharing of information within the features that will be used during the prediction, as well as between the features and the targets of the prediction, when the learning is supervised.
In order to measure the linear dependence between two sets of data, we used the \textit{Pearson correlation coefficient}, defined as the ratio between the covariance of the two datasets, and the product of their standard deviations. Constructed in this way, this coefficient always shows up with value between -1 and 1, pointing to the maximum of linear correlation when $\rho=1$ or $\rho=-1$, and to the absence of linear correlation when $\rho=0$. Note that, since Person coefficient computes linear correlation, it works only as a first indicator, and it does not replace the full analysis.

\begin{figure*}
	\includegraphics[width=0.75\textwidth]{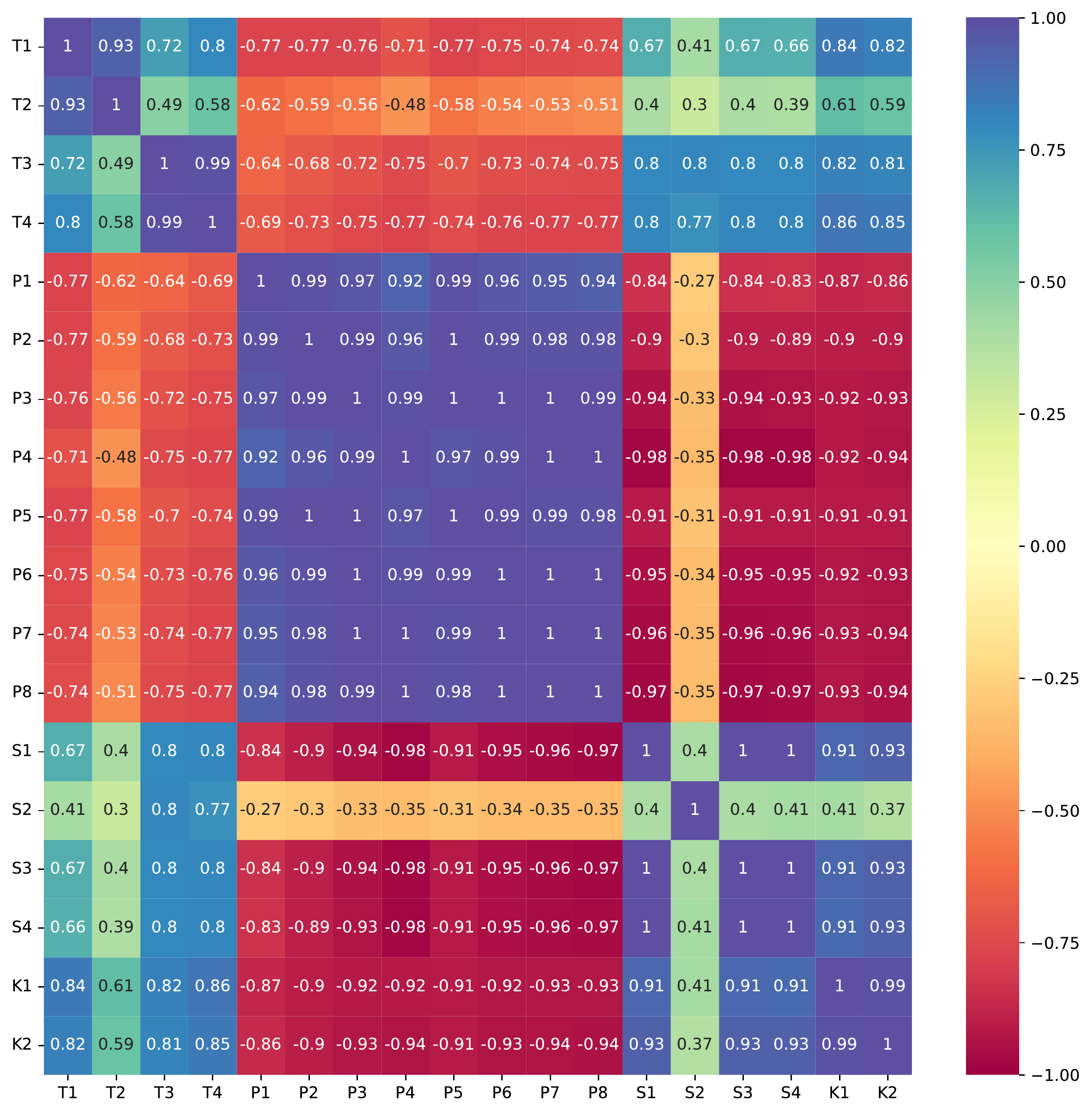}
	\caption{Correlation matrix for all features and targets.
    The colour associated with each element of the matrix is based on the value of the showed Pearson coefficient calculated for all combinations of target and features taken in pairs.
	The entries are organized in the following sequence: the 4 targets (denoted by T*), the rest-frame absolute magnitudes for the 8 photometric bands (denoted by P*), the 4 structural features (denoted by S*) and the 2 kinematical features (denoted by K*). See \S~\ref{ssec:features} and \S~\ref{ssec:targets} for more details.}
    \label{fig:corr}
\end{figure*}

In Fig.~\ref{fig:corr}, we start by analysing the correlation matrix for the targets and the features we use. Here, targets are labeled with a T* index, features of photometric type are labeled as P*, features of structural type are labeled as S* and kinematical features are labeled as K*, as defined in \S\ref{ssec:targets} and \S\ref{ssec:features}. The elements of the correlation matrix are the Pearson correlation coefficient computed for all combinations of target and features taken in pairs. The correlation matrix is also colour-coded according to the value of the Pearson coefficient. In this figure, five groups of correlation are clearly distinguishable (in purple), they are T1-T2, T3-T4, P1-P2-P3-P4-P5-P6-P7-P8, S2, S1-S3-S4-K1-K2. .

In the target domain, T1 is tightly correlated with T2, while T3 is strongly correlated with T4, but these two groups are mildly correlated between them. This is quite surprising, as it is expected, that total DM matter (T1) should be correlated to the mass inside some given radii (e.g., DM mass within the stellar half mass radius, T3), as the larger is the total mass, the larger should be the mass at all radii. On the other hand, the simulations seem to show that there is a more tight correlation between the total DM and the half DM radius, denoting the presence of a size-mass scaling relation for the DM component. We also note that, since there is a weak correlation between this latter with the comoving radius containing half of the stellar mass (S1), we can interpret the weaker correlation between T1 and both T3 and T4,
as a consequence of the weak correlation between the dark (T2) and stellar (S2) half mass radii. This latter is interesting as it may reflect the variance due to the combination of different galaxy populations, e.g. early and late-type galaxies (see e.g. \citealt{2017ApJ...838....6H,2020MNRAS.492.1671Z}), which we have not tried to separate in this work, but will investigate in the next paper (Wu et al., in preparation).

The group of the {\it Photometric} features presents strong positive correlation between them, which is obvious, as they are all connected via galaxy spectral energy distribution (SED), defining their stellar populations (see e.g. \citealt{BC03}; \citealt{Vazdekis+12}).
For the same reason, we see a strong anti-correlation between the photometry features and stellar masses\footnote{Note that the sign of the correlation between masses and magnitude is always reversed because more massive galaxies are also more luminous, which means that they have more negative magnitudes, by definition.} S3 and S4, which encodes the stellar mass-to-light ratios of the stellar populations. On the other hand, the correlation between photometry and the stellar half mass radius is weaker ($\sim-0.3$), but similar in magnitude to the one between the stellar masses and the same S2 ($\sim 0.4$). This means that the well known size-luminosity and the size-mass relations (see e.g., \citealt{Baldry2012}, \citealt{Roy2018}) are reproduced by the simulations, but their correlations are less significant than other ones. One reason for a poorer correlation size-mass correlations is that these have different slopes in differnt galaxy types. Since we have not separate early-type from late-type systems in the current analysis, then the overall correlation can be weakened (similarly to what commented for the dark and luminous radii).  
Finally, the kinematic features are strongly correlated between them, and are tightly positively correlated with masses (S1, S3 and S4) and photometry. This means that simulations can reproduce other classical scaling relations for galaxies, correlating luminosity (and masses) with the rotation velocity of late-type galaxies (the so-called Tully-Fisher relation, see e.g., ref ; also in its baryonic form, see e.g., \citealt{McGaugh12,McGaugh_Schombert15}) or luminosity and velocity dispersion (the so called Faber-Jackson relation, see e.g., \citealt{FJ76,HB09_curv}). We stress here that we are not interested to verify that the simulations are correctly reproducing the observed relations, as this would imply to have a full set of realistic mock observations of galaxy properties, that will be addressed in future analyses. 
Here we want to check whether, given some observables defined in the simulations and correctly reproducing the existence of correlations in real galaxies, it is possible to make the right predictions of the unknown DM properties of galaxies, consistently. 

We conclude this first visual inspection of the correlations among features and targets, by noticing that no features show a very tight correlation with targets. If such a tight correlation would exist, then one could use this single correlation to make direct predictions of the DM in galaxies. On the other hand, we expect that combining the information from different features, a ML algorithm can ``interpolate'' these correlations and obtain accurate value for the different targets.

In order to assess the relevance of each group in the prediction of the DM features, we will perform four analyses. First, we analyze how each group individually can be used for predicting DM features, and finally, we combine all groups in a joint analysis. 

\section{Method}
\label{sec:method}

In this work, we used supervised learning approach, a common branch of ML (\citealt{aaron}). In this learning category, we aim to build a numerical function that allows us to associate objects in the $X$ domain (features) to objects in the $Y$ domain (targets) $ f: X \to Y $, which can be seen as an approximation to a possible underling exact solution $\tilde{f}$. In other words, we seek to explain $Y$ with $X$, mapping inputs to outputs.

\subsection{Training and Test samples}
\label{ssec:tandt}

To build this approximate function, we use only a part of data sample called \textit{training set}. The best approximation for $f$ is achieved through the minimization of a cost function. The remaining sample, called \textit{test set}, is used to measure the performance, comparing the \textit{true} (\textit{simulated} in our case) values  with the predicted ones. 
The final goal is to optimize the model which performs equally well on the training and the test samples (i.e. a model that generalizes well). In order to avoid a bias due to the choice of the \textit{training set} and the \textit{test set}, a cross-validation resampling procedure is performed to evaluate the model. Such a model is expected to make predictions on other data that should be well represented by the test set.

The generalization is the property that differentiates an ML approach from a conventional optimization solution, typically using the whole dataset indistinguishably.

The \textit{training set}, denoted here by $A$, consists of an n-upla $(\textbf{x}_{i},\textbf{y}_{i})$ (input, output) as vector $\textbf{x}_i$ and $\textbf{y}_i$ belonging to the vector space $X$ and $Y$ respectively:
\begin{eqnarray}
A = ((\textbf{x}_{1}, \textbf{y}_{1} ), (\textbf{x}_{2}, \textbf{y}_{2} ), ...,(\textbf{x}_{n}, \textbf{y}_{n} )) .
\end{eqnarray}
with $X$ $\in$ $\mathbb{R}^c$ and $Y$ $\in$ $\mathbb{R}^d$. For regression tasks, the elements of $\textbf{y}_i$ belong to reals. For classification problems, the elements of $\textbf{y}_i$ are integers, but it is not our case. 

In this work, $X$ is represented by all the characteristics of the light sector, and $Y$ is represented by the target of the dark sector, and together form our dataset, that we split randomly in $80\%$ for the training sample and $20\%$ for the test sample.

\subsection{Pipeline}
\label{ssec:pipeline}

There are a plethora of ML methods that can be used to solve regression problems (\citealt{hastie2009elements}), then:
 $i)$  \textit{how to find, if any, the best ML method for a given problem?;}
 $ii)$ \textit{what are the best feature engineering operations to be applied before the ML analysis?}
 
Here, in order to explore and evaluate the full space of possibilities, we decide to use a python automated machine learning tool called Tree-based Pipeline Optimization Tool~(\citealt{OlsonGECCO2016}, \texttt{TPOT} in short).
It is worth mentioning that, more than a single regression method, \texttt{TPOT} returns the "best" pipeline, which can be made of a single method or a combination of different algorithms. 
A more detailed description of the \texttt{TPOT} analysis performed in this paper is found in the Appendix~\ref{sec:tpotperf} and the best pipelines reported in Tab.~\ref{tab:pipelines}.

\subsection{Performance metrics}
\label{ssec:metrics}

As anticipated, in this work, we focus on the accuracy that ML tools can reach in predicting some specific DM properties of galaxies. As we will use the same simulated galaxies as a test case, this experiment is rather ideal, and the accuracy we find has to be considered an upper limit for future applications to real galaxies.
In particular, this test can give strong indications on the potential of the full dataset, made of photometry, spectroscopy and structural parameters, or even a subset of these data. As described in \S\ref{sec:data}, these provide a 0-order approximation of the typical dataset expected from future surveys.

To quantify how accurate the predicted values, issued by the pipelines presented in Tab.~\ref{tab:pipelines}, are as compared to the ``ground truth'' values form the test sample, we make use of some well-established statistical metrics. These metrics are also used to prevent  underfitting/overfitting in the analyses. 

The first statistical metric we use is the \textit{coefficient of determination} $R^2$, which is defined by,
\begin{eqnarray} \label{r2}
R^2=1-\frac{\sum(\textbf{y}_i-\hat{\textbf{y}}_i)^2}{\sum(\textbf{y}_i-\bar{\textbf{y}})^2}\,,
\end{eqnarray}
where $\textbf{y}_i$ denotes the data points, $\hat{\textbf{y}}_i$ denotes the predicted points and $\bar{\textbf{y}}$ is the mean value of the dataset. Note that Eq.~\ref{r2} can be interpreted as a standardized version of the sum of the Mean Squared Error (MSE). The closer $R^2$ is to 1 the better is the pipeline. For the perfect case, where $R^2=1$, the model fits perfectly to the data, which corresponds to MAE $=$ MSE $= 0$. Furthermore, in regression analysis evaluation, $R^2$ was shown to be more informative and reliable than other metrics \cite{metrics}.

The second statistic estimator is the \textit{Pearson correlation coefficient}, previously introduced and used for the correlation matrix (see S~\ref{ssec:corrmatrix}),  that, for the purpose of evaluating the accuracy of the prediction, is similarly defined as
\begin{eqnarray} \label{pearson}
\rho = \frac{cov(\hat{\textbf{y}}, \> \textbf{y})}{\sigma_{\hat{\textbf{y}}} \sigma_{\textbf{y}}}\,,
\end{eqnarray}

Here, we are interested in measuring the correlation between test data $\textbf{y}_{i}$ and the relative predicted values $\hat{\textbf{y}}_{i}$ associated with a same \textit{target}.

The results for the aforementioned statistical metrics are presented and discussed in \S~\ref{ssec:results}. Additionally, for the sake of completeness, we also present the Mean Absolute Error (MAE) and the MSE, which are respectively defined by,
\begin{eqnarray}
{\rm MAE} &=& \dfrac{\sum |\hat{\textbf{y}}_i-\textbf{y}_i|}{N} \,, \\
{\rm MSE} &=& \dfrac{\sum \left(\hat{\textbf{y}}_i-\textbf{y}_i\right)^2}{N} \,,
\end{eqnarray}
where $N$ is the total number of data points.
The MAE gives a measure of the deviation of the predicted values from the true ones, and hence it is expected to be relatively small for accurate predictions. The MSE is a measure of the scatter of the prediction around the true values, and hence this is smaller for more precise predictions.

In general, if on one hand $R^2$ and $\rho$ measure the strict correlation between predicted and true values, the MAE and MSE measure the presence or not of systematic offset and the overall uncertainties of the predicted values.

\section{Analyses}
\label{sec:analyses}

In this section, we show the results about predictions from the ML pipelines selected by {\tt TPOT} (see \S\ref{ssec:pipeline} and Appendix \ref{sec:tpotperf}) have returned for the four targets, and discuss their accuracy and scatter. This allows us to draw conclusions about the viability of the ML methods to derive DM properties of galaxies, starting from the observational data of their luminous matter only. 
For each of the four targets, we perform predictions with four distinct input data as following:
\begin{itemize}
    \item[$i$)] Considering only the {\it Photometric} features;
    \item[$ii$)] Considering only {\it Structural} features;
    \item[$iii$)] Considering only {\it Kinematic} features;
    \item[$iv$)] Combining all features in a joint analysis.
\end{itemize}

This makes a total of 16 outputs, four predictions for the four targets. The predictions from the last of the listed analyses (iv) represents the main result of the paper, but the analyses i), ii) and iii) are useful to understand the predictive ability of the different kinds of observations separately.

For the joint analyses, we also present the importance of the subgroups of features in the regressions for all targets.
\begin{table*}
\centering
\caption{Table with the results of the statistical metrics for all analyses. We show the results for $R^{2}$ statistics, Pearson statistics, MAE and MSE, being all computed for both the training set and the test set. The results of the Joint analyses are highlighted in bold. See \S\ref{ssec:metrics} for more details.}
\label{tab:statistics}
\begin{tabular}{lcccccccc}
\hline\hline
\multicolumn{1}{c|}{\multirow{2}{*}{\textbf{FEATURES}}} & \multicolumn{2}{c}{$R^{2}$ statistics} & \multicolumn{2}{c}{Pearson statistics} & \multicolumn{2}{c}{MAE} & \multicolumn{2}{c}{MSE} \\ \cline{2-9} 
\multicolumn{1}{c|}{} & train & test & train & test & train & test & train & test \\ \hline
\multicolumn{9}{c}{\textbf{TARGET 1: Total DM matter}}  \\ \hline
\multicolumn{1}{l|}{Photometric}  & $0.916$ & $0.855$ & $0.958$ & $0.925$ & $0.140$ & $0.191$ & $0.043$ & $0.075$ \\
\multicolumn{1}{l|}{Structural}      & $0.723$ & $0.646$ & $0.851$ & $0.804$ & $0.660$ & $0.758$ & $0.747$ & $0.965$ \\
\multicolumn{1}{l|}{Kinematic}    & $0.823$ & $0.794$ & $0.907$ & $0.891$ & $0.210$ & $0.230$ & $0.090$ & $0.106$ \\
\multicolumn{1}{l|}{\textbf{Joint analysis}} & $\textbf{0.983}$ & $\textbf{0.917}$ & $\textbf{0.992}$ & $\textbf{0.958}$ & $\textbf{0.057}$ & $\textbf{0.131}$ & $\textbf{0.009}$ & $\textbf{0.043}$ \\ \hline
\multicolumn{9}{c}{\textbf{TARGET 2: Comoving radius containing half of the DM mass}} \\ \hline
\multicolumn{1}{l|}{Photometric}  & $0.831$ & $0.748$ & $0.915$ & $0.866$ & $0.129$ & $0.159$ & $0.033$ & $0.049$ \\
\multicolumn{1}{l|}{Structural}      & $0.565$ & $0.421$ & $0.755$ & $0.650$ & $0.227$ & $0.264$ & $0.085$ & $0.112$ \\
\multicolumn{1}{l|}{Kinematic}    & $0.613$ & $0.535$ & $0.784$ & $0.732$ & $0.201$ & $0.220$ & $0.075$ & $0.090$ \\
\multicolumn{1}{l|}{\textbf{Joint analysis}} & $\textbf{0.926}$ & $\textbf{0.865}$ & $\textbf{0.964}$ & $\textbf{0.930}$ & $\textbf{0.080}$ & $\textbf{0.111}$ & $\textbf{0.014}$ & $\textbf{0.026}$ \\ \hline
\multicolumn{9}{c}{\textbf{TARGET 3: DM mass within the stellar half mass radius}} \\ \hline
\multicolumn{1}{l|}{Photometric}  & $0.946$ & $0.883$ & $0.974$ & $0.940$ & $0.075$ & $0.114$ & $0.012$ & $0.026$ \\
\multicolumn{1}{l|}{Structural}      & $0.941$ & $0.937$ & $0.970$ & $0.968$ & $0.164$ & $0.180$ & $0.069$ & $0.075$ \\
\multicolumn{1}{l|}{Kinematic}    & $0.804$ & $0.765$ & $0.897$ & $0.875$ & $0.146$ & $0.164$ & $0.043$ & $0.053$ \\
\multicolumn{1}{l|}{\textbf{Joint analysis}} & $\textbf{0.986}$ & $\textbf{0.981}$ & $\textbf{0.993}$ & $\textbf{0.990}$ & $\textbf{0.037}$ & $\textbf{0.037}$ & $\textbf{0.003}$ & $\textbf{0.004}$ \\ \hline
\multicolumn{9}{c}{\textbf{TARGET 4: DM mass within twice the stellar half mass radius}}  \\ \hline
\multicolumn{1}{l|}{Photometric}  & $0.939$ & $0.899$ & $0.970$ & $0.948$ & $0.078$ & $0.107$ & $0.013$ & $0.023$ \\
\multicolumn{1}{l|}{Structural}      & $0.932$ & $0.915$ & $0.965$ & $0.957$ & $0.073$ & $0.088$ & $0.015$ & $0.019$ \\
\multicolumn{1}{l|}{Kinematic}    & $0.858$ & $0.821$ & $0.927$ & $0.906$ & $0.124$ & $0.144$ & $0.031$ & $0.040$ \\
\multicolumn{1}{l|}{\textbf{Joint analysis}} & $\textbf{0.993}$ & $\textbf{0.987}$ & $\textbf{0.997}$ & $\textbf{0.994}$ & $\textbf{0.022}$ & $\textbf{0.033}$ & $\textbf{0.002}$ & $\textbf{0.003}$ \\ \hline\hline
\end{tabular}
\end{table*}

\subsection{Results}
\label{ssec:results}

As discussed in \S\ref{ssec:tandt}, during the {\tt TPOT} process, the dataset is divided into two samples: \textit{training} and \textit{test} samples. Whereas the \textit{training sample}, which consist of 80\% of the total sample, is the data used by {\tt TPOT} to determine the best pipeline, the \textit{test sample} represents the data we use for making the predictions and assessing the accuracy of the pipeline found. This latter task is made by applying the performance metrics, discussed in \S~\ref{ssec:metrics}, to the \textit{test sample}: the better are the metrics, the better is the model. It is also important to evaluate the performance metrics for the \textit{training set} because they can be used to assess overfitting. Overfitting occurs when the metrics computed with the \textit{test sample} are significantly worse than the ones obtained with the \textit{training sample}. In this case, even though the model is able to fit part of the total data, it is considered unrealistic because the fitted pipeline cannot predict properly the data that were not used in the learning process. For a model with physical meaning, it is expected that the performances metrics computed with the \textit{test set} do not significantly differ from the values obtained with \textit{training set}.
\begin{figure*}
	\includegraphics[width=\columnwidth]{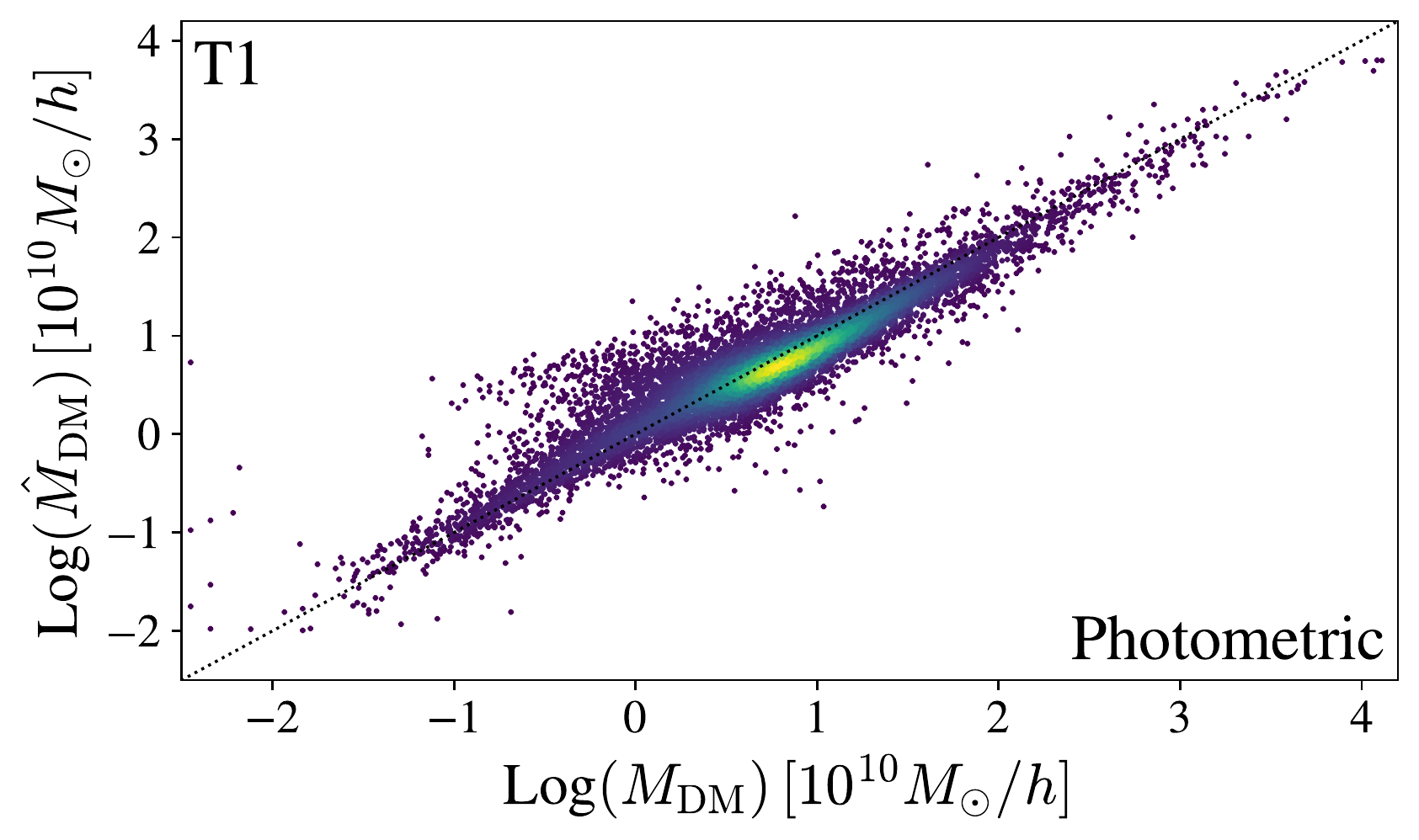}
	\includegraphics[width=\columnwidth]{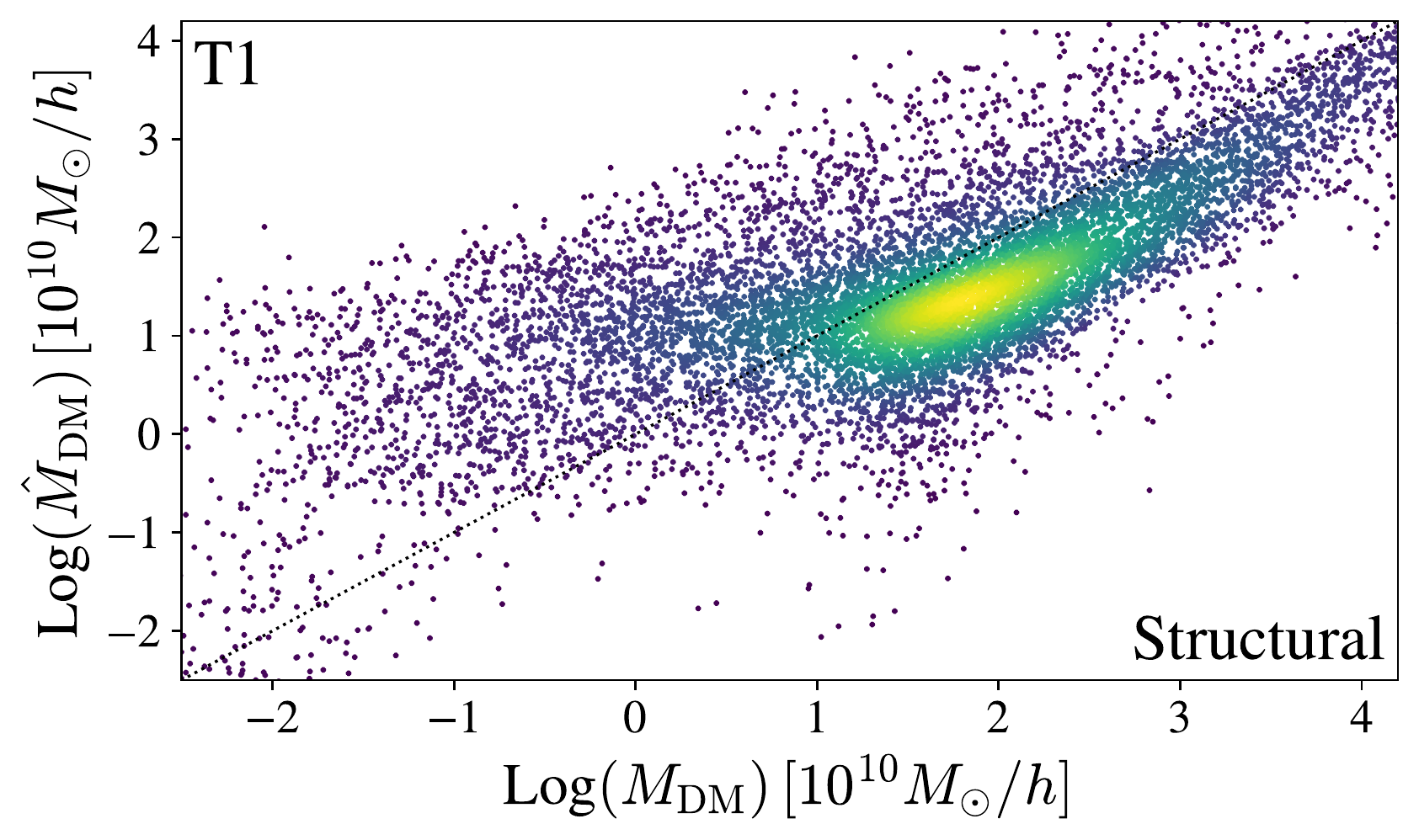}
	\includegraphics[width=\columnwidth]{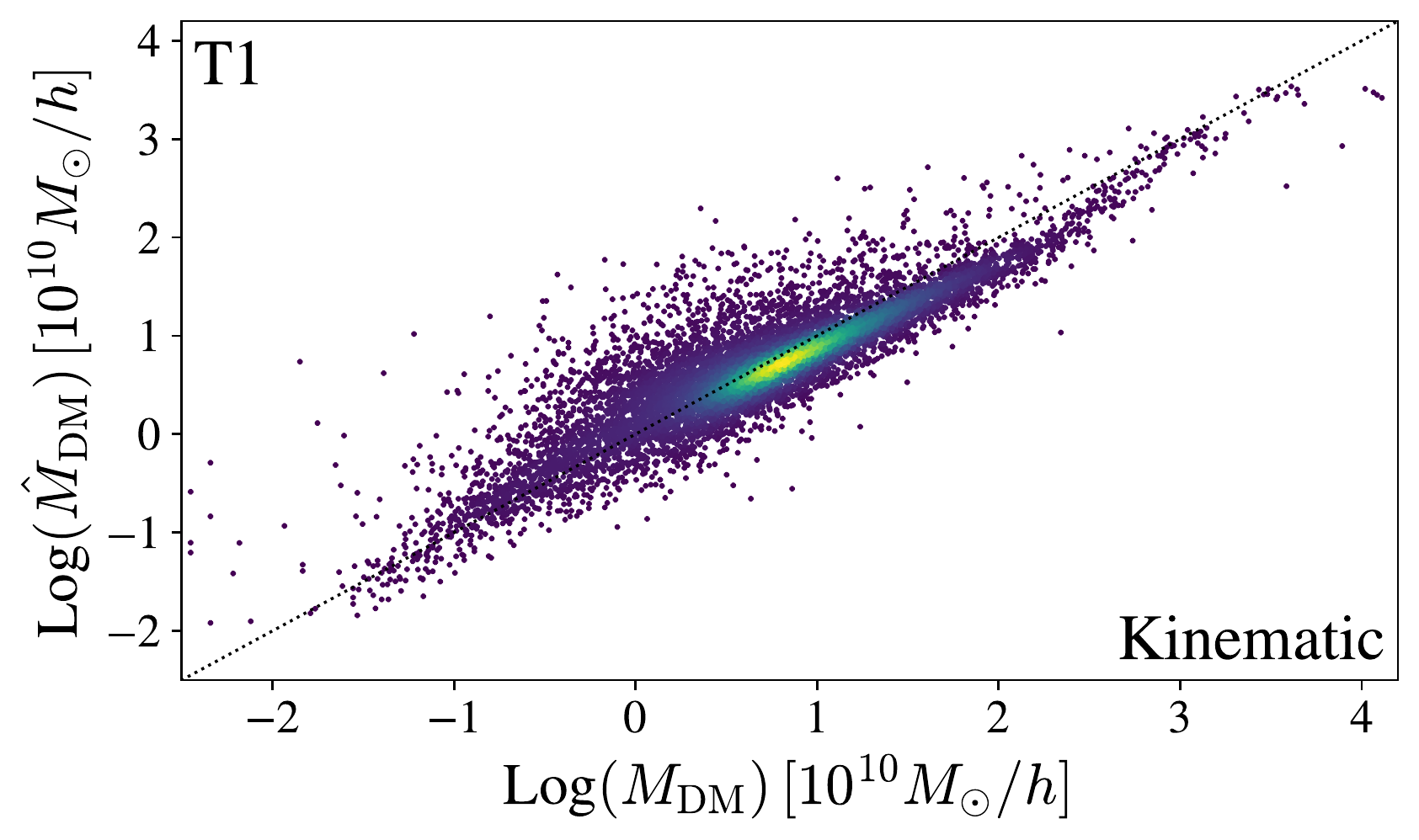}
	\includegraphics[width=\columnwidth]{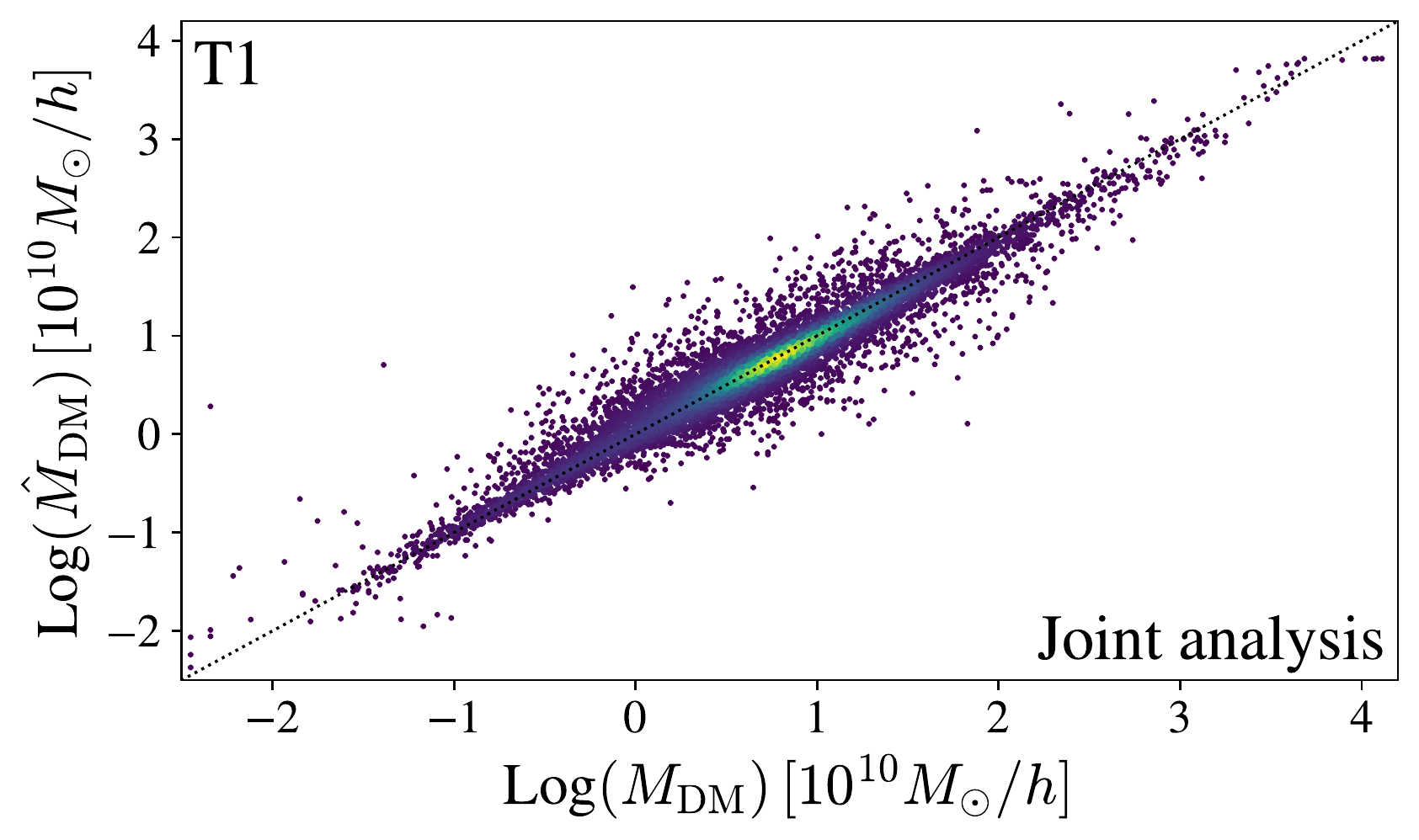}
    \caption{Comparison between the real values of T1 (Total DM matter, given in units of $10^{10} M_{\sun}/h$), denoted by $M_{\rm DM}$, and their respective predictions, denoted by $\hat{M}_{\rm DM}$. The figures are color-coded according to the point density. \textbf{Top-left panel: }Photometric features only. \textbf{Top-right panel: }Structural features only. \textbf{Bottom-left panel: }Kinematic features only. \textbf{Bottom-right panel: }Joint analysis.}
    \label{fig:t1}
\end{figure*}
\begin{figure*}
	\includegraphics[width=\columnwidth]{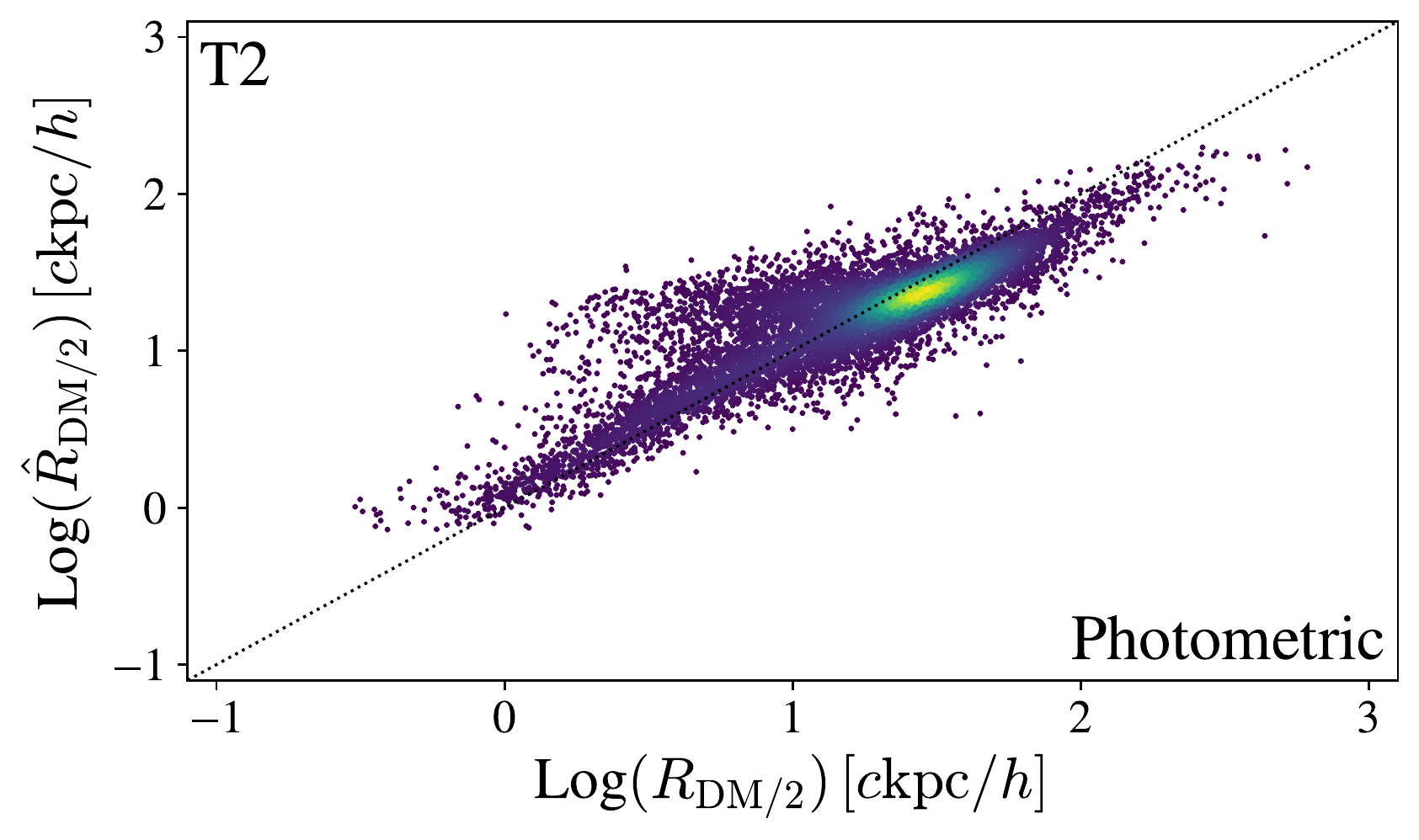}
	\includegraphics[width=\columnwidth]{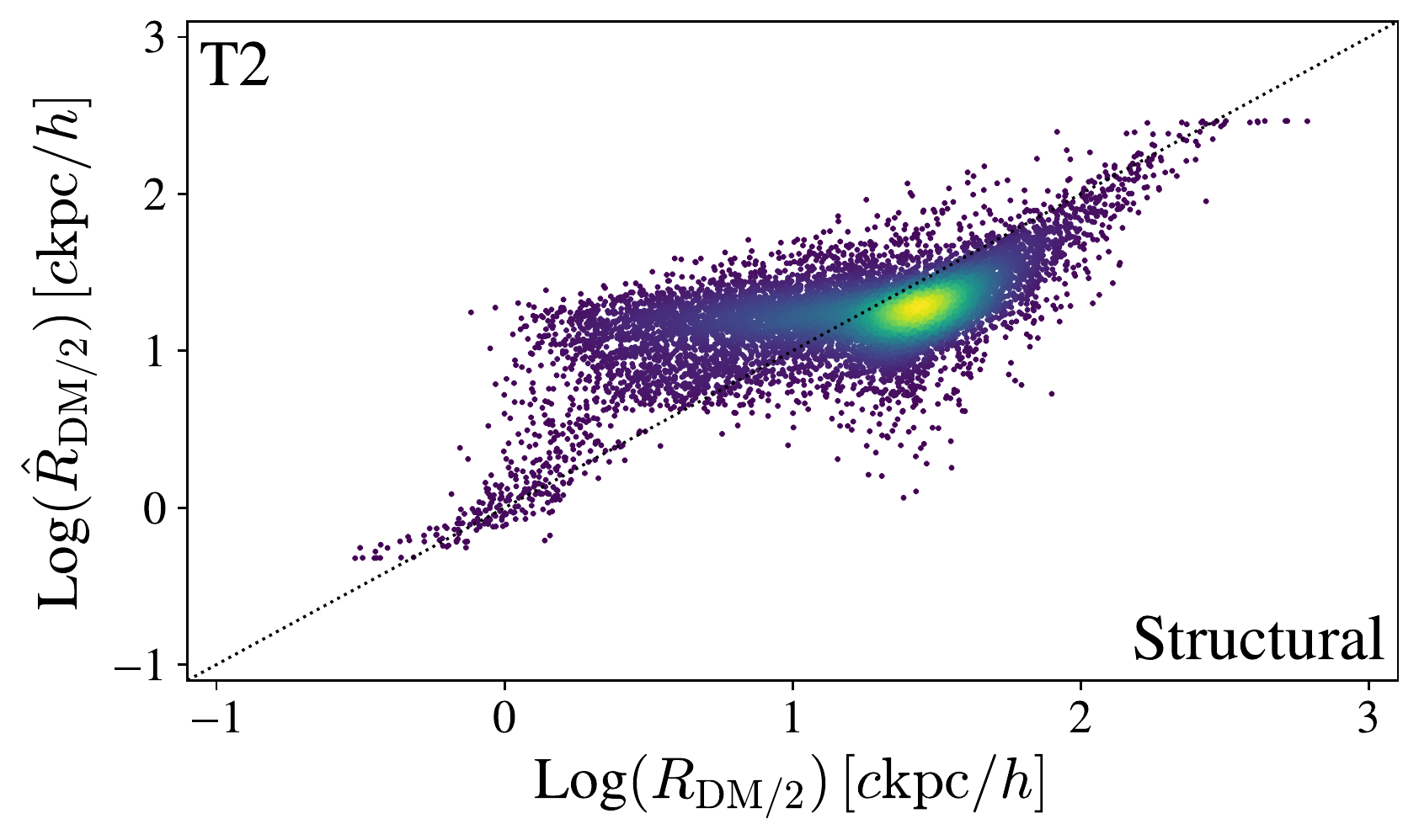}
	\includegraphics[width=\columnwidth]{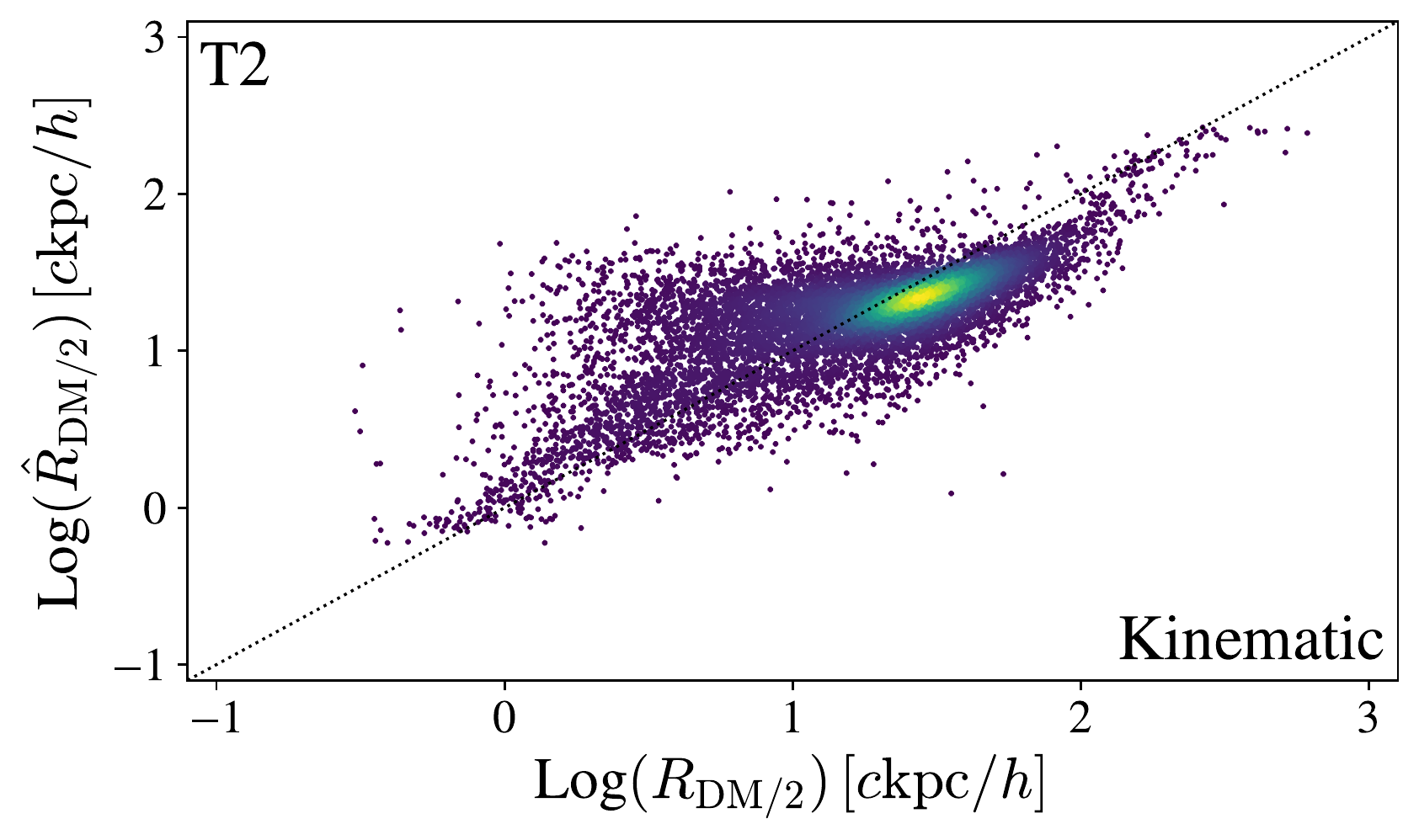}
	\includegraphics[width=\columnwidth]{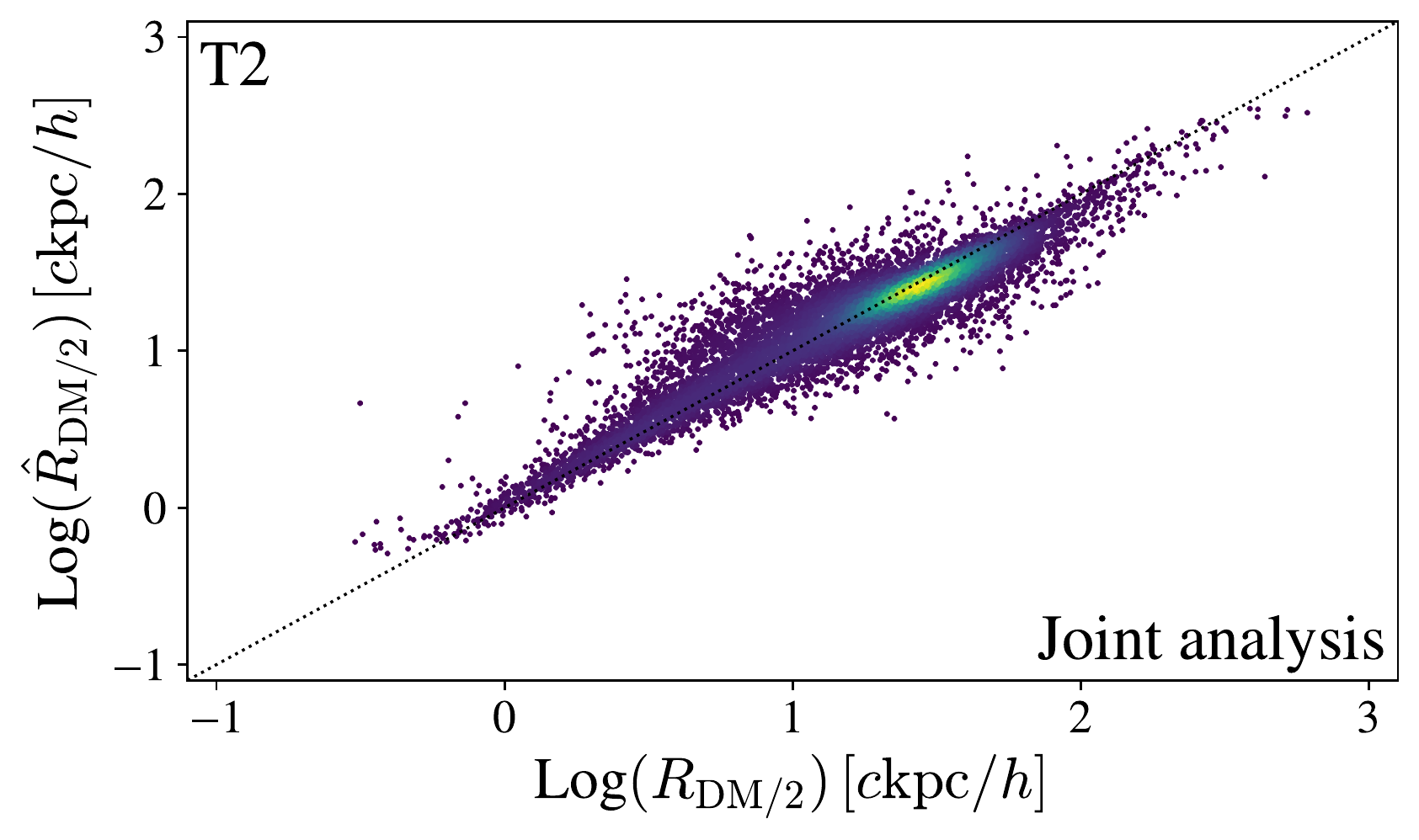}
    \caption{Similar to Fig.~\ref{fig:t1}, but for T2 (Comoving radius containing half of the DM mass, given in units of $c$kpc$/h$), denoted by $R_{\rm DM/2}$, and their respective predictions, denoted by $\hat{R}_{\rm DM/2}$. \textbf{Top-left panel: }Photometric features only. \textbf{Top-right panel: }Structural features only. \textbf{Bottom-left panel: }Kinematic features only. \textbf{Bottom-right panel: }Joint analysis.}
    \label{fig:t2}
\end{figure*}
\begin{figure*}
	\includegraphics[width=\columnwidth]{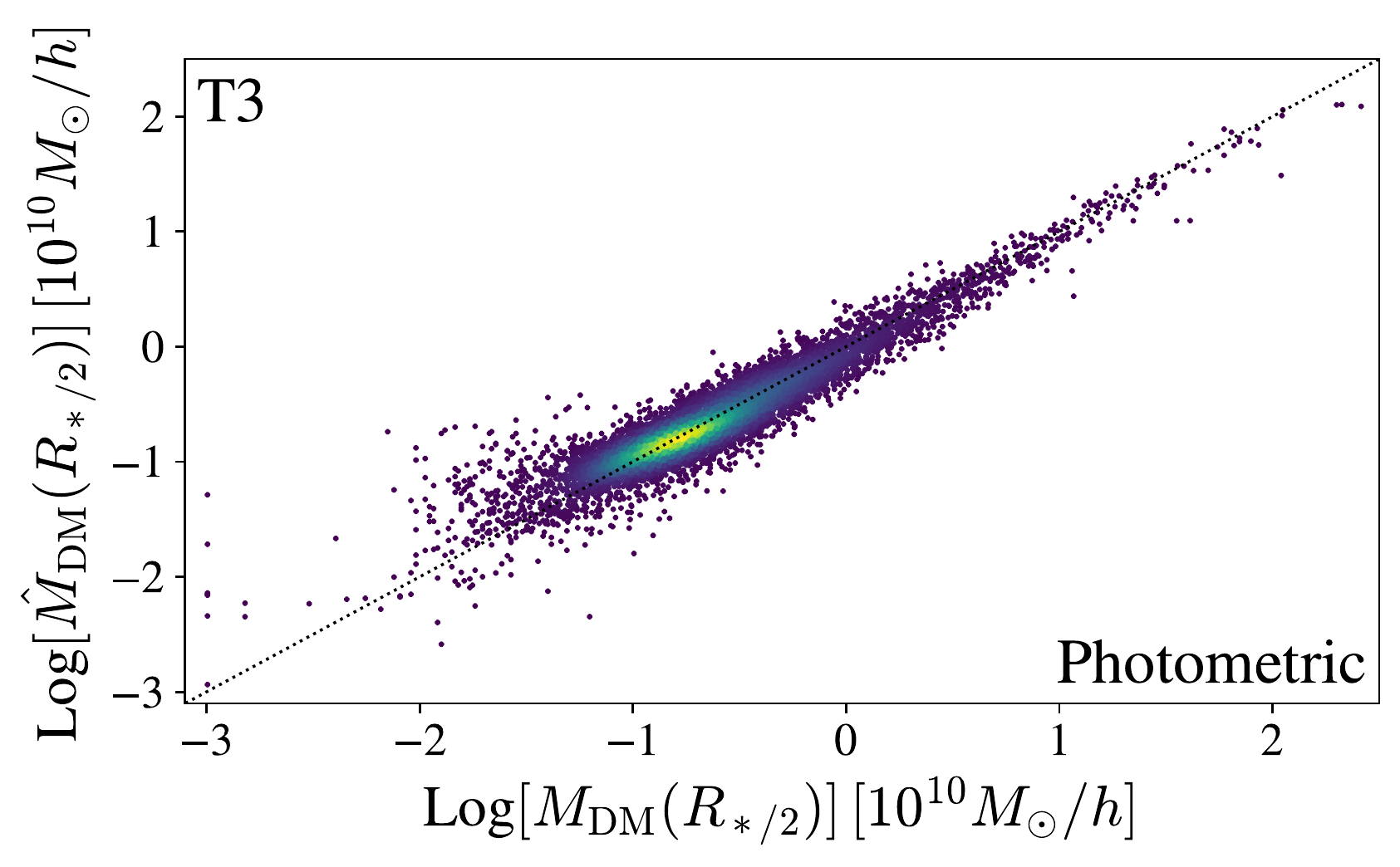}
	\includegraphics[width=\columnwidth]{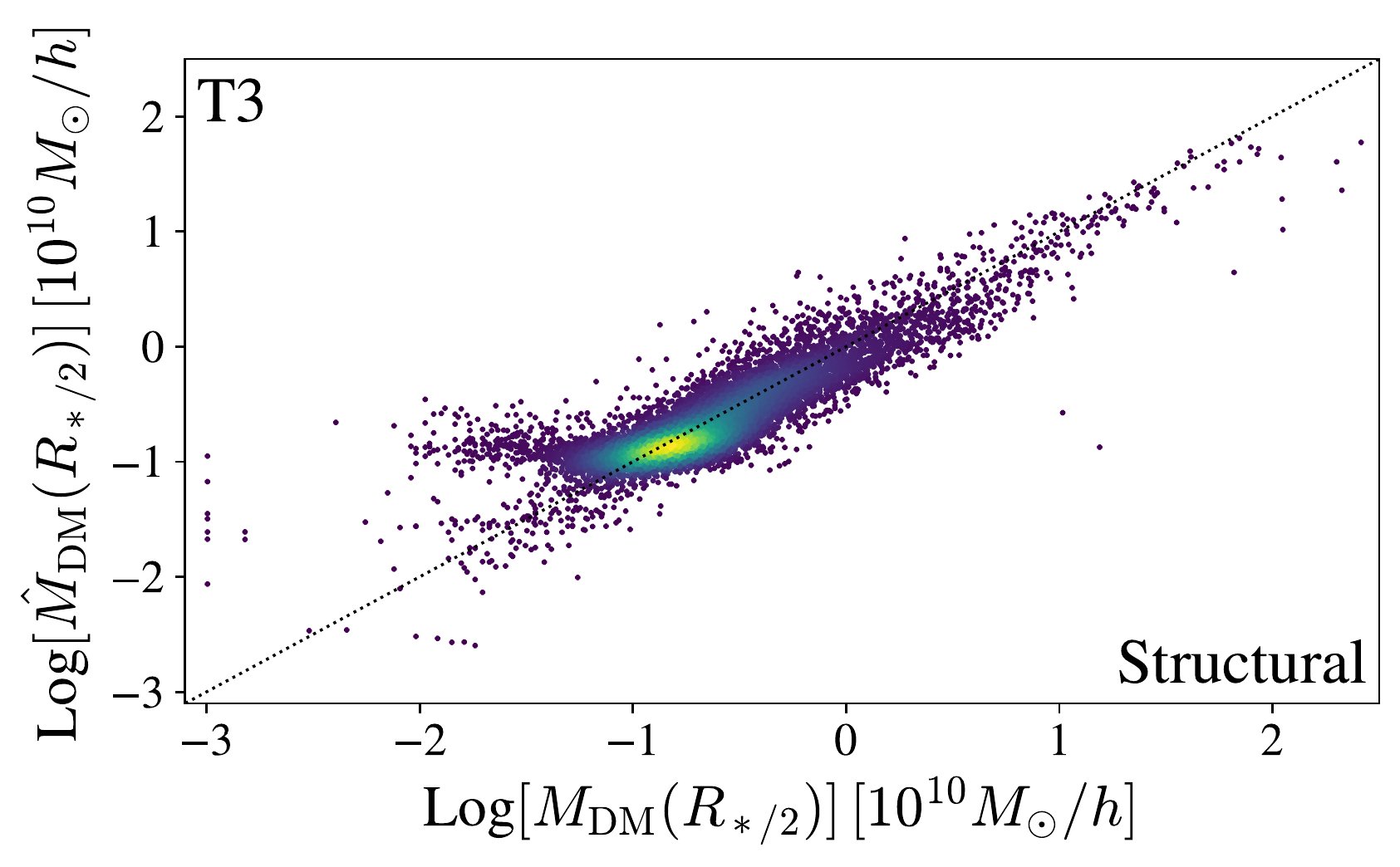}
	\includegraphics[width=\columnwidth]{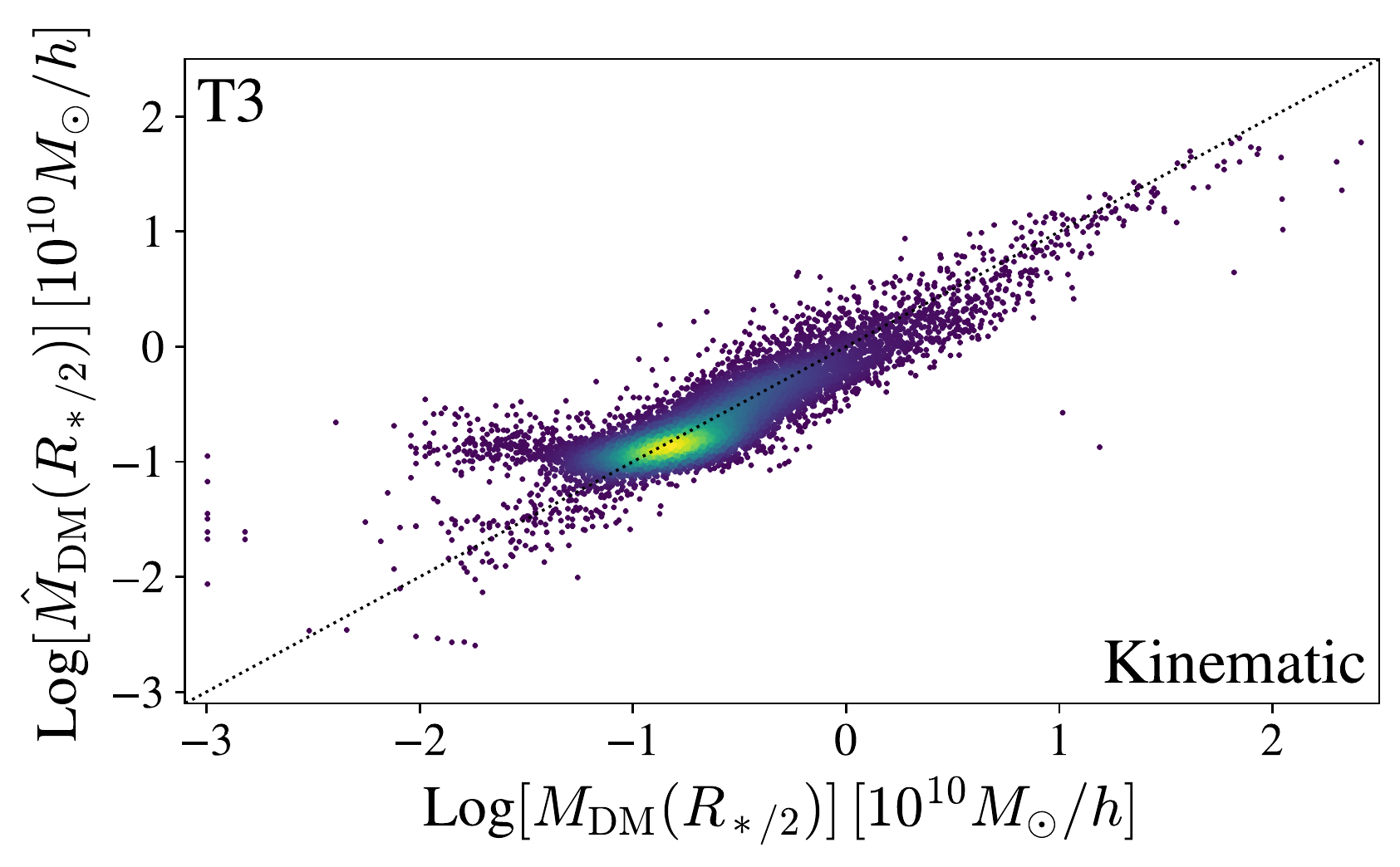}
	\includegraphics[width=\columnwidth]{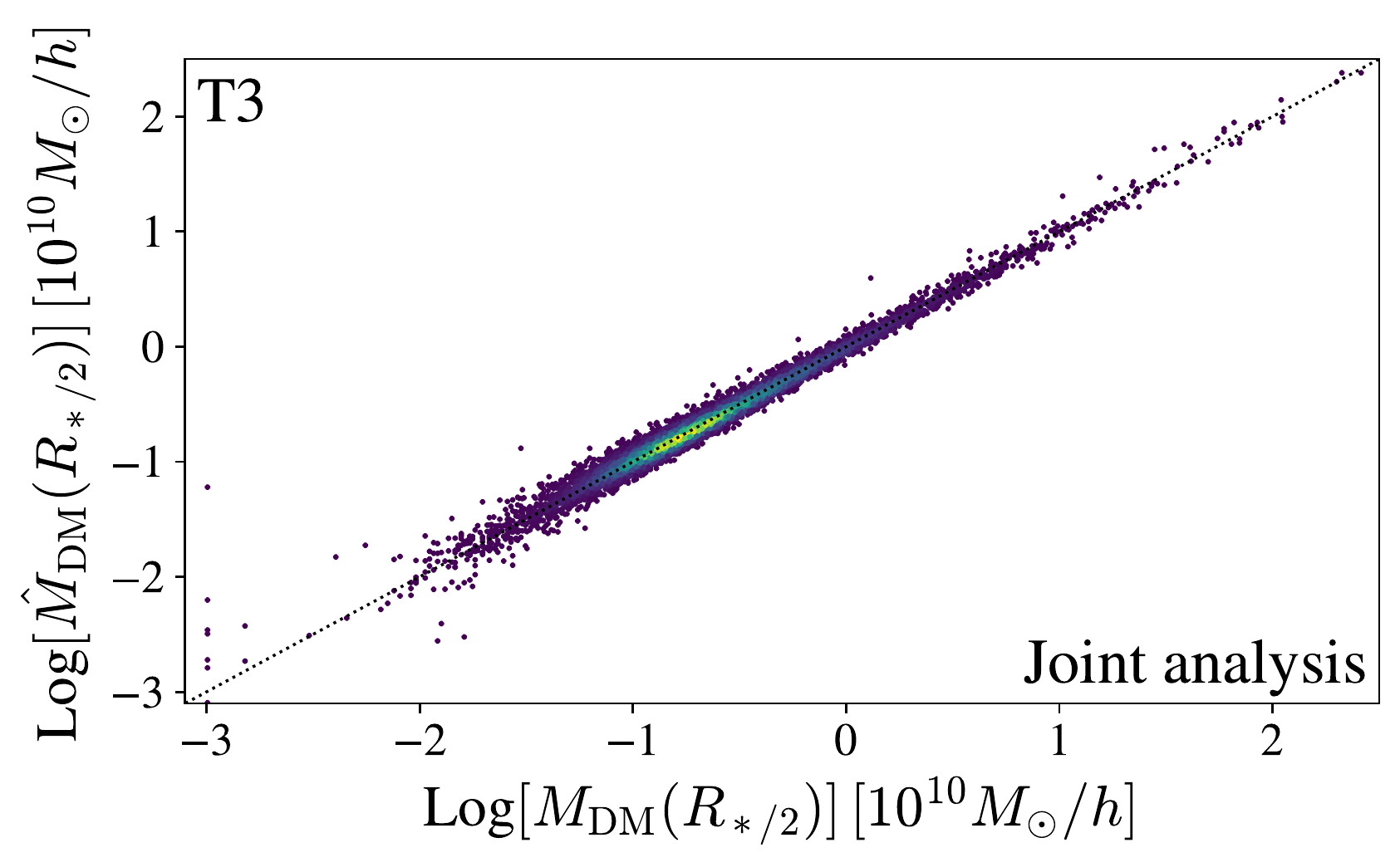}
    \caption{Similar to Fig.~\ref{fig:t1}, but for T3 (DM mass within the stellar half mass radius, given in units of $10^{10} M_{\sun} /h$), denoted by $M_{\rm DM}(R_{\rm*/2})$, and their respective predictions, denoted by $\hat{M}_{\rm DM}(R_{\rm*/2})$. \textbf{Top-left panel: }Photometric features only. \textbf{Top-right panel: }Structural features only. \textbf{Bottom-left panel: }Kinematic features only. \textbf{Bottom-right panel: }Joint analysis.}
    \label{fig:t3}
\end{figure*}
\begin{figure*}
	\includegraphics[width=\columnwidth]{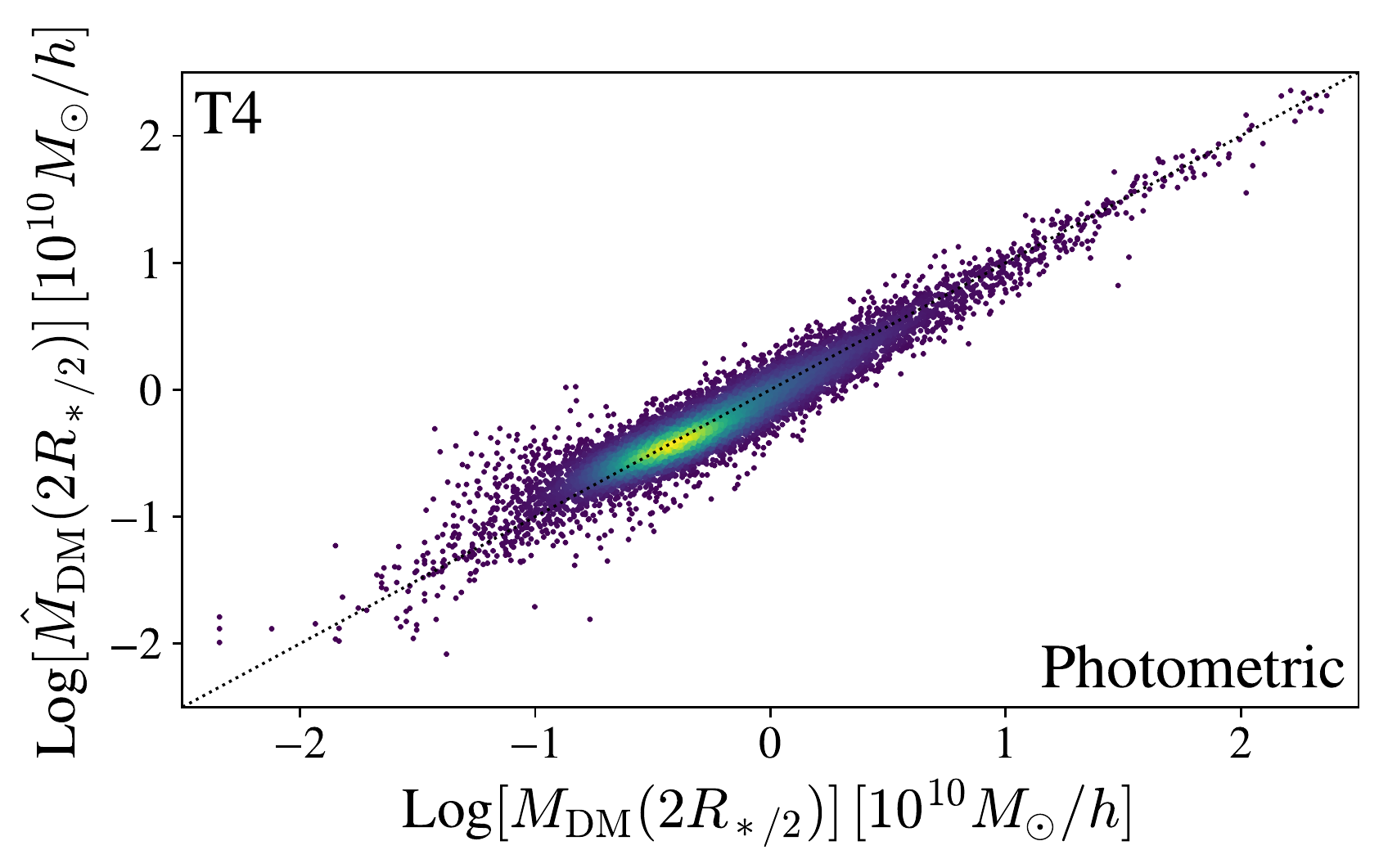}
	\includegraphics[width=\columnwidth]{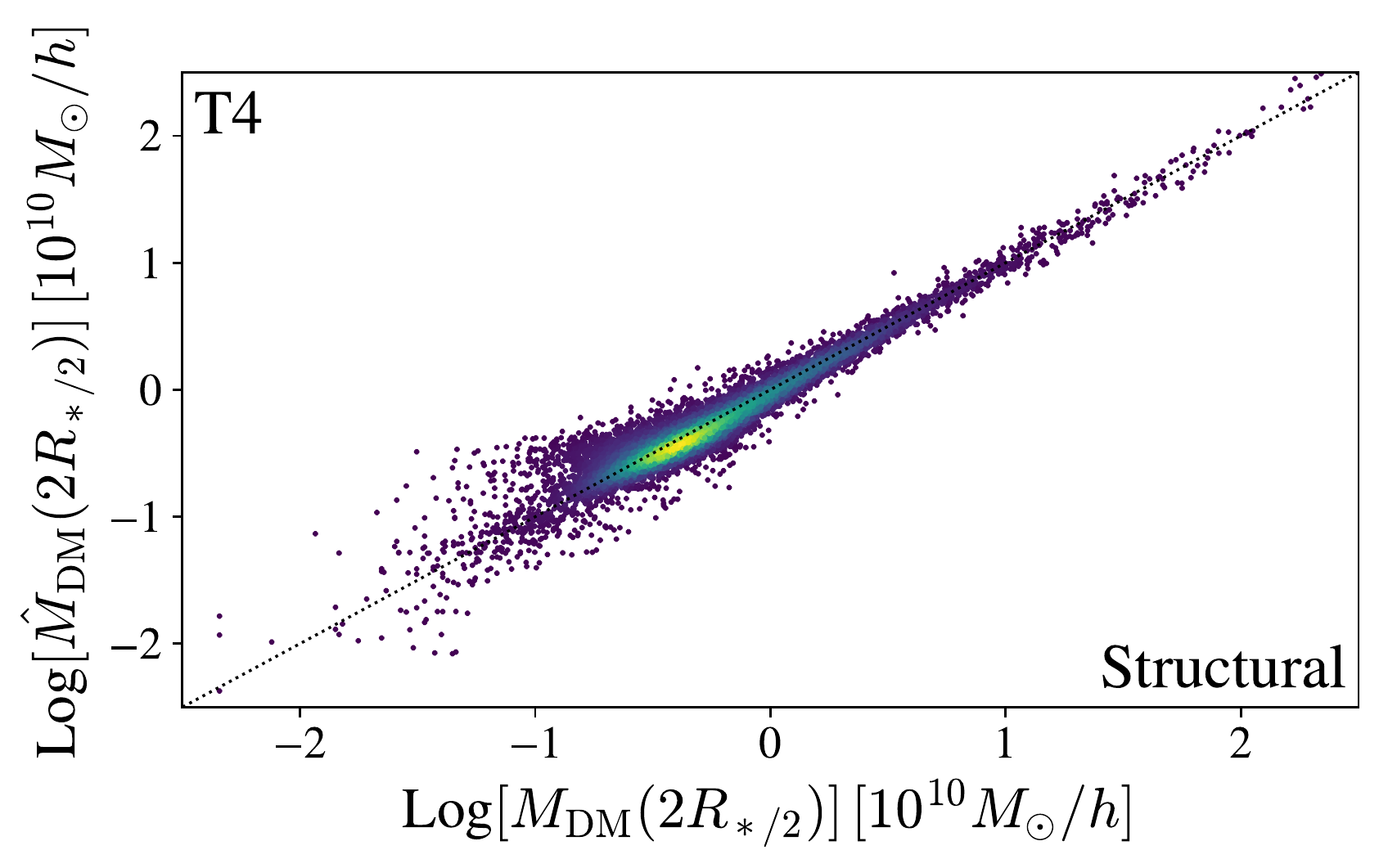}
	\includegraphics[width=\columnwidth]{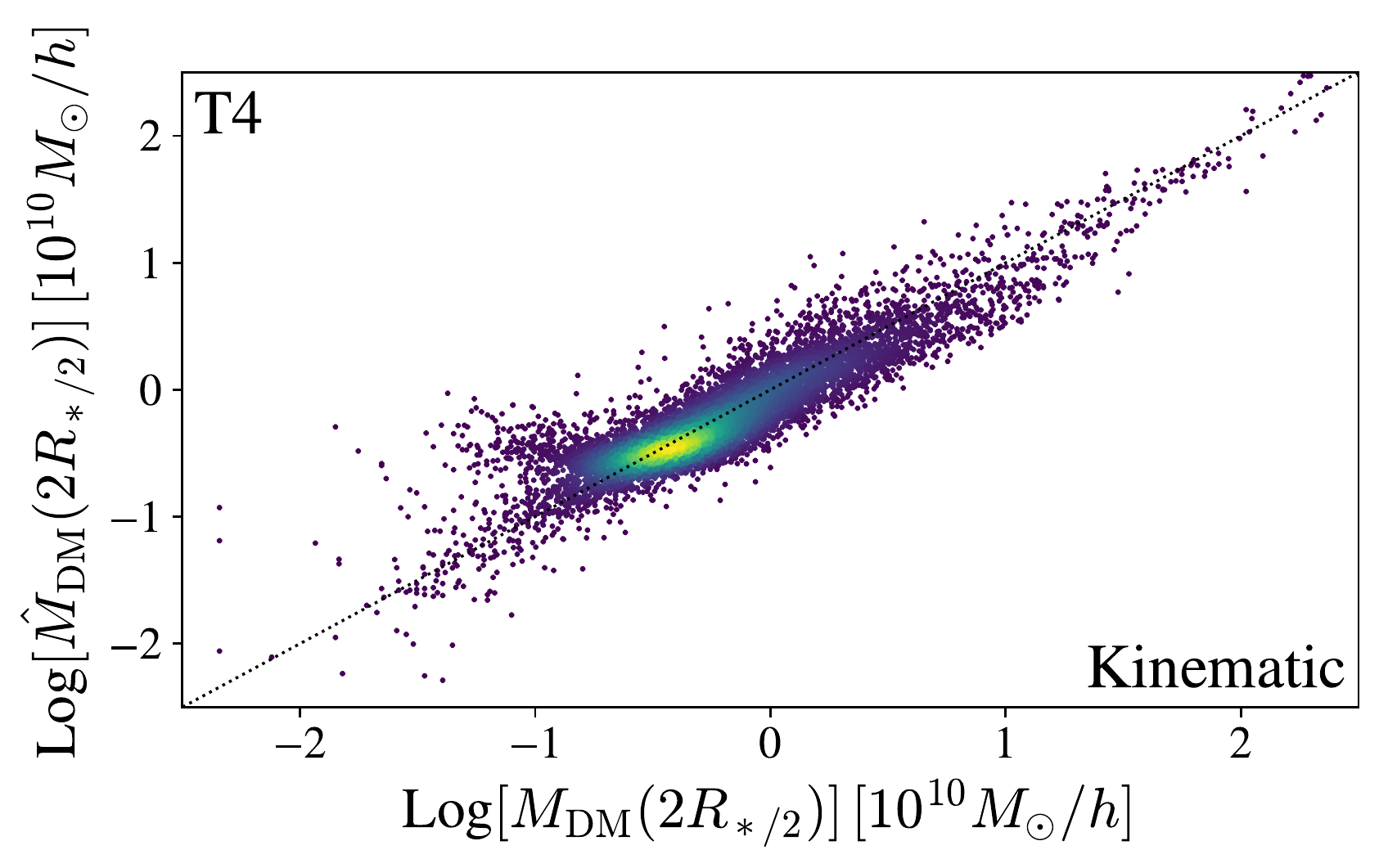}
	\includegraphics[width=\columnwidth]{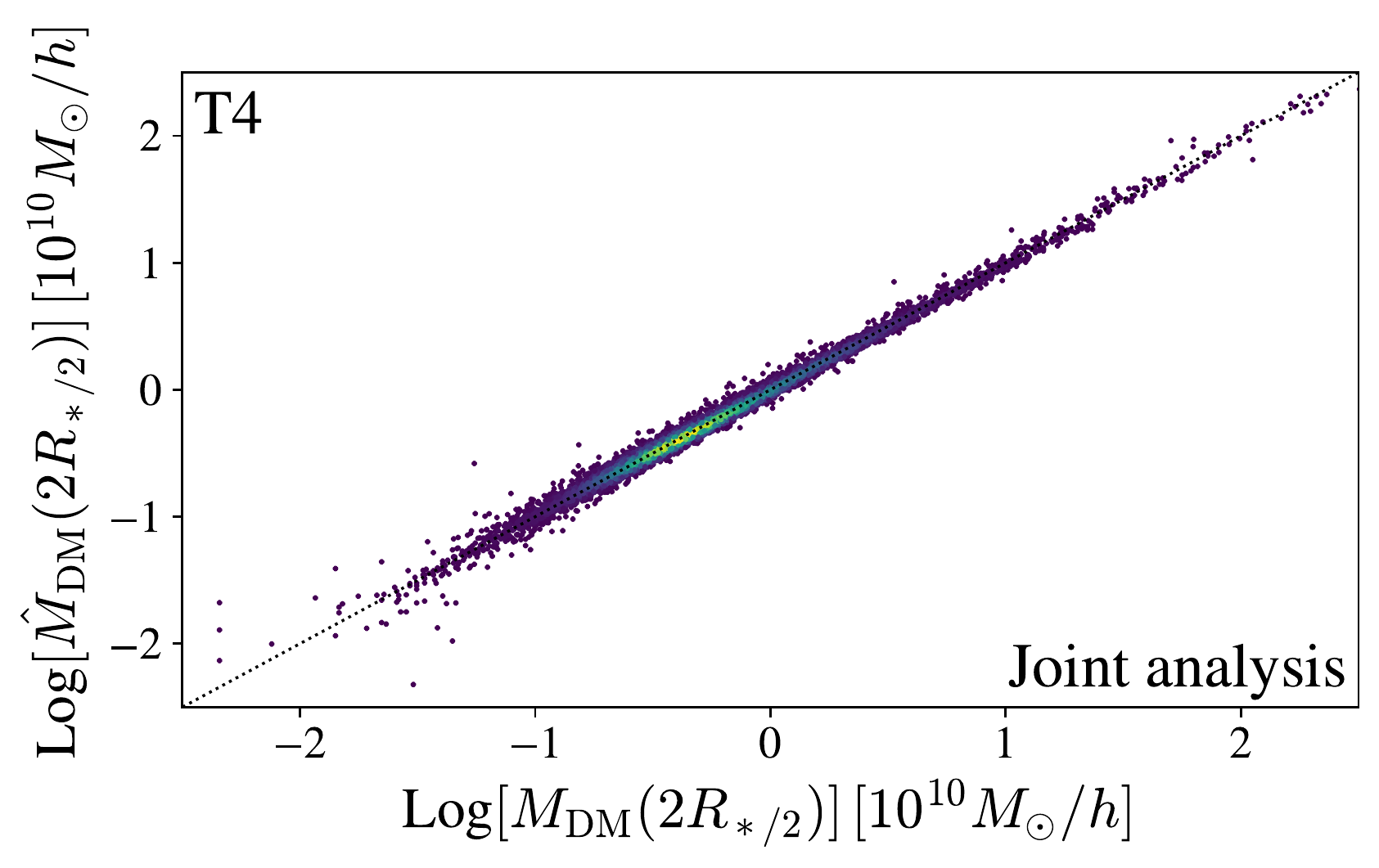}
    \caption{Similar to Fig.~\ref{fig:t1}, but for T4 (DM mass within twice the stellar half mass radius, given in units of $10^{10} M_{\sun} /h$), denoted by $M_{\rm DM}(2R_{\rm*/2})$, and their respective predictions, denoted by $\hat{M}_{\rm DM}(2R_{\rm*/2})$. \textbf{Top-left panel: }Photometric features only. \textbf{Top-right panel: }Structural features only. \textbf{Bottom-left panel: }Kinematic features only. \textbf{Bottom-right panel: }Joint analysis.}
    \label{fig:t4}
\end{figure*}

All the results for the performance metrics discussed in \S~\ref{ssec:metrics} (i.e., $R^2$,  $\rho$, MAE and MSE), computed for both \textit{training sample} and \textit{test sample}, are presented in Tab.~\ref{tab:statistics}. In general, we consider that none of the analyses suffer from overfitting, since all results obtained with \textit{training sample} and \textit{test sample} are compatible with each other. 

In Fig.~\ref{fig:t1} we show the results regarding T1 (total DM matter). One can see that Photometric and Kinematic features alone can generally provide better predictions in comparison to Structural features. In particular, Photometric features provide the best results ($R^{2}=0.855$, for the test sample), while Kinematic features perform slightly worse ($R^{2}=0.794$, for the same sample). These results are illustrated in the top-left and bottom-left panels of Fig.~\ref{fig:t1}, where we compare the real values of the targets, obtained from TNG100 simulation, with their respective predictions, denoted with a hat. Looking at the Photometric and Kinematic features, one can see that, even though they seem to have similar dispersion, the Kinematic features present a larger scatter for galaxies with larger DM. 
One can also see in the top-right panel of Fig.~\ref{fig:t1} that Structural features indeed provide a poorer fit for T1 ($R^{2}=0.646$ and $\rho=0.660$, for the test sample), with a large dispersion. This is mirrored by both a large MAE (0.758) and a large MSE (0.965) which indeed show the presence of both systematics and large scatter in the predictions. 
As expected, the accuracy of the predictions are considerably increased when all features are combined ($R^{2}=0.917$, for the training sample in Table \ref{tab:statistics}). The Joint analysis result is shown in the bottom-right panel of Fig.~\ref{fig:t1}. Compared to the single group analyses, we see that the joint analysis predictions provide a smaller dispersion and a more consistent result for all T1 values. 
This is also quantified by the smaller MAE (0.131) and MSE (0.043), showing a rather good accuracy.

For T2 (comoving radius containing half of the DM mass) no separated feature group alone is able to provide predictions with $R^{2}>0.8$, however, when combined in the joint analysis all features together return a satisfactory prediction ($R^{2}=0.865$, for the test sample). As it is shown in the top-left, top-right and bottom-left panels of Fig.~\ref{fig:t2}, one can see that all single group analyses (Photometric, Structural and Kinematic), show a sensitive scatter particularly in the range $0\lesssim {\rm Log}R_{\rm DM/2} \lesssim 1$, where we also observe an ``horizontal tail'' in all cases. The $R^{2}$ values computed for the training sample for the Photometric, Structural and Kinematic analyses are $0.748$, $0.421$ and $0.535$, respectively. Only the Photometric case shows a reasonable accuracy (MAE=0.159 -- test sample) and precision (MSE=0.049 -- test sample), even better than the ones shown for T1. For the Joined feature analysis, these parameters (MAE=0.111 and MSE = 0.026 -- test sample) are slightly better than the Joined feature analysis of T1 (0.131 and 0.043 respectively).

As for T3 (DM mass within the stellar half mass radius), this seems to be a quantity well predicted from visible matter features. Concerning individual groups, the analysis with the Structural features is particularly noteworthy ($R^{2}=0.937$, for the test set). This excellent performance is illustrated in the top-right panel of Fig.~\ref{fig:t3}.
From the physical point of view, a strong correlation between the central DM mass and the effective radius has been pointed out first in observations (e.g., \citealt{NRT10}; \citealt{SPIDER-VI}; \citealt{Alabi+16}), then in  high-resolution zoom-in galaxy scale hydrodynamical simulations (e.g. \citealt{Wu+14}; \citealt{Remus+17}). Interestingly, this correlation seems encoded also in lower resolution hydrodynamical simulations, like the TNG100 here adopted, as demonstrated by the correlation matrix in Fig. \ref{fig:corr} where there is a fair correlation between T3 and the S* features ($\sim 0.8$). This is important in the perspective to make DM predictions for real galaxies, using ML algorithms trained over hydrodynamical large-scale simulations.

Furthermore, the predictions from both Photometric and Kinematic features also provides high-performance metrics. In particular, the Photometric features show slightly better $R^{2}$ (0.883) and $\rho$ (0.940) for the test sample than the results from Kinematic features ($R^{2}=0.765$ and $\rho=0.875$, for the test set). Their results are presented in the top-left and bottom-left panels of Fig.~\ref{fig:t3}. Beyond the good performances in the individual analyses, the most remarkable of this case is that, when combined, the complementarity of features leads to an excellent prediction ($R^{2}=0.981$ and $\rho=0.990$, for the training sample). 
It is worth mentioning that, in the joint analysis, the $R^{2}$ computed with the test sample is virtually identical to the value obtained using the training set ($R^{2}=0.986$), which corroborates the conclusion that this is a physically consistent result. The joint analysis result is presented in bottom-right panel of Fig.~\ref{fig:t3}. The tight and narrow scatter plot between predicted and true values is quantified by the MAE and MSE, 0.037 and 0.004 respectively, that give the measure of the good accuracy and precision, much higher than the results show by T1 and T2. 

Finally, for T4 (DM mass within twice the stellar half mass radius) we have a result very similar, to T3, which is expected given their tight correlation (see Fig.~\ref{fig:corr}). Furthermore, these two quantities are physically similar since they are computed within a radius which is twice for T4 with respect to T3\footnote{Note that the typical DM scales are much larger that the typical luminous matter scales. Hence, we have small variation of DM mass going to the stellar half-mass radius of T3 to twice the stellar half-mass radius.}. However, despite the tight connection between the two, the correlations of T3 and T4 with the features are not identical, and this implies a different predictive power of the features and even a different pipeline to optimize the results. This is shown in Tab.~\ref{tab:pipelines}, where all the best pipelines for the corresponding analysis for T3 and T4 are different. 

Despite these differences, generally speaking, the performance metrics of T4 are slightly better than the ones found for T3. The Structural group of features is the one  providing the best performance if used alone to predict T4 ($R^{2}=0.915$, for the test sample), which is a bit smaller than the result of this group for T3. On the other hand, both Photometric ($R^{2}=0.899$, for the test sample) and Kinematic ($R^{2}=0.821$, for the test sample) features provides a slight improvement in the predictions for T4. The scatter plots are presented in top-left, top-right and bottom-left panels of Fig.~\ref{fig:t4}. Here we also see a complementarity between the feature groups, as the joint analysis achieves an almost perfect performance level ($R^{2}=0.987$, for the training set, and $R^{2}=0.993$, for the test set). The joint result is shown in the bottom-right panel of Fig.~\ref{fig:t4}. Visually this looks as tight as the T3 plot, which is also demonstrated by the MAE=0.033 and MES=0.003 (for the test sample), which are almost identical to the same values from T3 (0.037 and 0.004 respectively).

To conclude this section, we can affirm that a major result of this analysis is that ML tools have the predictive power to put constraint on the DM properties of galaxies. Given the simple input observational parameter coming from the bulk of the stellar component, the most accurate predictions are confined in the central regions of the galaxy, and regard the DM mass in one and two the half-mass radii. This is possibly the most important result of this paper, but also not surprising, as these are all quantities physically connected and with no need of any extrapolation. This also suggests that, to extend the predictions to other more ``global'' DM properties, like the total mass or the characteristic scales of the DM halos, we might need in the future to consider larger scale observables (e.g., large radii kinematics like HI rotation curves for spirals: \citealt{Lelli+16_SPARC}, or planetary nebulae and globular clusters: \citealt{Napolitano+14}, \citealt{Pota+15_SLUGGS}; and X-rays: \citealt{kin_xrays_ell}, for elliptical galaxies).

In the next section we will quantify the importance of each group (i.e. Photometric, Structural and Kinematic) for the joint analyses. 
In particular, this will allow us to understand whether it is reasonable to expect a direct connection between the ML analysis with a single feature group as presented here and the corresponding features importance, in the sense that, we aim to investigate if the better the ML performance of a given group is, the more important it would be in the joint analysis.

\subsection{Feature Importance}
\label{ssec:featimp}

The analysis of the ``feature importance'' is a common practice in ML applications. This is used to figure which of the features plays the most decisive role in the ML algorithms to predict the targets.
This has nothing to do with accuracy, as the feature importance does not select the features that return the best accuracy, but only their impact on the final predictions.

For our feature importance analysis, we use the method \texttt{permutation\_importance} (\citealt{breiman2001}), provided by the \texttt{scikit-learn} software (\citealt{scikit-learn}). 

\begin{figure*}
	\includegraphics[width=\columnwidth]{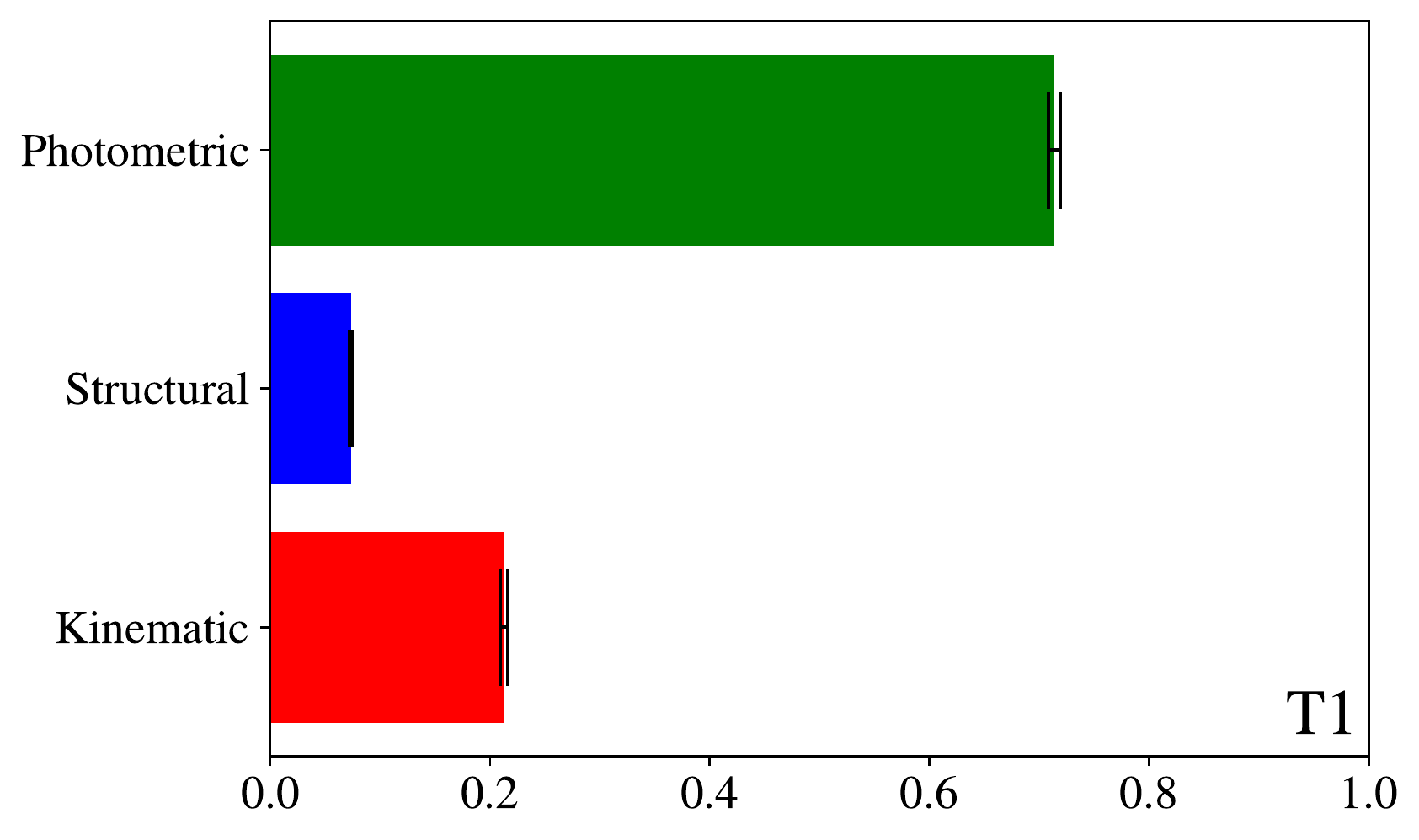}
	\includegraphics[width=\columnwidth]{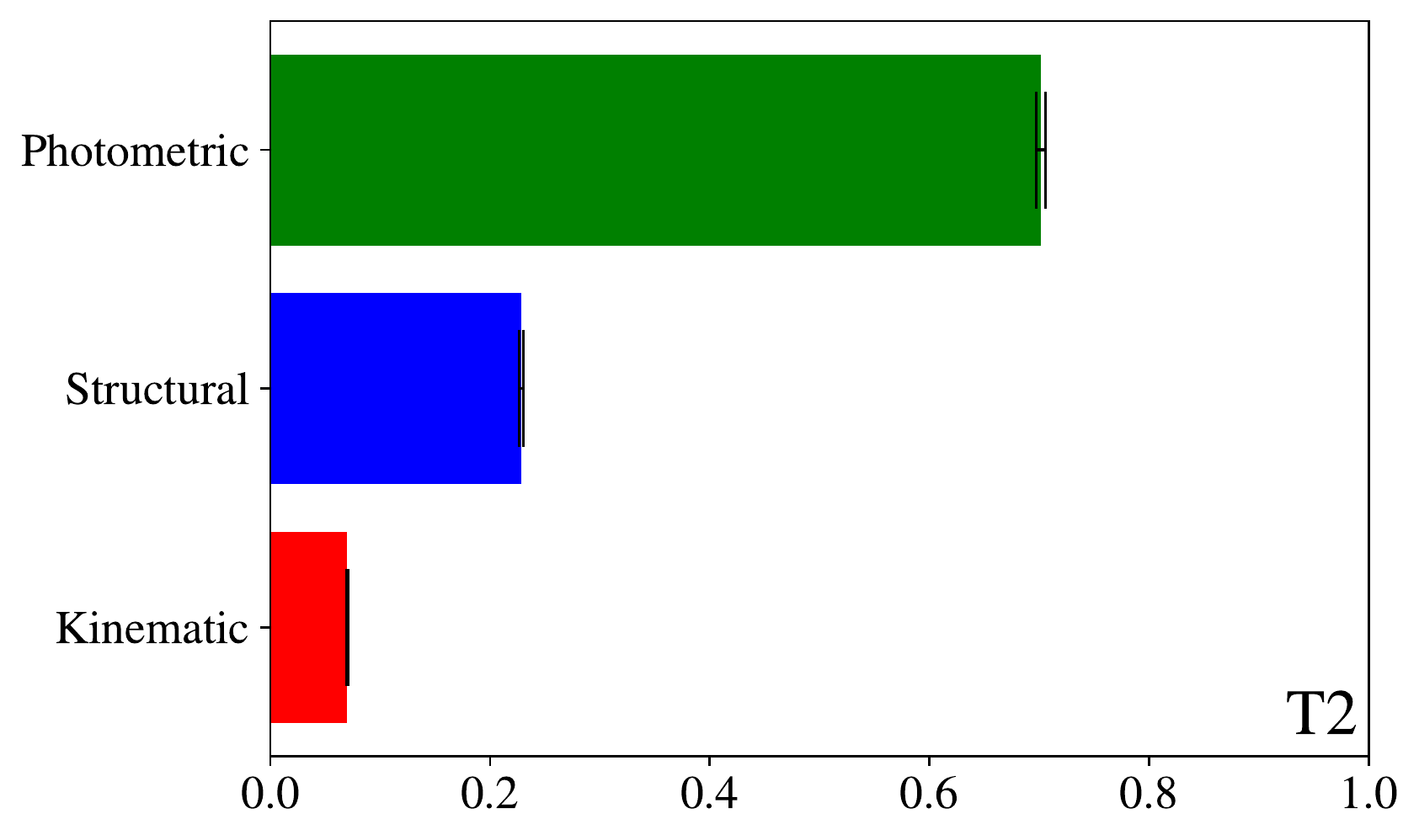}
	\includegraphics[width=\columnwidth]{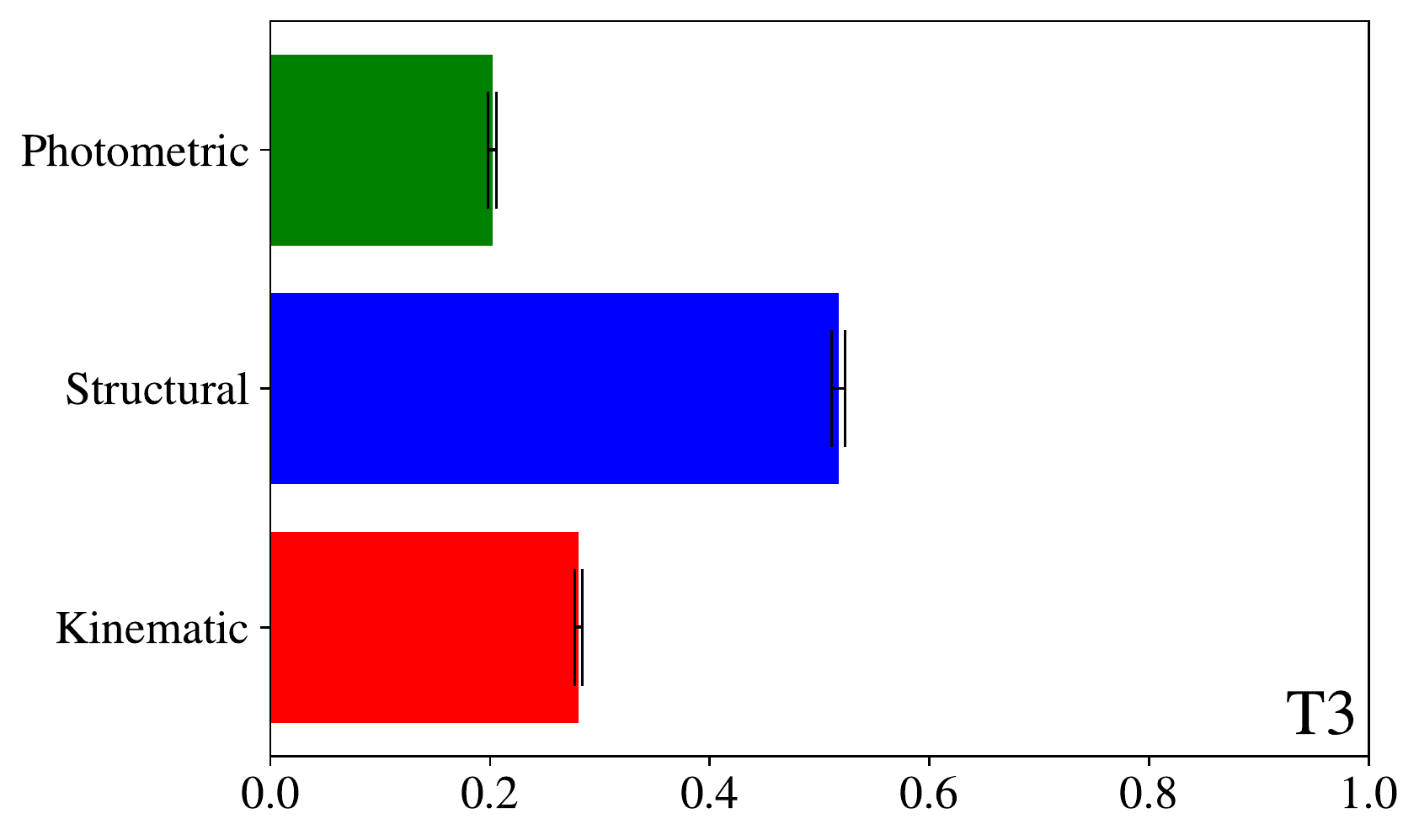}
	\includegraphics[width=\columnwidth]{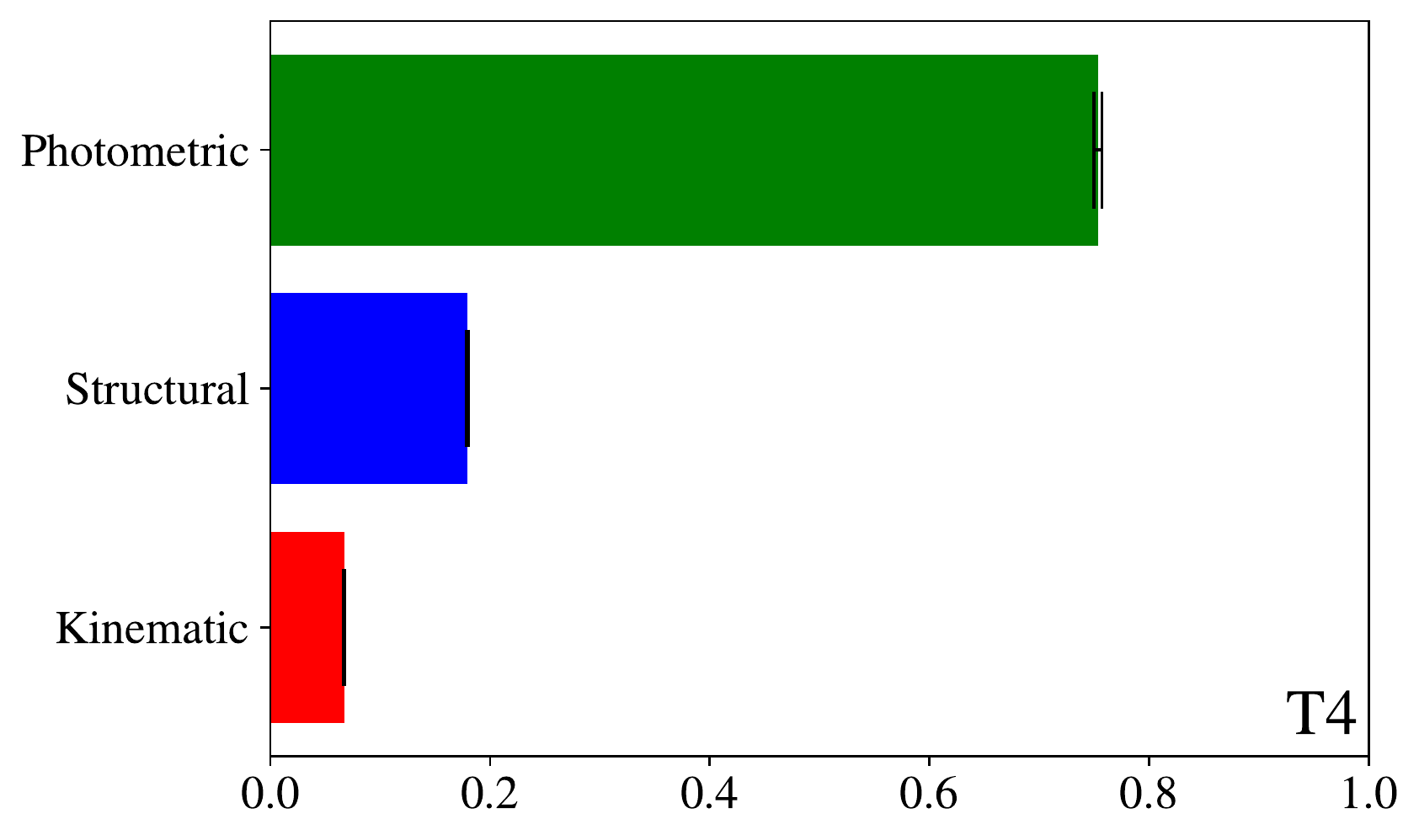}
    \caption{Result of the feature importance analysis. Here, because of the correlation between the features, instead of considering all individual features, we combine all features in the groups Photometric, Structural and Kinematic. See \S\ref{ssec:featimp} for more details.}
    \label{fig:feature2}
\end{figure*}

This procedure evaluates how much the performance of the model drops when one of the features has been mixed. In fact, each time a characteristic is corrupted with a shuffle, the relationship of that particular characteristic with the target is broken, and the loss in the performance denotes how important the feature is. We can choose the number of times a feature is randomly shuffled with the hyperparameter \texttt{n\_repeats}. For our experiments, we took  \texttt{n\_repeats} $ = 100$, after we have explored the range of $20-100$ without finding appreciable differences. In Fig.~\ref{fig:feature2} we show the results we have obtained with this setting. In particular, we show the results for the joint analysis only, where we can cumulatively show the feature importance values for each group of features suitably normalized to the unity. This is rather useful for two main reasons: 
\begin{itemize}
    \item[$i$)] It helps to figure the physics behind the suitability of the ML approach, which is somehow encrypted in the scaling relations among the different group of features discussed in \S\ref{sec:intro} (e.g., size--mass, Tully-Fisher and Faber-Jackson relations, Fundamental Plane etc.);
    \item[$ii$)] It also helps to draw strategies for data acquisition and analysis, especially if time-consuming, (what kind of data? photometry or spectroscopy? what kind of parameter to extract from these data? do we need structural parameters for all galaxies?);
\end{itemize}

For this first paper we do not enter into the details of the relative importance of the features in each given group (e.g. optical vs. NIR bands in Photometry, stellar vs. gas mass in Structural, etc.) because this might be partially function of the galaxy types, which we also do not distinguish in this analysis. These aspects will be part of the next step of the project (Wu et al., in preparation), thus, hereafter, we will only discuss the relative importance of the different groups.
Following the order of the targets in Fig.~\ref{fig:feature2} we can see that Photometric features are the ones driving more the predictions of the Total DM matter (T1), half DM mass comoving radius (T2), and DM mass within twice the stellar half mass radius (T4), while for DM mass within the stellar half mass radius (T3), the most important features are the Structural features (stellar half mass radius and total stellar masses within one and two the stellar half mass radii, in particular). The kinematics is the second important feature for T1 and T3, while surprisingly is the least important for T4, which, according to the discussion in the previous section, should not be dissimilar from T3, given the close definition of the two quantities. 

To understand the T4 behavior and, in general, to move to the physical interpretation of these ``feature importance'' results, we need first to stress what happens when two or more features are highly correlated. In this case, permuting one feature will have tiny impact on the ML predictions because the model can get the same information from a correlated feature that is not permuted. Fig.~\ref{fig:corr} shows some strong correlations among features of different groups. For instance, the photometry and the stellar mass have correlation indexes $>0.9$, due to their tight physical connection. Indeed, the mass inside the half-light radius ($R_h$), $M_{\rm star}(R_h)=M/L_{ x}\times L_{x}/2$, where $M/L_{x}$ and $L_{x}$ are the mass-to-light ratio and the total luminosity in the $x$-band, respectively.

With this necessary premise, we can try to interpret the difference between T3 and T4 feature importance. Photometry, by containing the information about the stellar mass as well, is an important feature to predict the dark mass, as the stellar mass and the dark mass are strongly correlated (see e.g., \citealt{Moster+10}). Hence, photometry is important for both T3 and T4. However, T3 is the mass strictly within the half-light radius, hence this structural parameter (S1) counts as much as the stellar mass. Then for T3, overall, the structural parameters (radii and stellar masses) take over the Photometry.

The importance of the half-light radius might suggest that Structural parameters might count more on the T2 (DM half-mass radius). This would be true if there is a strong correlation between the half-mass radii of the DM and the stellar matter. As already mentioned in \S\ref{ssec:corrmatrix}, according to the Correlation matrix in Fig. \ref{fig:corr}, there is no strong correlation between T2 and S2 (0.3).
Hence, the strong dependence of T2 in Photometry parameters is reasonable, as these latter (combined with the stellar mass inside the structural parameters) give the larger constraints on the dark mass and the DM half-mass radius is driven by its correlation with the DM mass.

Overall, form Fig. \ref{fig:feature2} we can conclude that there is no particular group of observable that we can discard when making dark matter predictions in galaxies. Photometry is unmissable, as this provides information on the luminosity and stellar masses of galaxies. But this is rather obvious and imaging surveys are always the first step to catalog galaxies. Kinematic parameters are less obvious. Despite internal kinematics is recognized to be major information about galaxies, they are difficult to measure, so these are often left aside in large spectroscopic surveys where the primary objectives remain the galaxy redshift measurements (but see SDSS: \citealt{thomas_kin_sdss} ; LAMOST: \citealt{Napolitano2020_lamost}). Future surveys are designed to list these parameters (see WEAVE: \citealt{Costantin+19_steps}; WAVES: \citealt{2019Msngr.175...46D}). Hence these will be crucial for the ML analyses we are proposing here.

{\it We have anticipated that the most important result of this paper is that luminous observables can give accurate predictions of the DM inside the half-light radius}. For these analyses, the galaxy structural parameters are crucial measurements to collect, despite these are rather difficult and computationally demanding (see e.g. \citealt{2011MNRAS.414.1625Y}, \citealt{Baldry2012}). However, ML tools are being lately developed to perform also this kind of analyses and make them possible for future large sky surveys (\citealt{2018MNRAS.475..894T}, \citealt{Li+20_KiDS}).

To conclude this section, we want to emphasize the differences between the ``feature importance'' and the statistical metrics of the separated analysis of the Photometric, Structural and Kinematic groups reported in Table \ref{tab:statistics}. Although these have an important connection (as aforementioned), they do not carry the same information, but rather, they represent complementary datasets.
In particular, by comparing the ``feature importance'' and the statistical metrics results, we can see how the performances of the different groups changes when we allow interplay between them. For instance, for T1, the order of contribution in the feature importance analysis is the same as the order of performance from the statistical metrics as $R^2$ or $\rho$ scale with the same ranking as in Fig. \ref{fig:feature2}. On the contrary for T2 and T3, although the most important group of features remains the same, the other two permute their order of importance. This means that in the Join analysis for T3 the Kinematic parameters count more than Structural parameters to improve the accuracy, even if, individually, Kinematics alone has a poorer $R^2$ and $\rho$ than Photometry, and also poorer MAE and MSE.

For T4, the group of Photometric features takes the lead role in the joint analysis over the group of Structural features, for the reasons discussed above.
This leads us to conclude that the feature importance is not a quantitative metrics for a given dataset, as it is not deterministically derived. 
Indeed, its outcome is dependent on the specific pipelines under use (see Appendix \ref{sec:tpotperf}), which have privileged one feature of a group of features over another, but this does not mean that another pipeline may give preference to the features in a different way.

However, the importance of this analysis remains to figure whether there are groups of variables that can be excluded from the analysis in virtue of giving little or no contribution to the target predictions. In particular, we have demonstrated that for the best target we can predict (the DM inside the half mass radius, T3), all features are almost equally important.

\section{Conclusions}
\label{sec:conclusions}

We have used the Illustris-TNG simulations to investigate whether Machine Learning tools can make solid prediction about the DM content in galaxies in a given Cosmology, with a given galaxy formation recipe, starting from simple observational parameters. The main goals of this work are still exploratory as we are using, for this project, ideal dataset where both the {\it training sample} and the {\it test sample} are extracted from simulations. In other words, the ``observational quantities'' themselves used as mock catalogs are still missing observational realism to allow us a full assessment of the accuracies one can expect employing these techniques. However, as a first step, we want to demonstrate that this is a viable direction to explore for future analyses of large galaxy samples as the one expected to be observed with future photometric and spectroscopic surveys. This is the first attempt to apply ML technique to galaxies, which follows previous attempts made to make dark matter predictions in galaxy clusters (e.g. \citealt{yan2020galaxy}).

Taking advantage of the public catalogs from Illustris-TNG, we have identified a series of observational parameters, representing the features we want to use to make DM predictions. These have been grouped in three major ensembles: {\it Photometry} parameters (i.e., the magnitudes in 8 different bands), {\it Structural} parameters (i.e., the stellar half-mass radius and three different baryonic masses), {\it Kinematic} parameters (i.e., 1D velocity dispersion of all particles and the maximum rotation velocity). 
We have also used Dark matter halo parameters `(total DM mass, DM half-mass comoving radius, DM mass inside the stellar half-mass radius and tho half-mass radii) as a series of ``targets'' for the ML to predict.
Due to the multiplicity of ML learning algorithm on the market, all having advantages and disadvantages in reaching the best accuracy in the target predictions, we have adopted a supervised learning where we explored a wide set of models with the help of \texttt{TPOT}. This is an automated tool for data transforms and machine learning algorithms that uses a genetic search procedure to efficiently discover a model pipeline that perform better for a given dataset.
We presented our results graphically in Figs \ref{fig:t1}, \ref{fig:t2}, \ref{fig:t3}, and \ref{fig:t4}, and numerically in Table \ref{tab:statistics}. The best pipelines, founded by the automated ML tool, provide different levels of accuracy of predictions depending on the DM target.  

We summarize the major results here below:
\begin{itemize}
    \item[$i$)] ML tools are a promising solution to make predictions of the DM content in galaxies starting from a series of simple photometric, spectroscopic and structural measurements. The  results obtained in this work, though, are base to idealized ``luminous observables'' and the accuracies reached in this first experiment are possibly too optimistic. This does not impact the conclusion about the applicability of these techniques, but encourages us to move further in the development of these tools using increased realism in the simulated datasets. These latter will be used to train ML tools to make prediction on DM content of real galaxies in next generation photometric and spectroscopic surveys;    

    \item[$ii$)] Structural and Photometric features are particularly effective for the prediction of DM of the stellar half mass radius ($0.88 \lesssim R^2 \lesssim 0.94$), and the prediction of DM within twice the stellar half mass ration ($0.90 \lesssim R^2 \lesssim 0.92$). Furthermore, Photometric features stands out to predict the Total DM matter ($R^2 \lesssim 0.86$).
    
    \item[$iii$)] The comoving radius containing half of the DM mass is the most difficult target to be determined and no group of observations alone succeeds in doing this ($R^2 \lesssim 0.75$), but the joint analysis improved noticeably the predictions giving to all the targets (including this radius too) an excellent result ($0.87 \lesssim R^2 \lesssim 0.98$). 
    
    \item[$iv$)] In Fig \ref{fig:feature2} we show the feature importance analysis for the join analysis only. 
    This analysis has demonstrated that the different groups of observables drive the predictions of the different targets according to the importance of the physical links among them. For example, Photometry is very important in most of the targets because the stellar masses are strongly linked to that. However, for the best predicted target, the DM mass inside the half-light radius, structural parameters like the half-light radius increase the significance of this group of observables that become primary for the best determination of the DM prediction.
    Finally, the feature importance showed that  photometry, structural parameters are all significant for DM predictions, although Kinematics might not help much in particular to have the inference on the DM halo size.
\end{itemize}

In the next work, we intend to move a step forward toward the use of ML technique in real galaxies. We will evaluate the predictive powers of these algorithms when separating galaxies in different classes and mass ranges  (spheroidal, disks and dwarf systems), having accurately selected these systems in the simulations closer following the observed scaling relations. We expect this allows starting testing ML algorithms trained on such a simulated sample on a first observational sample, e.g., to derive the DM fraction to compare with results based on standard dynamical analyses (e.g. \citealt{SPIDER-VI,Tortora+18_KiDS_DMevol,Cappellari2013_fDM}). 

Ultimately, we will aim at applying ML techniques to large sky survey observations to put constraints on the baryonic and the dark matter assembly in galaxies, in a unique ``universal'' model. 
By training ``supervised'' ML tools over different simulations, making predictions of observational ``features'' and dark matter ``targets'' in different cosmologies and galaxy formation recipes (see e.g. \citealt{Camels2021} as a prototype), we expect to derive the best ``universe'' interpreting survey data from different projects.

\section*{Acknowledgements}

RvM acknowledges support from the Programa de Capacitação Institucional PCI/ON/MCTI.
LC acknowledges financial support from CNPq (Conselho Nacional de Desenvolvimento Científico e Tecnológico) do Brazil, and P.O. Baqui for discussions.
NRN acknowledges financial support from the “One hundred top talent program of Sun Yat-sen University” grant N. 71000-18841229 and the European Union Horizon 2020 research and innovation programme under the Marie Skodowska-Curie grant agreement n. 721463 to the SUNDIAL ITN network. AC acknowledges the Simons Foundation (Grant n. 884966, AF), the Brazilian National Council for Scientific and Technological Development (CNPq) via Grant No. 311375/2020-0.

\section*{Data Availability}

The data underlying this article will be shared on reasonable request.







\appendix

\section{Distributions}
\label{sec:dist}

\begin{figure*}
	\includegraphics[width=0.5\columnwidth]{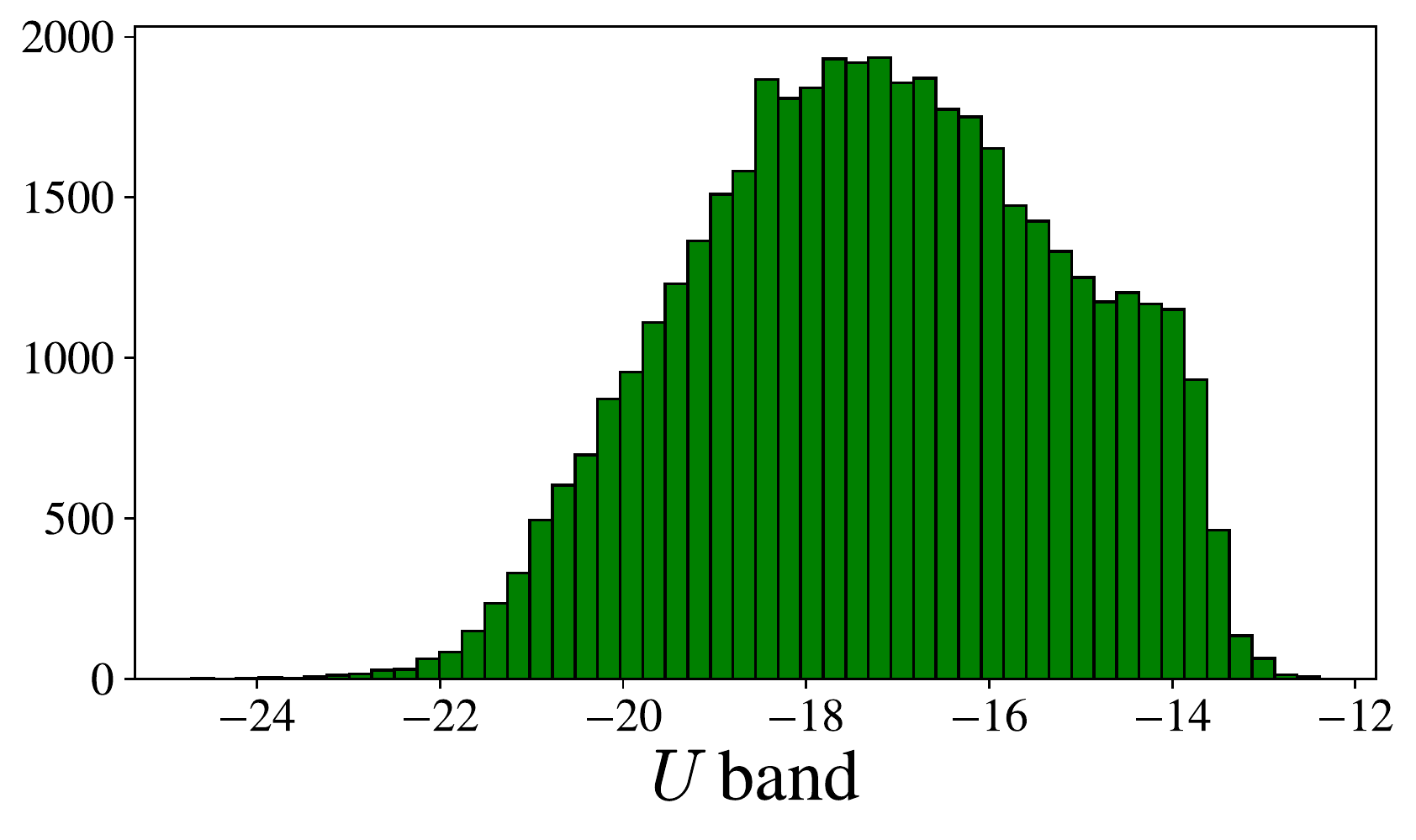}
	\includegraphics[width=0.5\columnwidth]{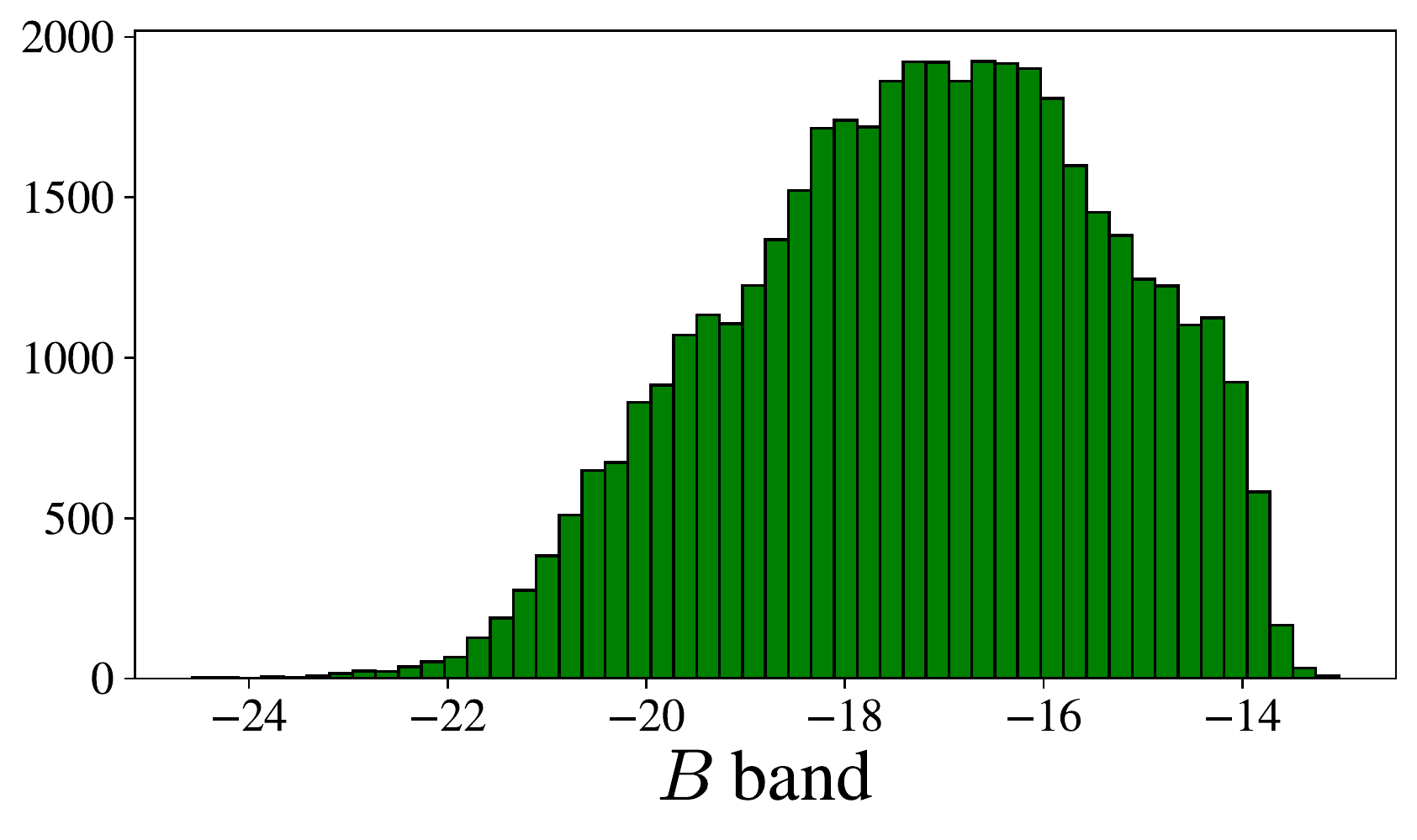}
	\includegraphics[width=0.5\columnwidth]{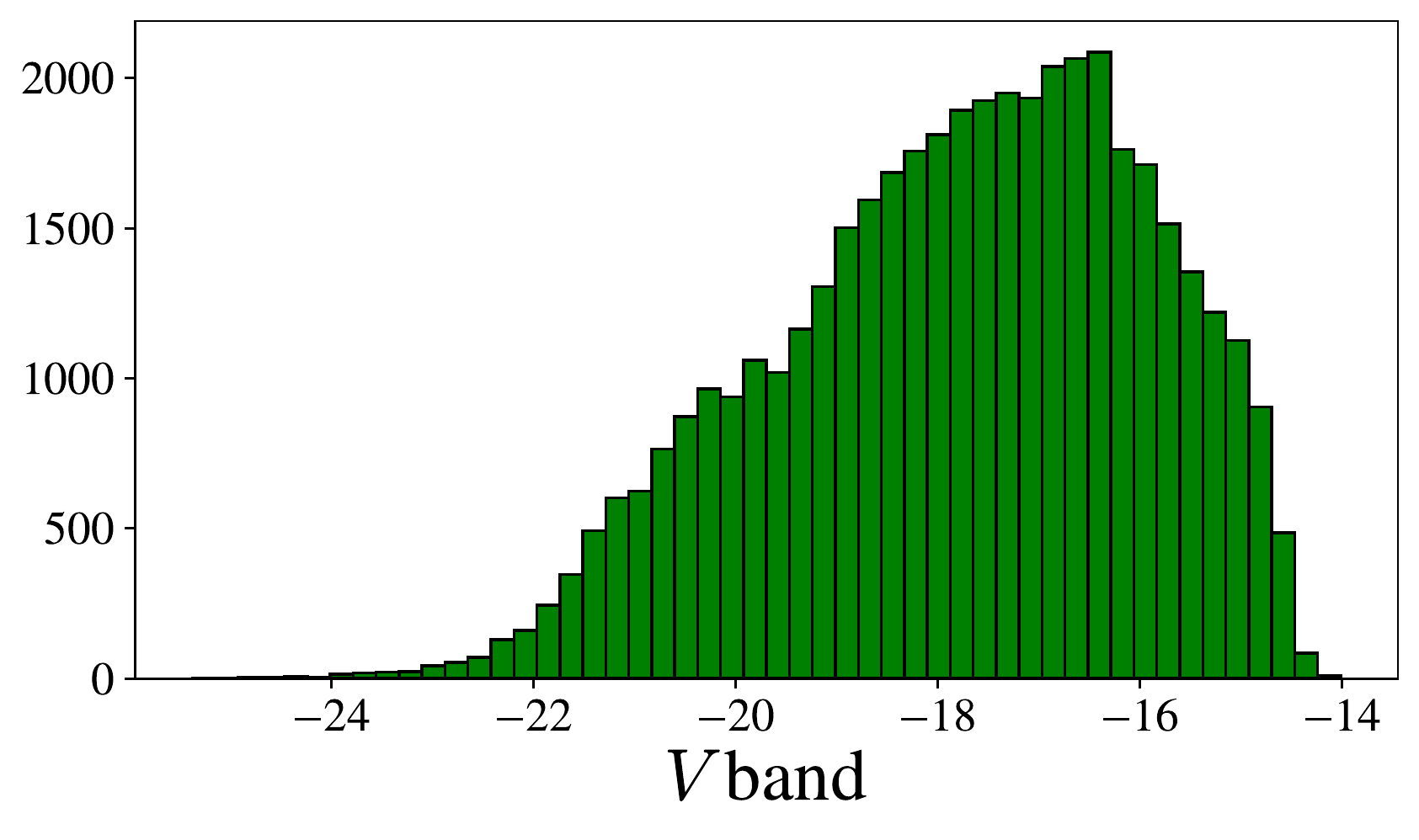}
	\includegraphics[width=0.5\columnwidth]{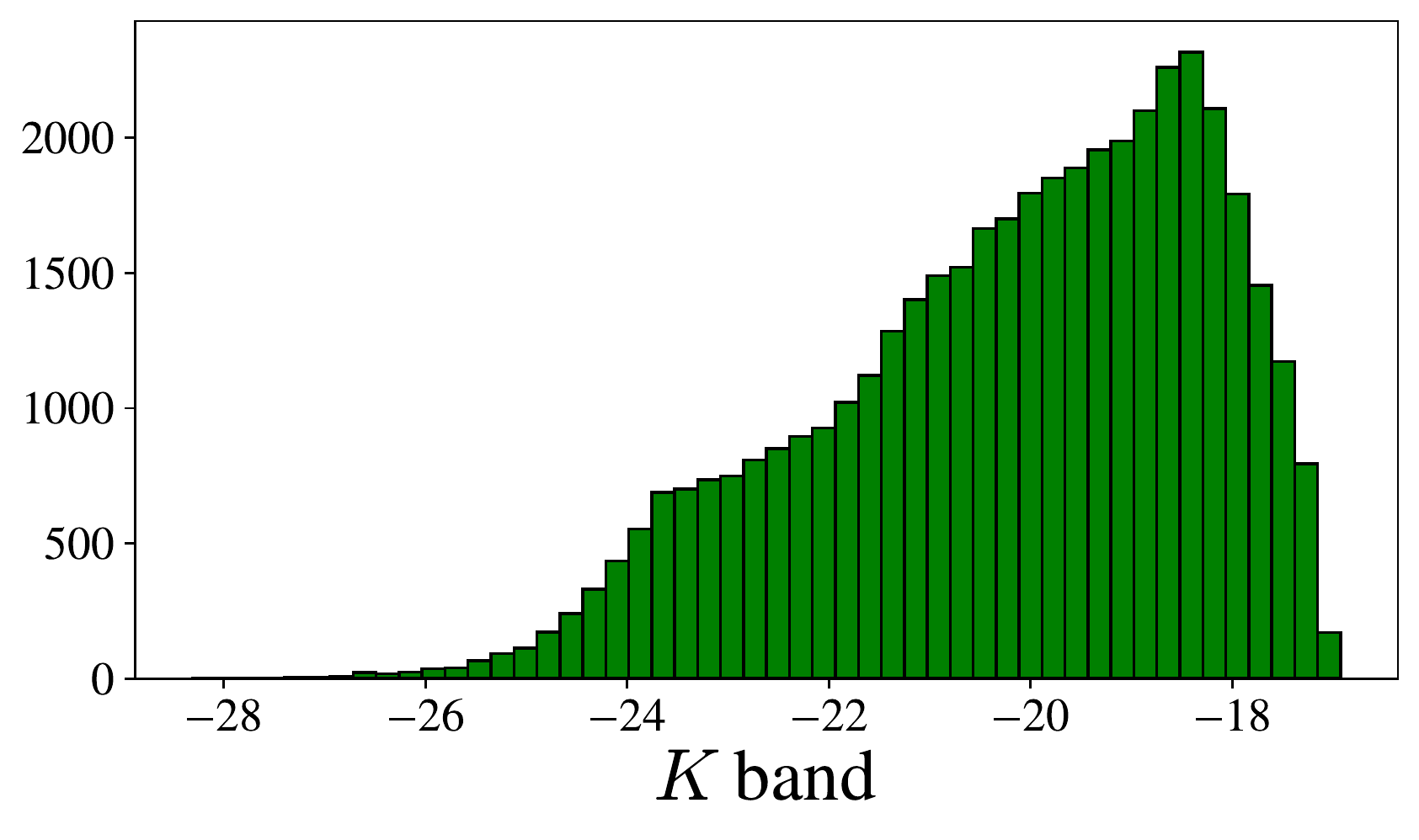}
	\includegraphics[width=0.5\columnwidth]{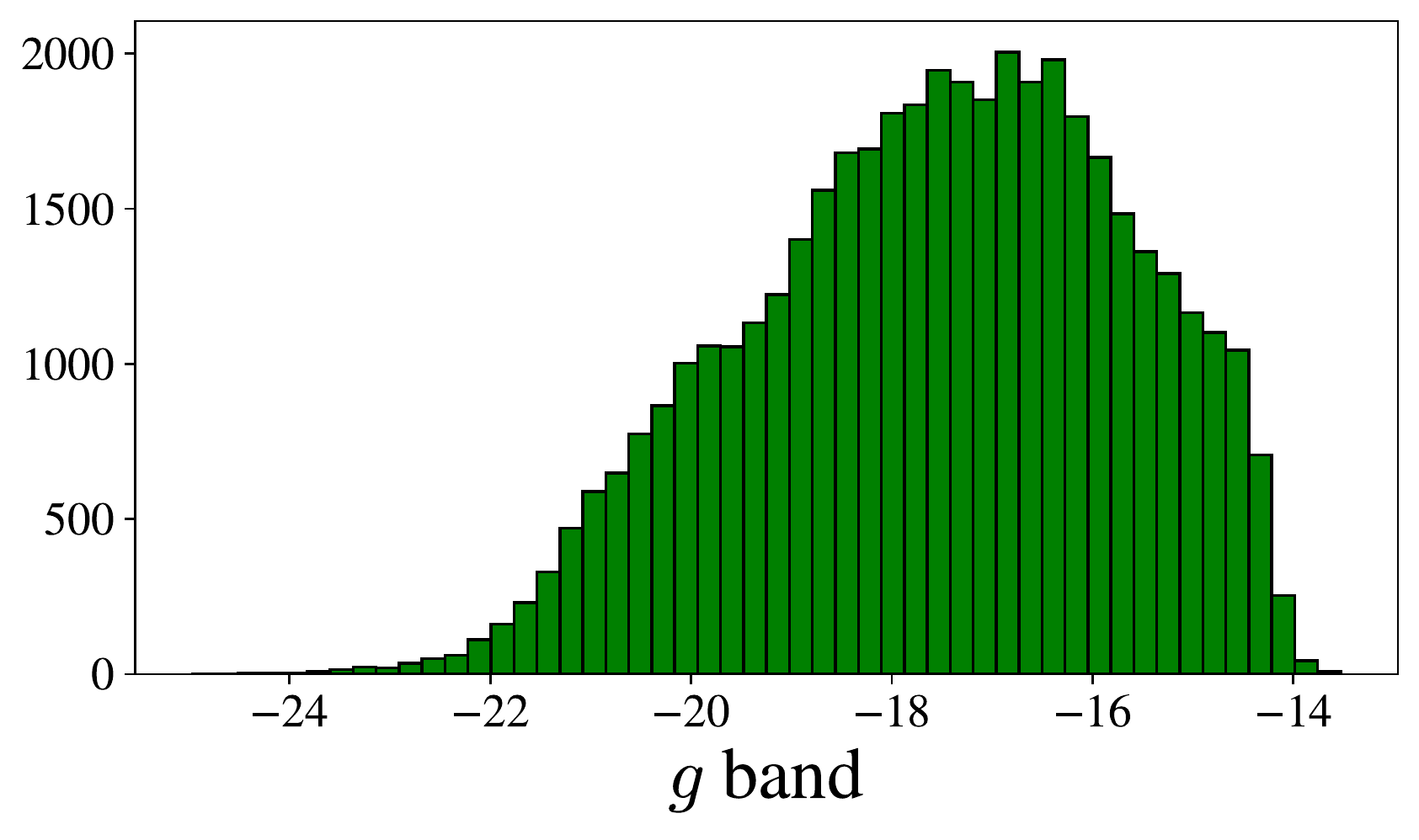}
	\includegraphics[width=0.5\columnwidth]{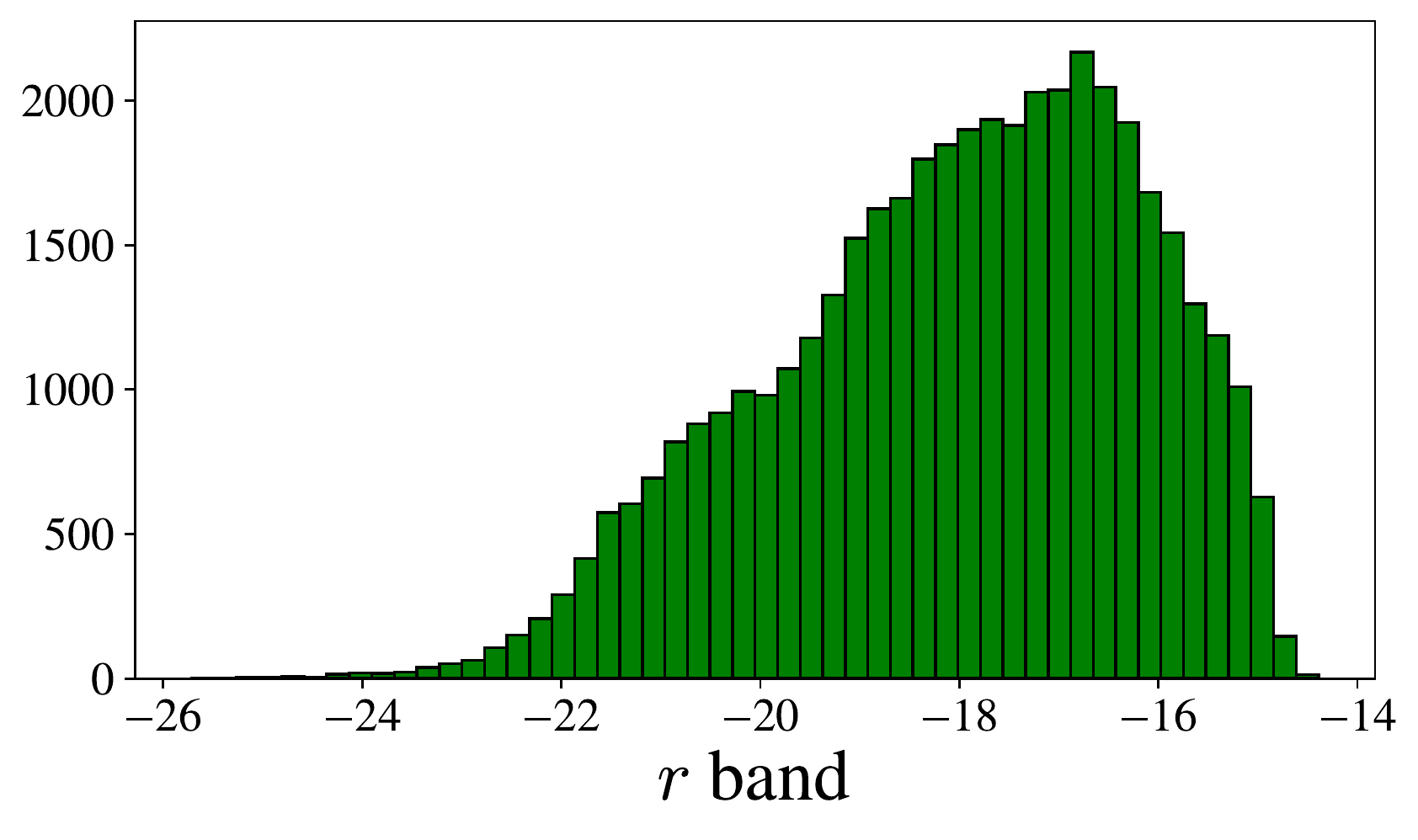}
	\includegraphics[width=0.5\columnwidth]{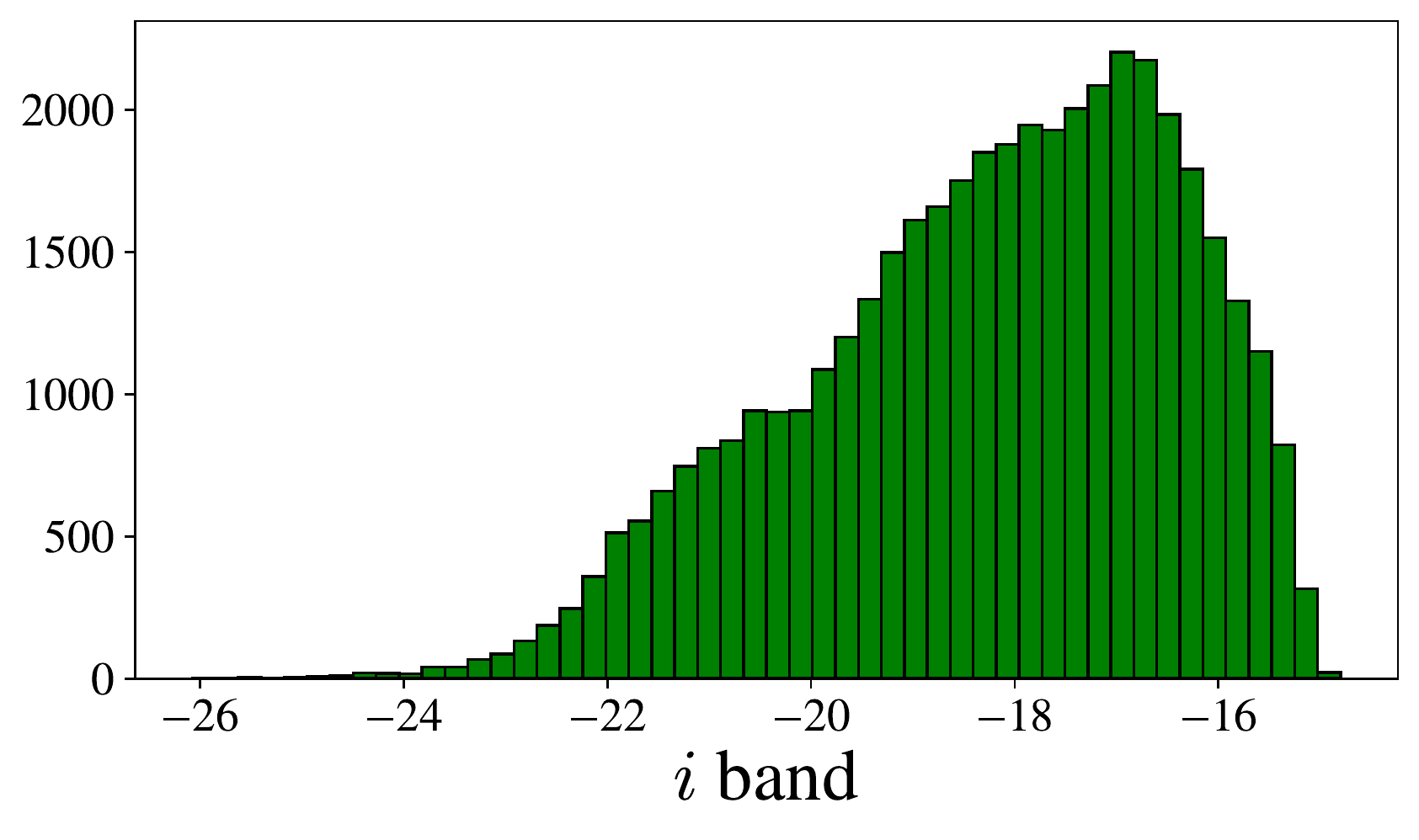}
	\includegraphics[width=0.5\columnwidth]{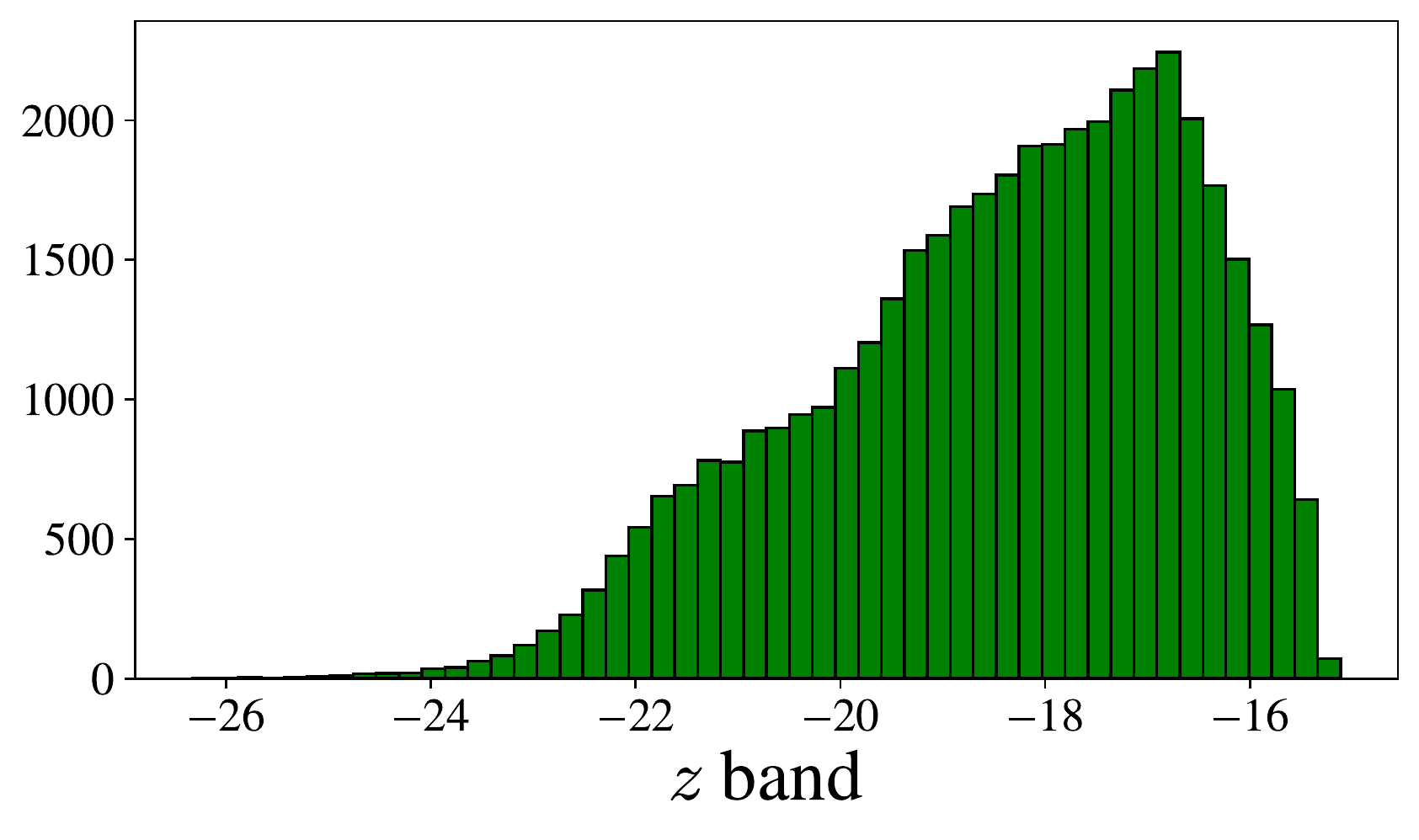}
    \caption{Distribution of the Photometric features. Going clockwise from the top left plot, the panels show: Johnson-Bessel $U$ band, Johnson-Bessel $B$ band, Johnson-Bessel $V$ band, Johnson-Bessel $K$ band, SDSS $z$ band, SDSS $i$ band, SDSS $r$ band and SDSS $g$ band.}
    \label{fig:photometrical}
\end{figure*}

\begin{figure*}
	\includegraphics[width=\columnwidth]{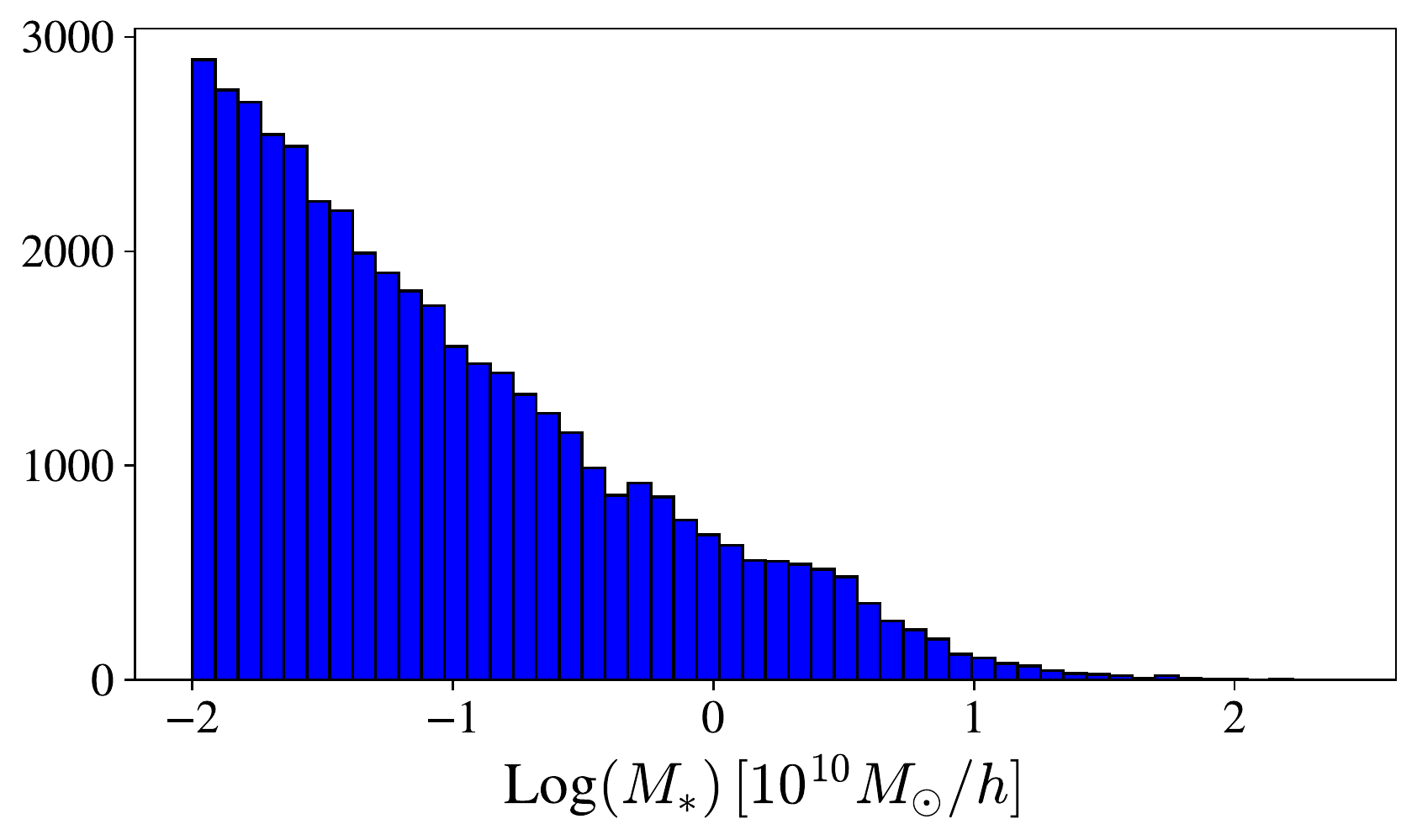}
	\includegraphics[width=\columnwidth]{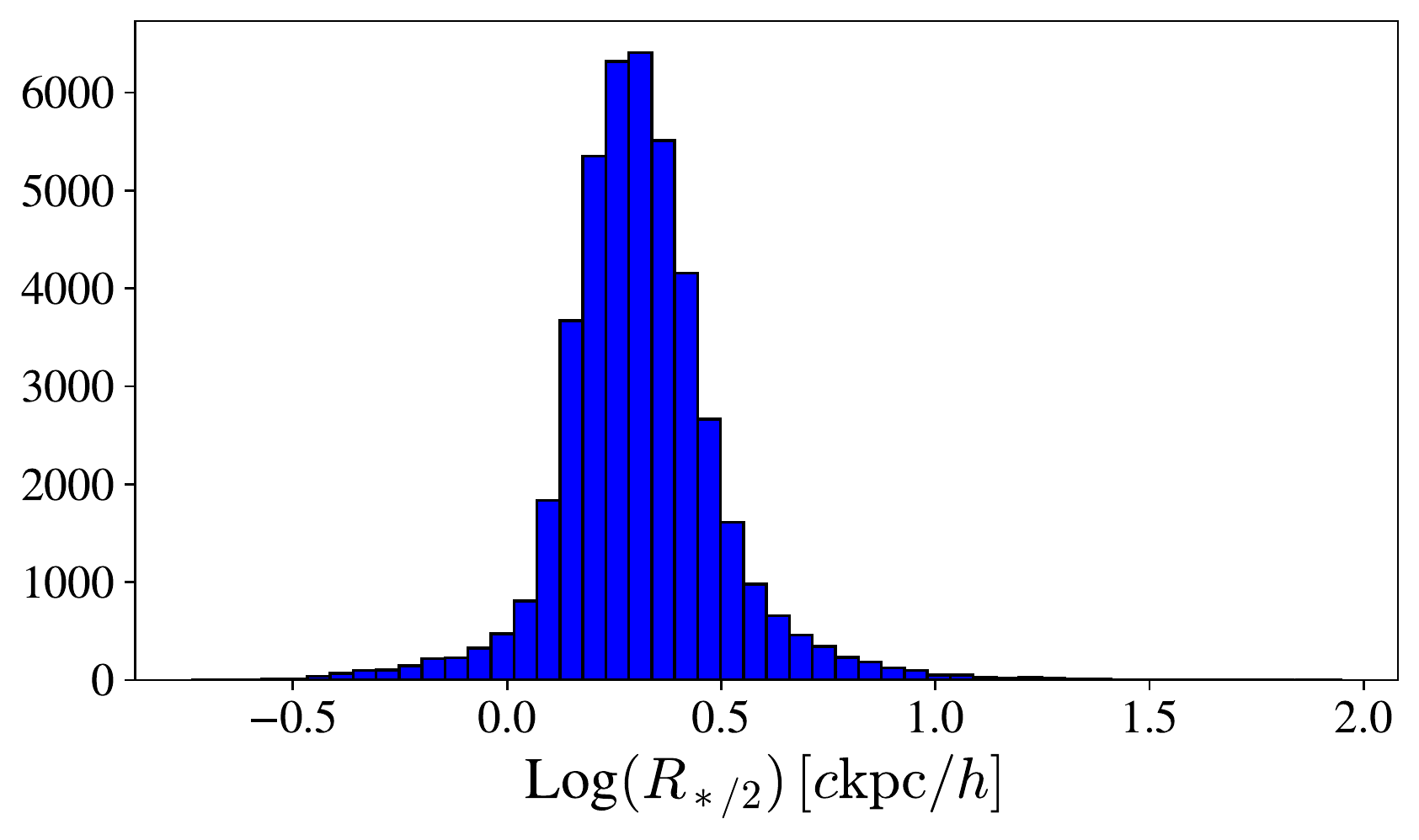}
	\includegraphics[width=\columnwidth]{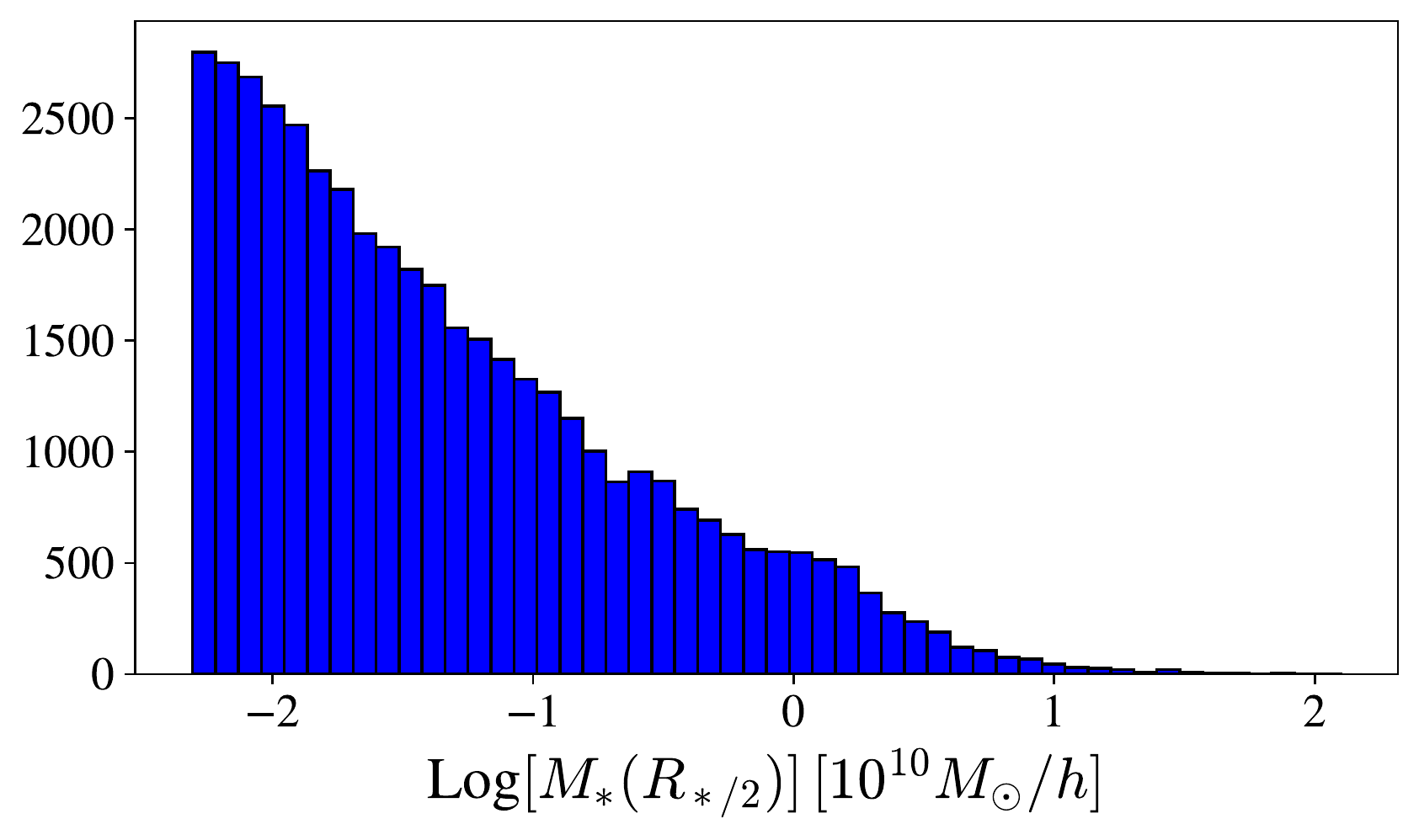}
	\includegraphics[width=\columnwidth]{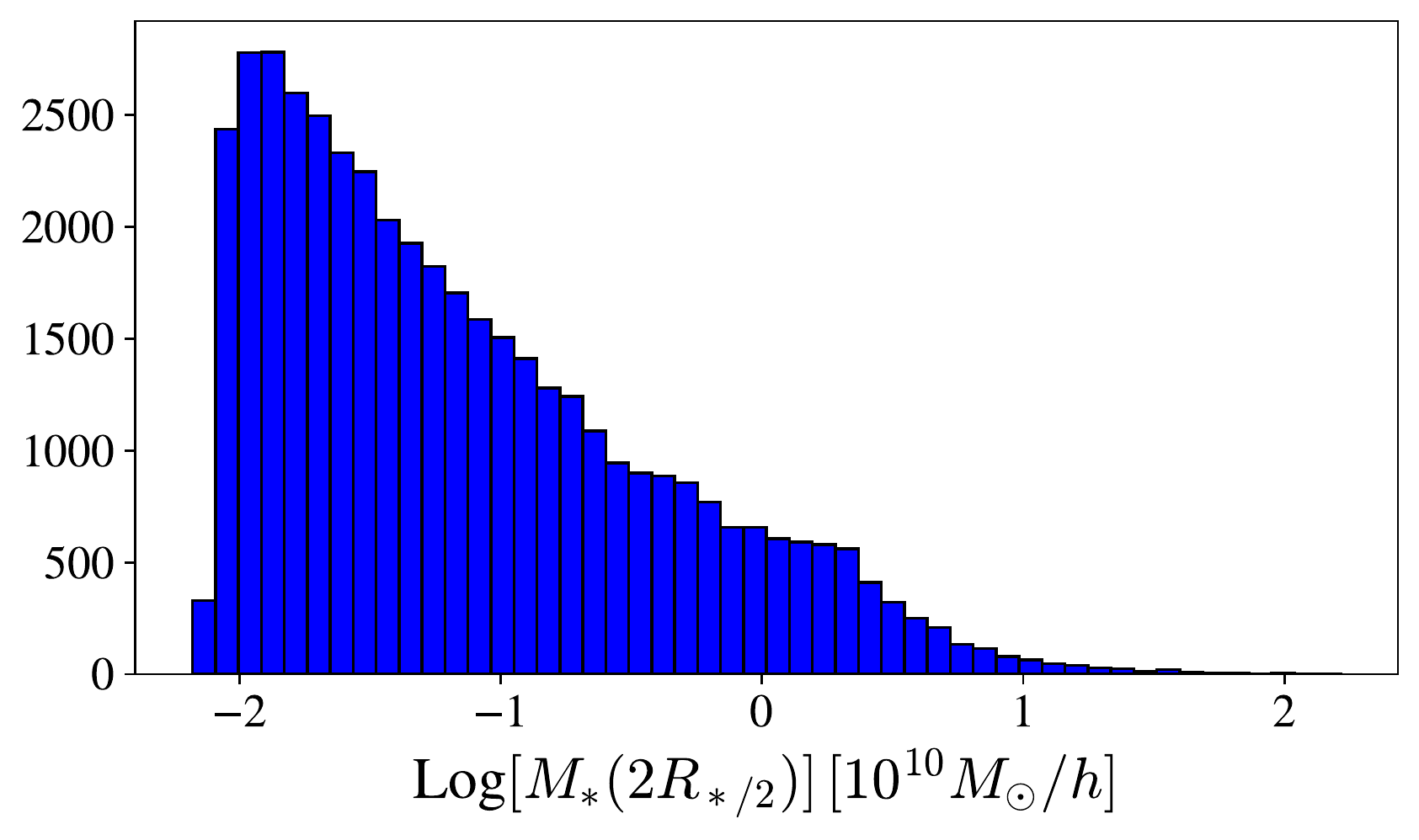}
    \caption{Distribution of the (Log of the) Structural features. \textbf{Top-left panel: }Total stellar matter. \textbf{Top-right panel: }Comoving radius containing half of the stellar mass. \textbf{Bottom-left panel: }Stellar mass within the stellar half mass radius. \textbf{Bottom-right panel: }Stellar mass within twice the stellar half mass radius.}
    \label{fig:dynamical}
\end{figure*}

\begin{figure*}
	\includegraphics[width=\columnwidth]{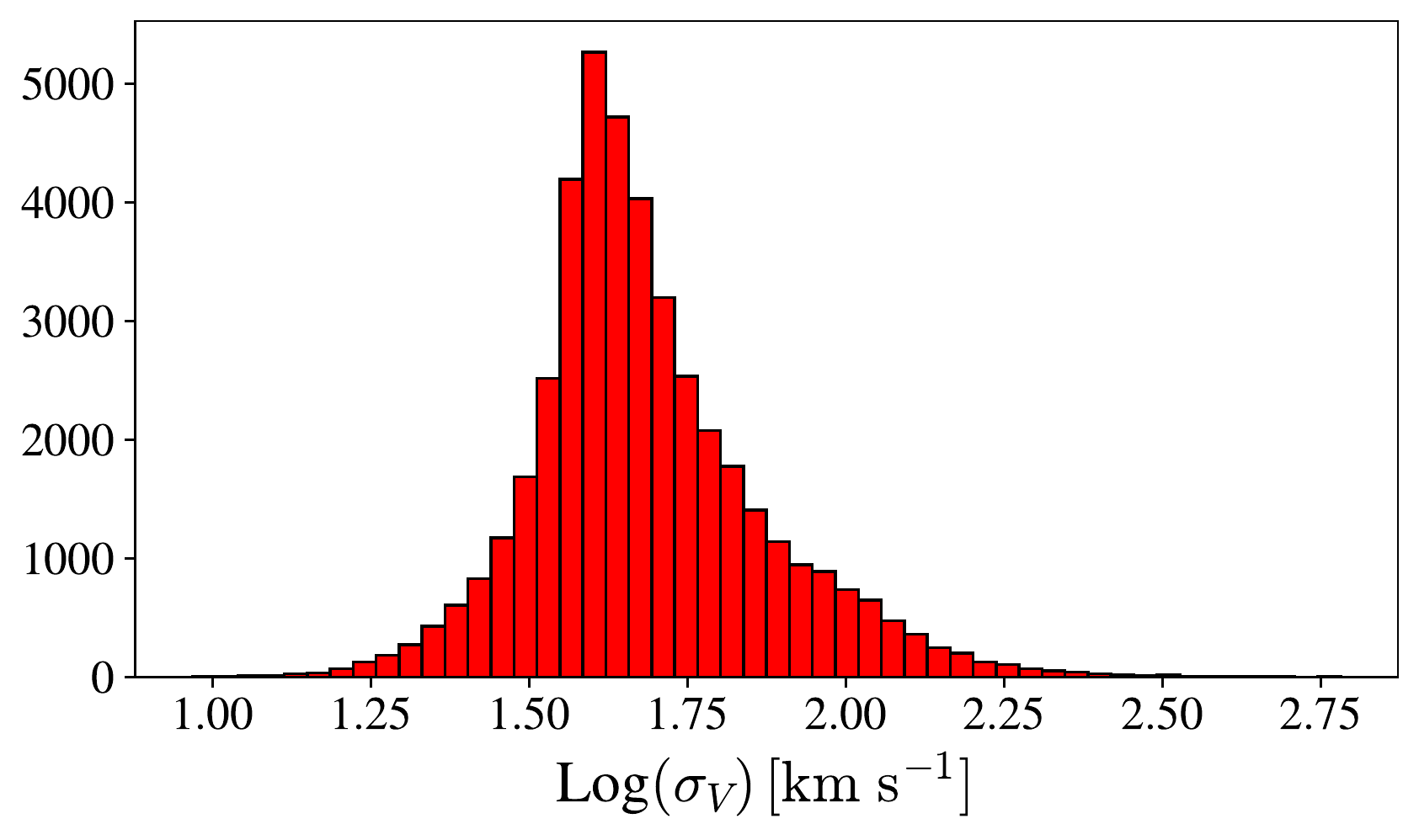}
	\includegraphics[width=\columnwidth]{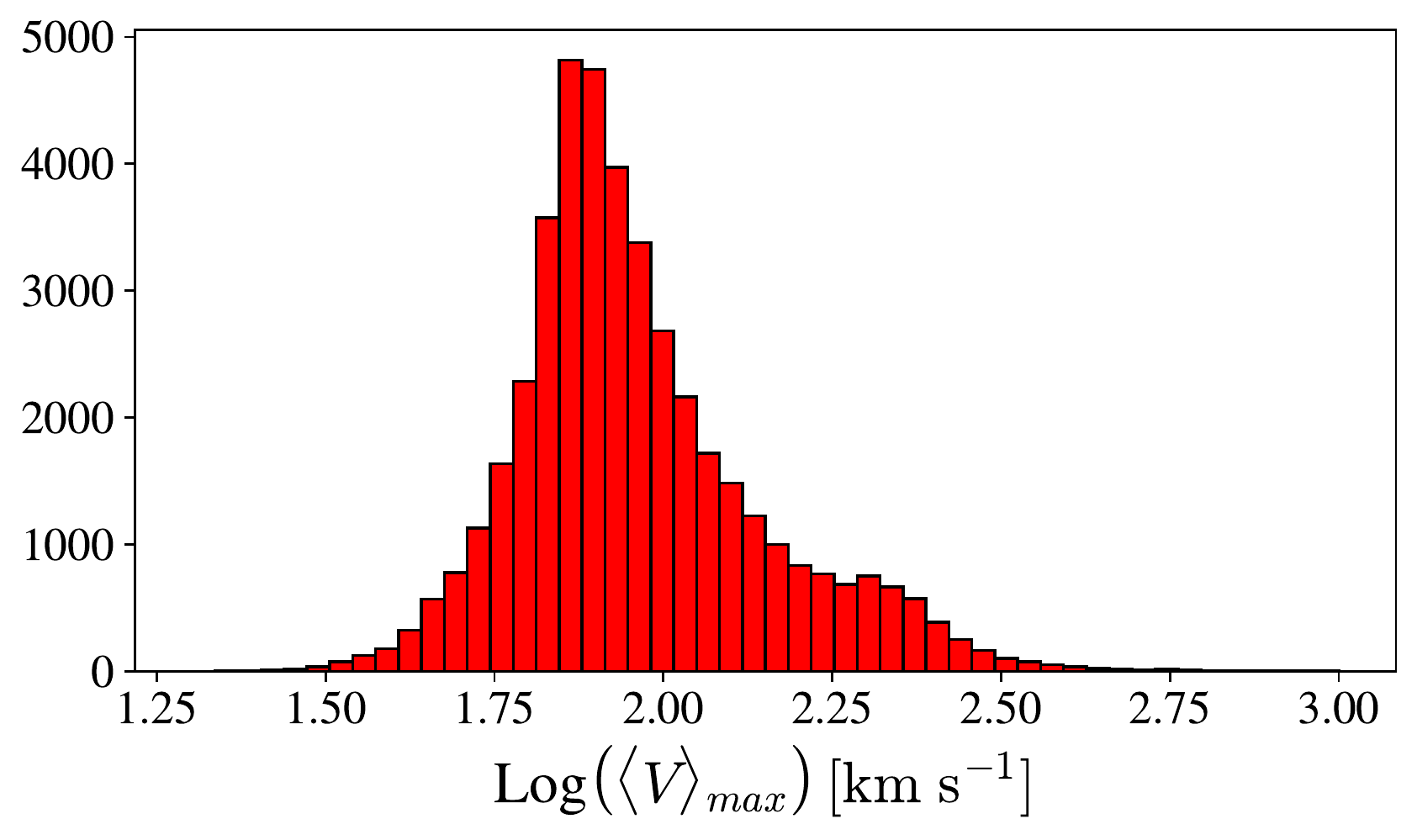}
	\caption{Distribution of the (Log of the) Kinematic features. \textbf{Left panel: }One-dimensional velocity dispersion. \textbf{Right panel: }Maximum value of the spherically averaged rotation curve.}
    \label{fig:kinematical}
\end{figure*}

\begin{figure*}
	\includegraphics[width=\columnwidth]{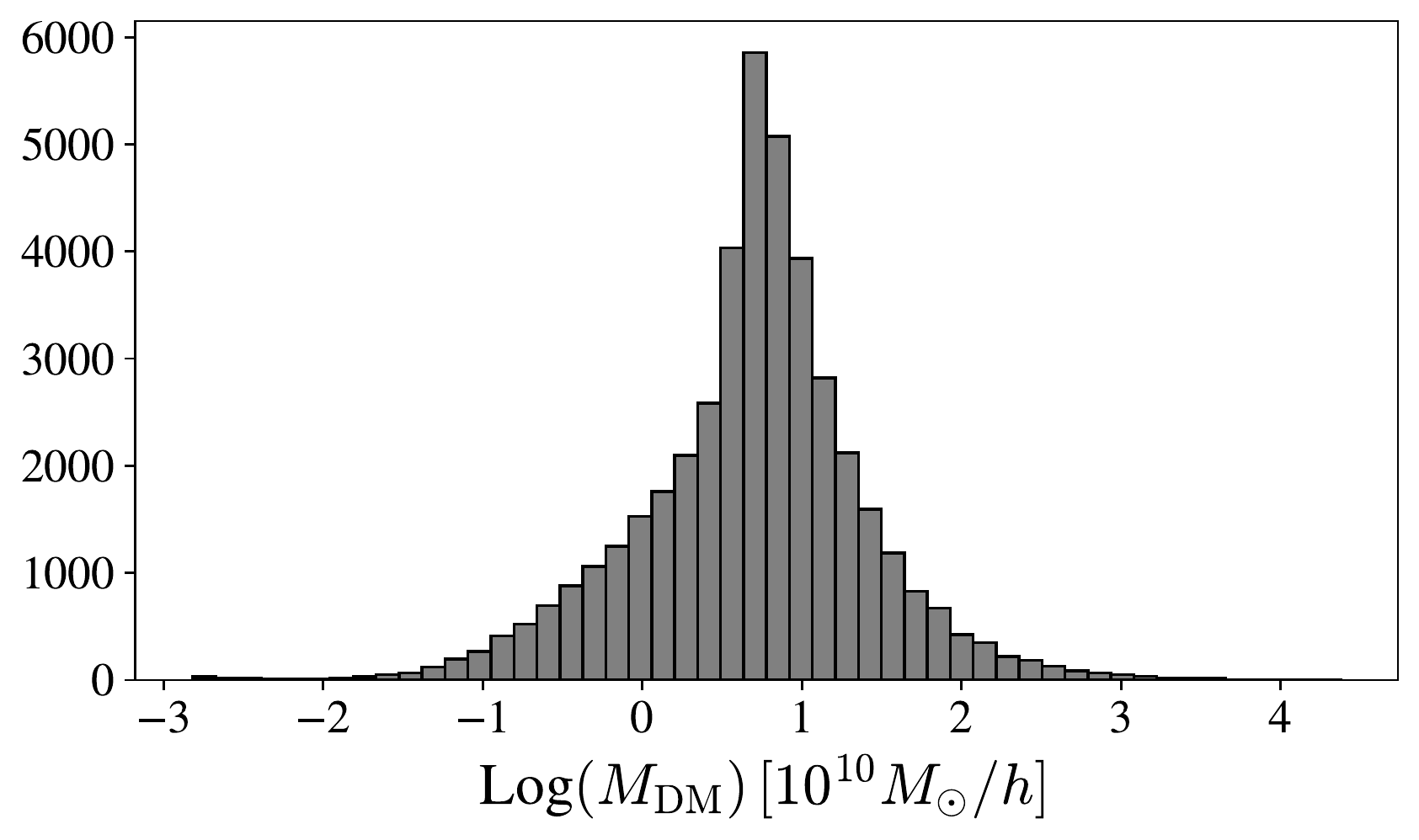}
	\includegraphics[width=\columnwidth]{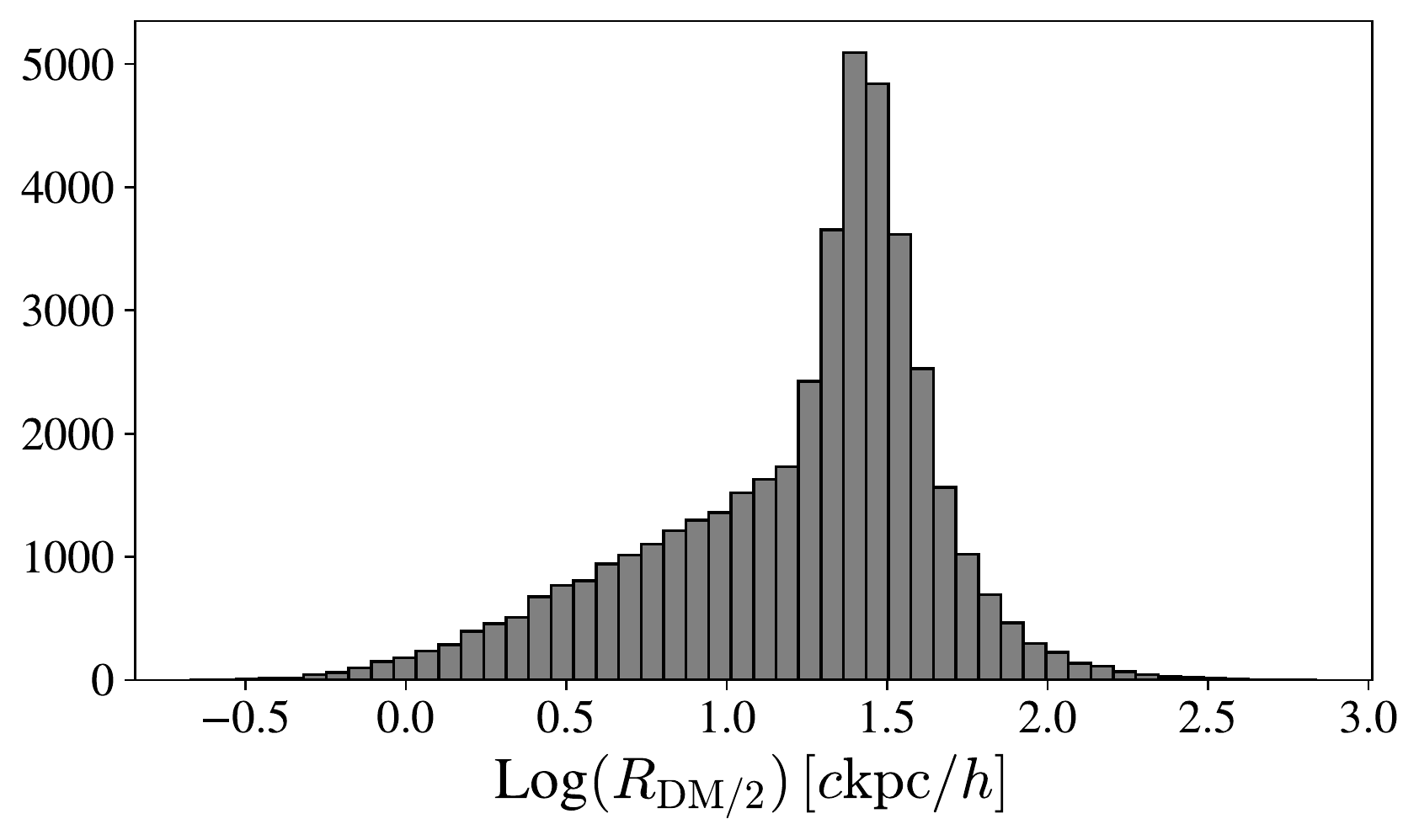}
	\includegraphics[width=\columnwidth]{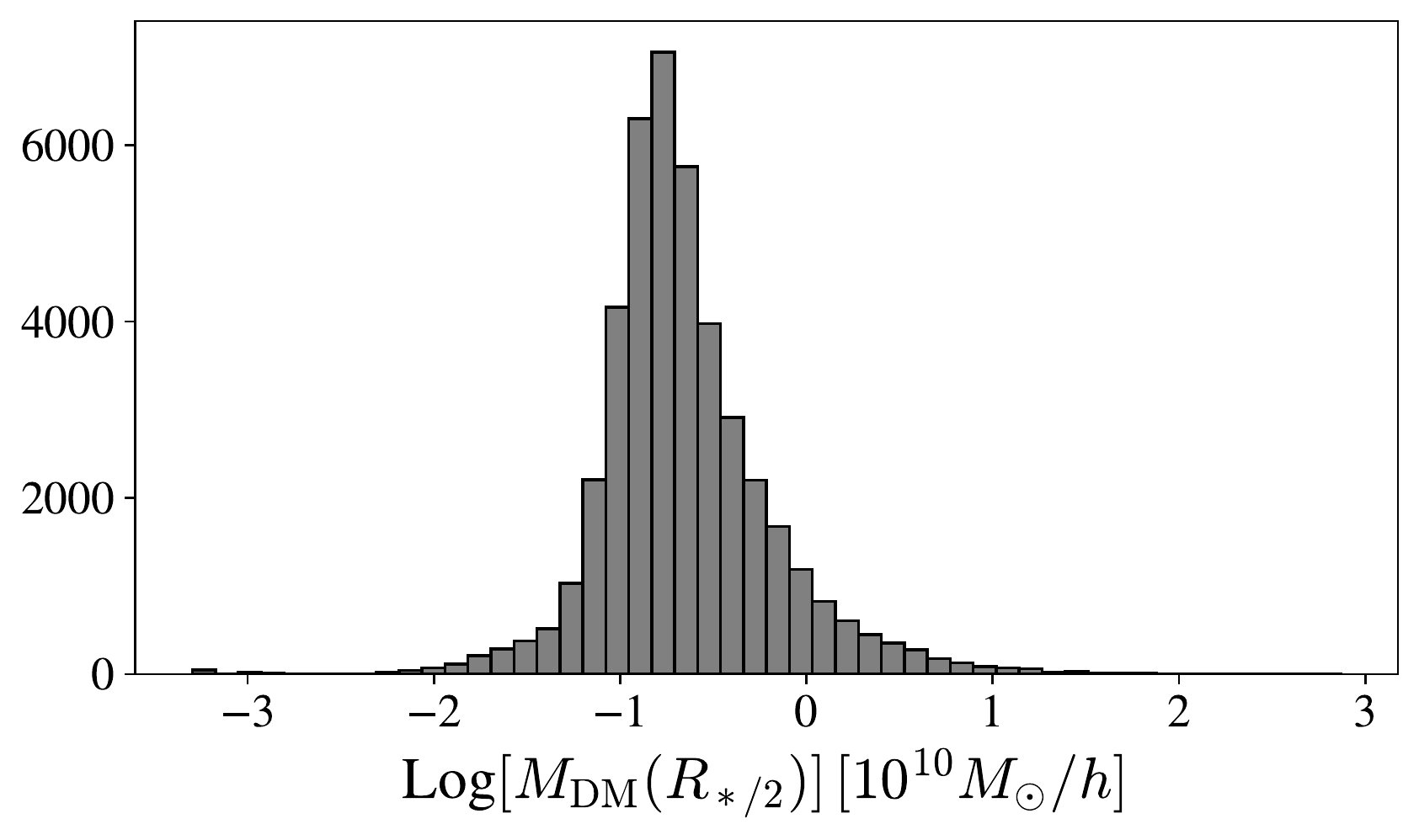}
	\includegraphics[width=\columnwidth]{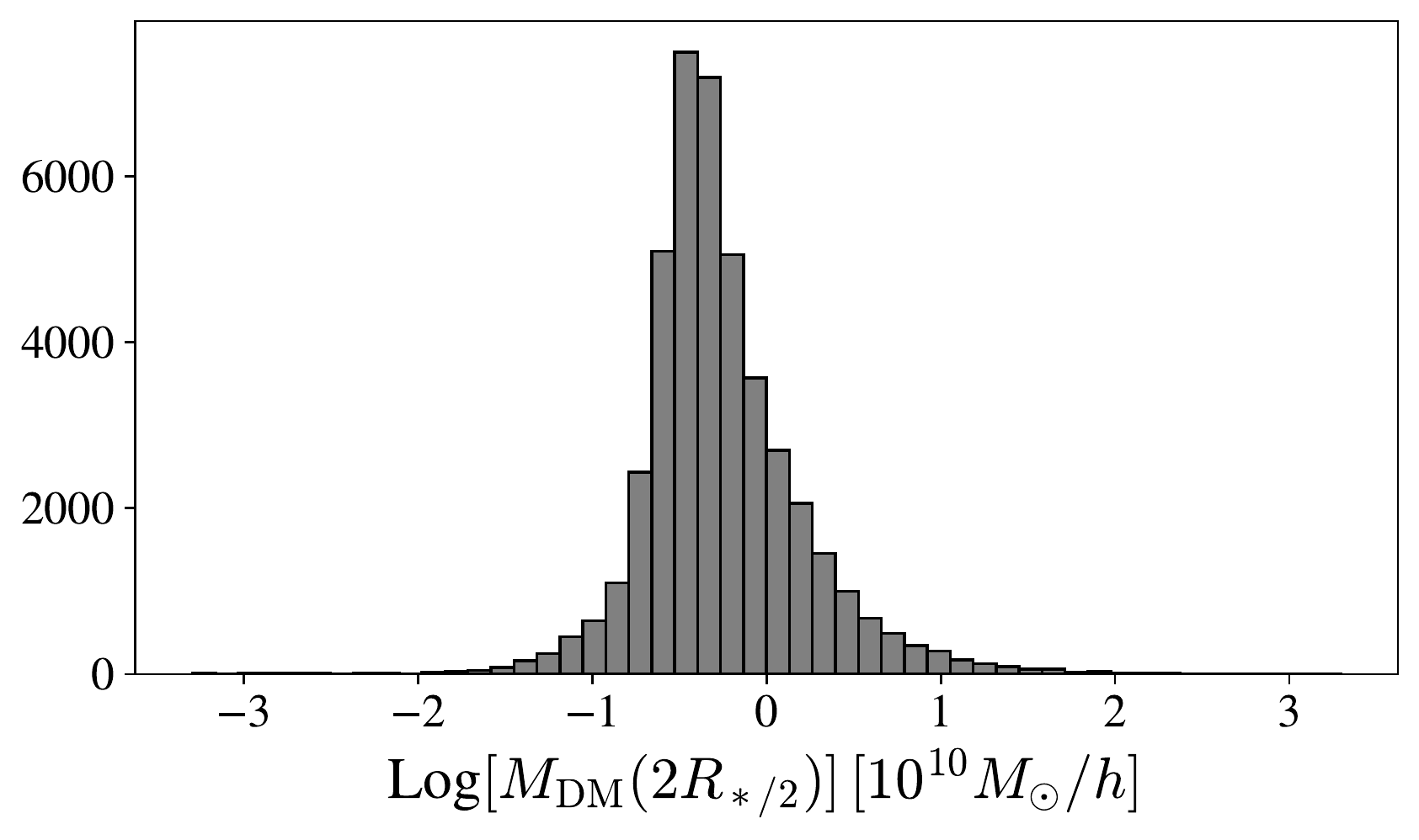}
    \caption{Distribution of the Log10 of the Targets. \textbf{Top-left panel: }Total DM matter. \textbf{Top-right panel: }Comoving radius containing half of the DM mass. \textbf{Bottom-left panel: }DM mass within the stellar half mass radius. \textbf{Bottom-right panel: }DM mass within twice the stellar half mass radius.}
    \label{fig:targets}
\end{figure*}

In this Appendix, we show the distributions of the features. 
Except for the Photometric features (see Fig. \ref{fig:photometrical}), we took the logarithm of the other features and targets in order to simplify the correlations among features and targets. Note that most of the galaxy scaling relation involving Photometry, Structural and Kinematic parameters are power laws that become linear correlations in log-log spaces. 
Looking at the distributions of these log quantities, the Kinematic features and S1, the comoving radius containing half of the stellar mass, show a characteristic bell shape (see Fig. \ref{fig:dynamical}, Fig. \ref{fig:kinematical}),  which also found for the targets  (Fig. \ref{fig:targets}).
All the three structural features concerning the stellar mass S1, S3 and S4, present a common decreasing profile because of the selection of the galaxies based on the minimum value of the stellar mass of $10^8M_{\odot}/h$. In Tab.~\ref{tab:describe}, we show a more detailed description of the statistical properties of all distributions.

\begin{table*}
\label{tab:describe}
\centering
\caption{Table with the statistical description of the distributions of the targets and features. The columns show, from left to right, the mean, the standard deviation, the minimum value, the value at 25\%, the value at 50\%, the value at 75\% and the maximum value.}
\begin{tabular}{l|lllllll}
\hline\hline
                                                                  & \multicolumn{1}{c}{mean}    & \multicolumn{1}{c}{std}   & \multicolumn{1}{c}{min}     & \multicolumn{1}{c}{25\%}    & \multicolumn{1}{c}{50\%}    & \multicolumn{1}{c}{75\%}    & \multicolumn{1}{c}{max}     \\ \hline
T1: Log$(M_{\rm DM})\, [10^{10} M_{\odot}/h]$                     & \multicolumn{1}{c}{0.688}   & \multicolumn{1}{c}{0.714} & \multicolumn{1}{c}{-2.819}  & \multicolumn{1}{c}{0.332}   & \multicolumn{1}{c}{0.733}   & \multicolumn{1}{c}{1.066}   & \multicolumn{1}{c}{4.375}   \\
T2: Log$(R_{\rm DM/2})\, [c{\rm kpc}/h]$                          & \multicolumn{1}{c}{1.221}   & \multicolumn{1}{c}{0.441} & \multicolumn{1}{c}{-0.669}  & \multicolumn{1}{c}{0.974}   & \multicolumn{1}{c}{1.348}   & \multicolumn{1}{c}{1.504}   & \multicolumn{1}{c}{2.835}   \\
T3: Log$[M_{\rm DM}(R_{\rm*/2})]\, [10^{10} M_{\odot}/h]$         & \multicolumn{1}{c}{-0.657}  & \multicolumn{1}{c}{0.470} & \multicolumn{1}{c}{-3.296}  & \multicolumn{1}{c}{-0.918}  & \multicolumn{1}{c}{-0.722}  & \multicolumn{1}{c}{-0.450}  & \multicolumn{1}{c}{2.867}   \\
T4: Log$[M_{\rm DM}(2R_{\rm*/2})]\, [10^{10} M_{\odot}/h]$        & \multicolumn{1}{c}{-0.260}  & \multicolumn{1}{c}{0.470} & \multicolumn{1}{c}{-3.296}  & \multicolumn{1}{c}{-0.520}  & \multicolumn{1}{c}{-0.331}  & \multicolumn{1}{c}{-0.048}  & \multicolumn{1}{c}{3.295}   \\ \hline
P1: $U$ band                                                      & \multicolumn{1}{c}{-17.194} & \multicolumn{1}{c}{1.980} & \multicolumn{1}{c}{-24.718} & \multicolumn{1}{c}{-18.631} & \multicolumn{1}{c}{-17.202} & \multicolumn{1}{c}{-15.689} & \multicolumn{1}{c}{-12.389} \\
P2: $B$ band                                                      & -17.224                     & 1.886                     & -24.570                     & -18.553                     & -17.123                     & -15.801                     & -13.030                     \\
P3: $V$ band                                                      & -17.779                     & 1.841                     & -25.376                     & -19.030                     & -17.591                     & -16.360                     & -14.011                     \\
P4: $K$ band                                                      & -20.183                     & 1.918                     & -28.309                     & -21.443                     & -19.865                     & -18.627                     & -16.926                     \\
P5: $g$ band                                                      & -17.562                     & 1.866                     & -24.966                     & -18.858                     & -17.432                     & -16.148                     & -13.530                     \\
P6: $r$ band                                                      & -18.018                     & 1.833                     & -25.716                     & -19.252                     & -17.803                     & -16.591                     & -14.389                     \\
P7: $i$ band                                                      & -18.252                     & 1.834                     & -26.073                     & -19.473                     & -18.002                     & -16.798                     & -14.802                     \\
P8: $z$ band                                                      & -18.417                     & 1.847                     & -26.329                     & -19.636                     & -18.148                     & -16.943                     & -15.115                     \\ \hline
S1: Log$(M_{*})\, [10^{10} M_{\odot} /h]$                         & -1.034                      & 0.752                     & -2.000                      & -1.650                      & -1.208                      & -0.577                      & 2.400                       \\
S2: Log$(R_{\rm */2})\, [c{\rm kpc}/h]$                           & 0.310                       & 0.188                     & -0.735                      & 0.206                       & 0.300                       & 0.402                       & 1.946                       \\
S3: Log$[M_{*}(R_{\rm */2})]\, [10^{10} M_{\odot}/h]$             & -1.336                      & 0.753                     & -2.307                      & -1.952                      & -1.510                      & -0.879                      & 2.099                       \\
S4: Log$[M_{*}(2R_{\rm */2})]\, [10^{10} M_{\odot}/h]$            & -1.139                      & 0.739                     & -2.180                      & -1.743                      & -1.313                      & -0.684                      & 2.214                       \\ \hline
K1: Log$(\sigma_{V})\, [{\rm km\ s^{-1}}]$                        & 1.684                       & 0.184                     & 0.968                       & 1.574                       & 1.653                       & 1.775                       & 2.781                       \\
K2: Log$\left(\langle V\rangle_{max}\right)\, [{\rm km\ s^{-1}}]$ & 1.962                       & 0.184                     & 1.302                       & 1.846                       & 1.925                       & 2.050                       & 3.000                       \\ \hline\hline
\end{tabular}
\end{table*}

\section{Tpot performance}
\label{sec:tpotperf}

In this Appendix, we discuss the use of \texttt{TPOT} \citep{OlsonGECCO2016} in more details. As mentioned in \S~\ref{ssec:pipeline}, \texttt{TPOT} helps to find strongly optimized pipelines that provides high performance in the prediction of the targets using the available features. In order to find these pipelines, \texttt{TPOT} performs a genetic programming to automatically design and optimize a series of data transformations (e.g., standardization) and machine learning methods. The first step of \texttt{TPOT} functioning is an inspection of the dataset, aiming to make it more suitable for fitting. At that level, we have three possibilities: ($i$) \textit{feature transformation}, which consist in rescale the features, e.g., standardization or PCA \citep{halko2011finding}; ($ii$) \textit{feature selection}, which consist in removing seemingly useless features; ($iii$) \textit{feature construction}, which consist in building new features from the existing ones. Afterwards, \texttt{TPOT} performs the \textit{model selection}, i.e., it identifies the best ML method for addressing the problem. In the standard version, \texttt{TPOT} considers all models available in \texttt{scikit-learn} \citep{scikit-learn}, e.g., Support Vector Machine (SVM) 
\citep{Smola04atutorial} or Random Forest (RF) \citep{breiman2001}; Finally, the last step is the \textit{parameter optimization}, where the meta-parameters of the chosen method are optimized, e.g., the number of hidden layers in the Artificial Neural Networks (ANN) \citep{ann2018}. The number of pipelines analyzed depend on the choice of three parameters: 
\begin{itemize}
    \item[$i$)] \textit{Generations: }Number of iterations to run the pipeline optimization process;
    
    \item[$ii$)] \textit{Population size: }Number of individuals to retain in the genetic programming population every generation;
    
    \item[$iii$)] \textit{Offspring size: }Number of offspring to produce in each genetic programming generation;
\end{itemize}

\begin{figure*}
	\includegraphics[width=\columnwidth]{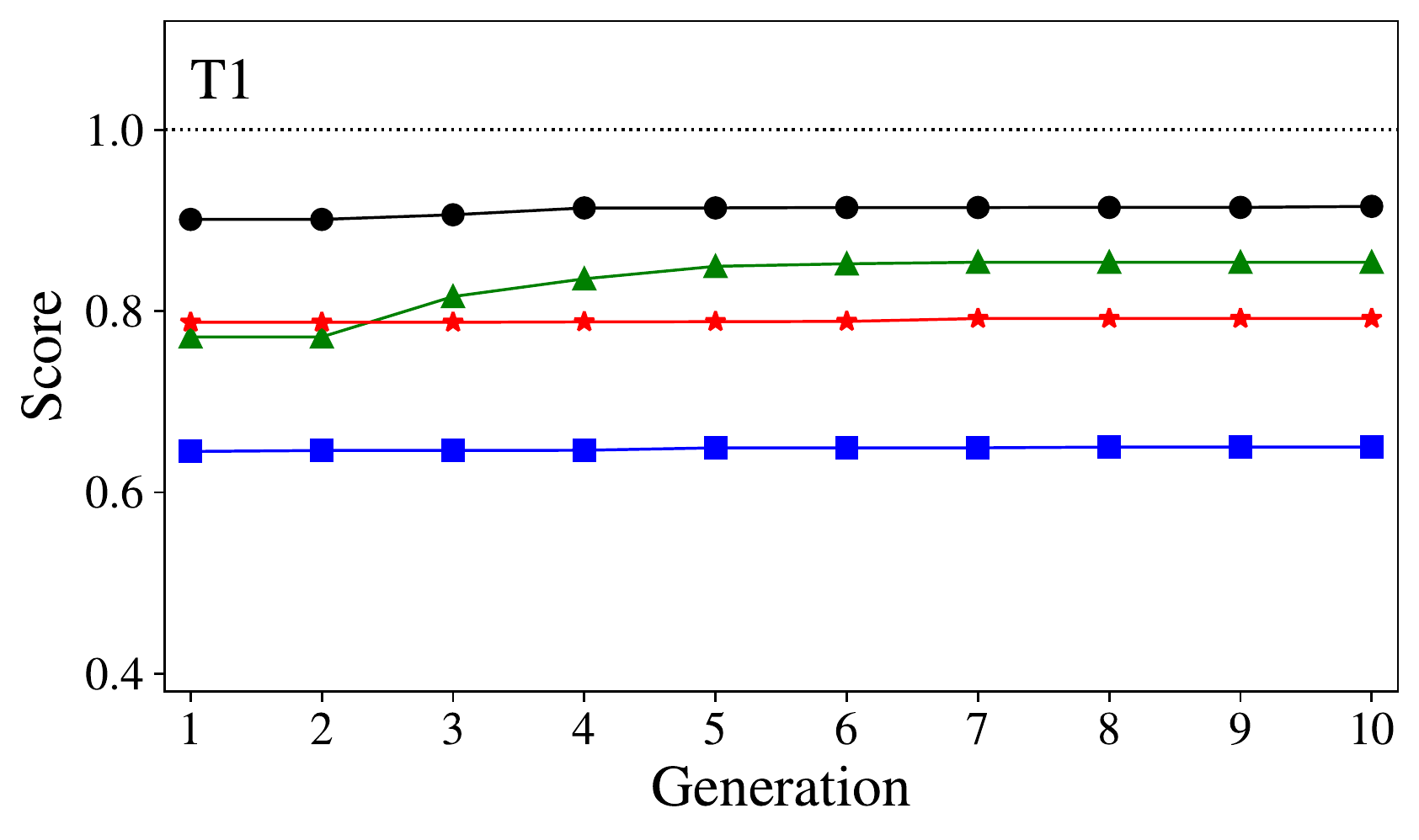}
	\includegraphics[width=\columnwidth]{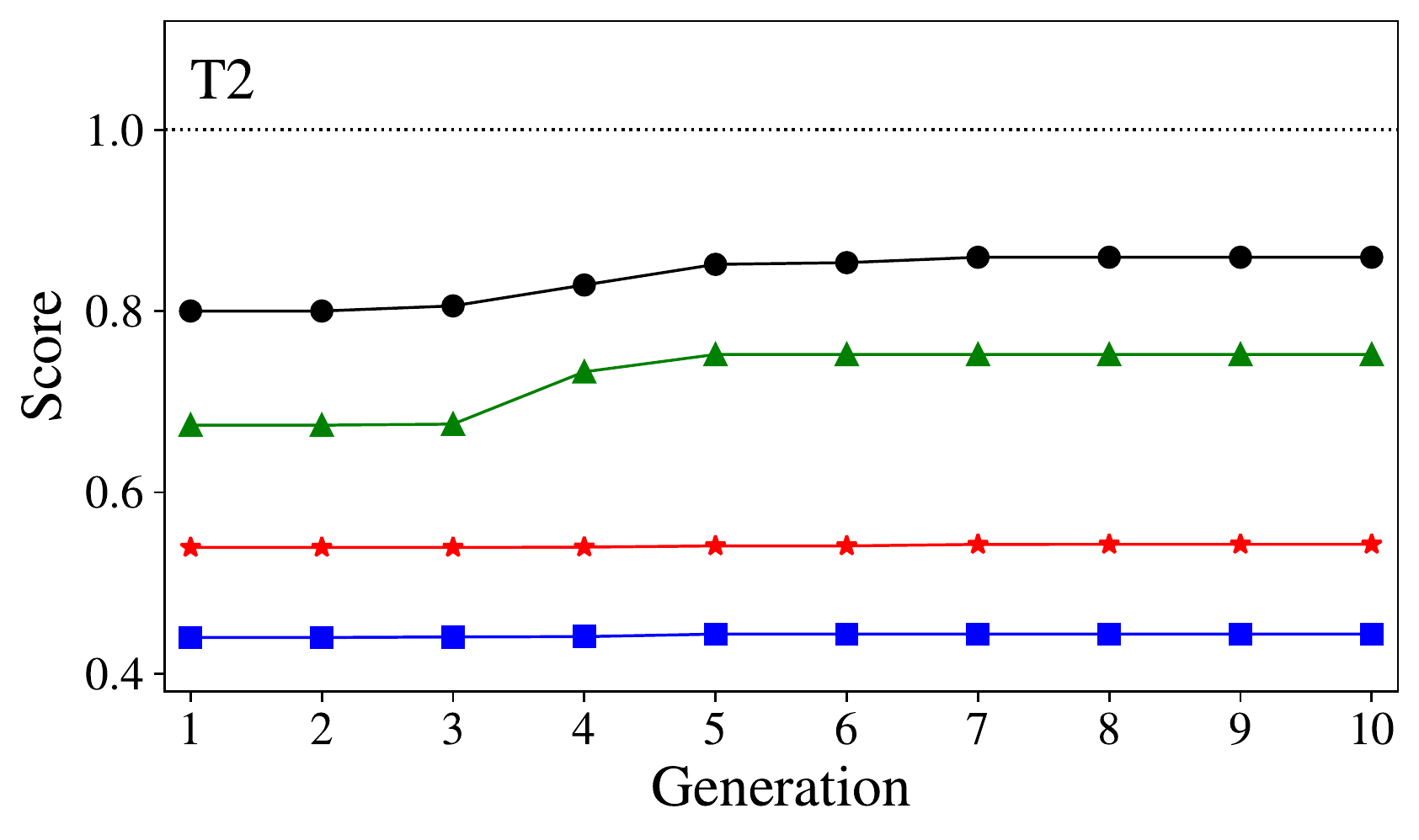}
	\includegraphics[width=\columnwidth]{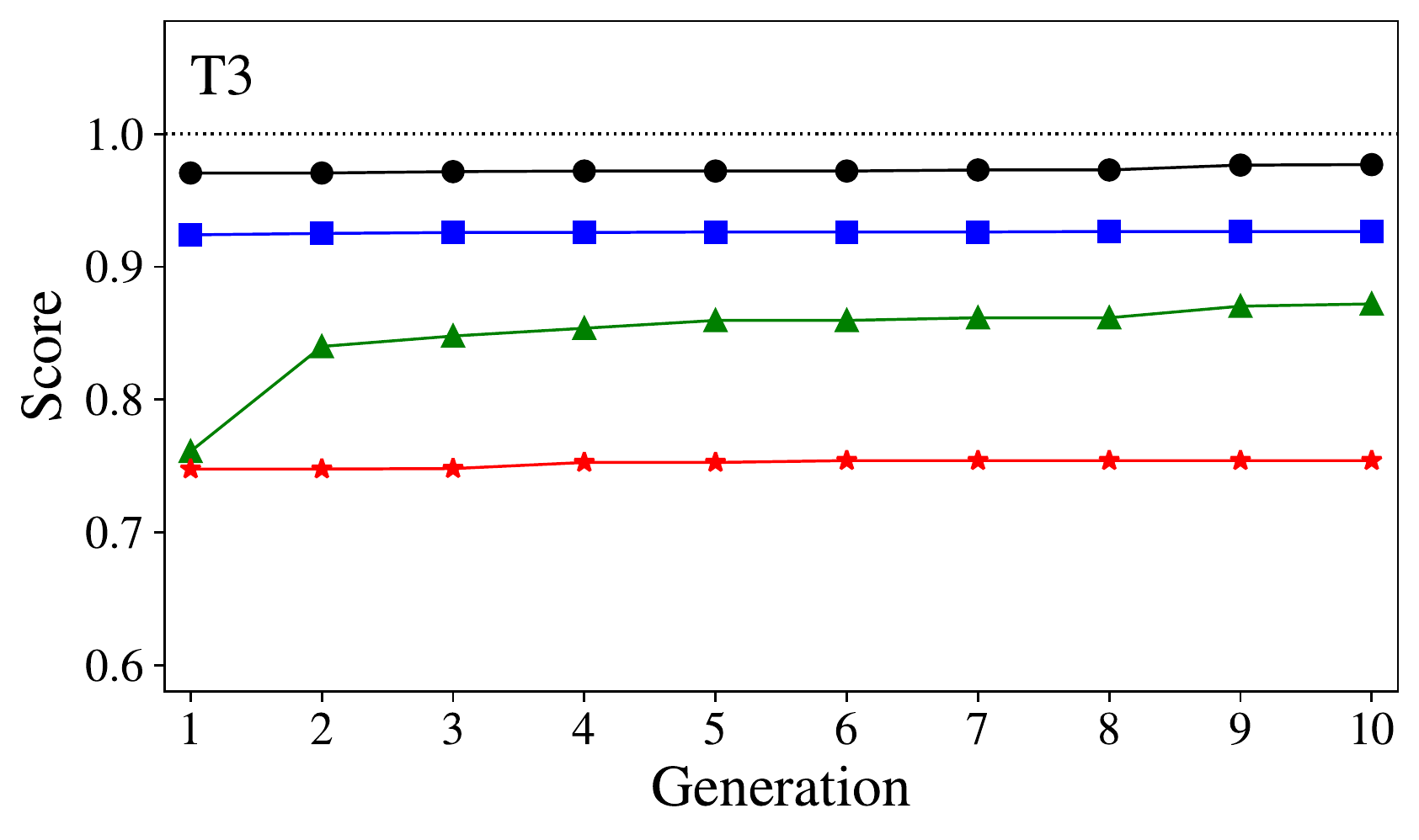}
	\includegraphics[width=\columnwidth]{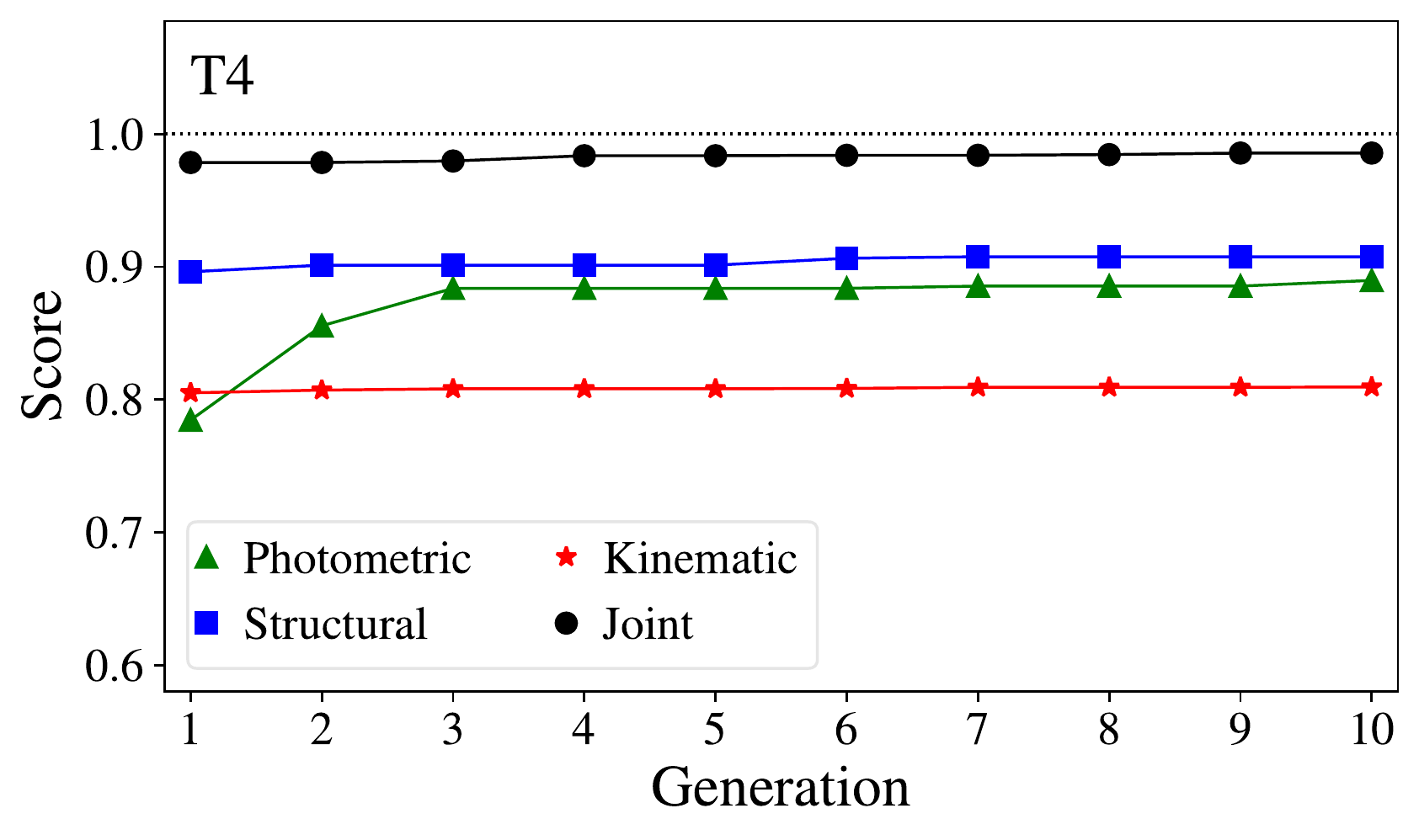}
    \caption{Evolution of the score during the \texttt{TPOT} running of all analyses.}
    \label{fig:score}
\end{figure*}

In our analysis, these three parameters are set to 10. The number of analyzed pipelines is given by \textit{Population size} $+$ \textit{Generations} $\times$ \textit{Offspring size}, which means that, for each case, 110 pipelines have been considered in our analysis. In order to avoid overfitting and any bias due to the choice of the training and test samples, each pipeline is evaluated by a cross-validated score based on the Mean Squared Error (MSE). At the end, \texttt{TPOT} returns the score of the best pipeline so far. The evolution of the score over the generations for all the analyses are shown in Fig.~\ref{fig:score}. As one can see, in all cases the evolution curve reaches a stable behavior. This fact indicates that our choice for the parameters \textit{Generations}, \textit{Population size} and \textit{Offspring size} are feasible. 

Finally, the found pipelines are described in Tab.~\ref{tab:pipelines}. One can note that Random Forest appears in all cases. Even though Random Forest seems to be a good choice for all analyzes, it is worth mentioning that \texttt{TPOT} processing was crucial to capture the individual peculiarities from the different groups. Beyond differences in the featuring transformation process and in the values of the meta-parameters, the Random Forest is stacked with other methods in some cases, for example, it is stacked with Elastic Net (EN) \citep{friedman2010regularization} in the analysis of the target T3 (DM mass within the stellar half mass radius) using the structural features only. Stacking different methods can be understood as an ensemble learning technique, where two or more regression methods are performed, and their outputs are used as input in a final (meta) regression. The possibility of stacking different methods is another benefit of using \texttt{TPOT}.
\begin{table*}
\centering
\caption{Best pipeline found with \texttt{TPOT} for all analyses.}
\begin{tabular}{l|l|l}
\hline\hline
\multicolumn{2}{c|}{Analysis}                      & \multicolumn{1}{c}{Best pipeline}                                                                                                                                       \\ \hline
\multirow{9}{*}{T1} & \multirow{2}{*}{Photometric} & \begin{tiny} {\tt make\_pipeline(make\_union(FunctionTransformer(copy), FastICA(tol=0.9)), RandomForestRegressor(bootstrap=True, max\_features=0.75, min\_samples\_leaf=8,} \end{tiny}                \\
                    &                              & \begin{tiny} {\tt min\_samples\_split=3, n\_estimators=100)) } \end{tiny}                                                                                                                             \\ \cline{2-3} 
                    & \multirow{2}{*}{Structural}  & \begin{tiny} {\tt make\_pipeline(FastICA(tol=0.25),StackingEstimator(estimator=ElasticNetCV(l1\_ratio=0.5, tol=0.1)),RandomForestRegressor(bootstrap=True, max\_features=0.75,  } \end{tiny}            \\
                    &                              & \begin{tiny} {\tt min\_samples\_leaf=17, min\_samples\_split=9, n\_estimators=100))  } \end{tiny}                                                                                                       \\ \cline{2-3} 
                    & \multirow{2}{*}{Kinematic}   & \begin{tiny} {\tt make\_pipeline(PCA(iterated\_power=7, svd\_solver="randomized"), StackingEstimator(estimator=RidgeCV()), RandomForestRegressor(bootstrap=True, max\_features=0.4, } \end{tiny}       \\
                    &                              & \begin{tiny} {\tt min\_samples\_leaf=16, min\_samples\_split=14, n\_estimators=100)) } \end{tiny}                                                                                                     \\ \cline{2-3} 
                    & \multirow{3}{*}{Joint}       & \begin{tiny} {\tt make\_pipeline(StackingEstimator(estimator=RandomForestRegressor(bootstrap=True, max\_features=0.4, min\_samples\_leaf=5, min\_samples\_split=9, n\_estimators=100)), } \end{tiny}    \\
                    &                              & \begin{tiny} {\tt StackingEstimator(estimator=RidgeCV()), RandomForestRegressor(bootstrap=True, max\_features=0.4, min\_samples\_leaf=9, min\_samples\_split=14, } \end{tiny}                           \\
                    &                              & \begin{tiny} {\tt n\_estimators=100)) } \end{tiny}                                                                                                                                                      \\ \hline
\multirow{8}{*}{T2} & \multirow{2}{*}{Photometric} & \begin{tiny} {\tt make\_pipeline(FastICA(tol=0.25), StackingEstimator(estimator=ElasticNetCV(l1\_ratio=0.4, tol=0.01)), RandomForestRegressor(bootstrap=True, } \end{tiny}                              \\
                    &                              & \begin{tiny} {\tt max\_features=0.05, min\_samples\_leaf=7, min\_samples\_split=13, n\_estimators=100)) } \end{tiny}                                                                                    \\ \cline{2-3} 
                    & \multirow{2}{*}{Structural}  & \begin{tiny} {\tt make\_pipeline(PCA(iterated\_power=6, svd\_solver="randomized"), StackingEstimator(estimator=LassoLarsCV(normalize=True)),RandomForestRegressor(bootstrap=True, } \end{tiny}          \\
                    &                              & \begin{tiny} {\tt max\_features=0.8, min\_samples\_leaf=17, min\_samples\_split=9, n\_estimators=100)) } \end{tiny}                                                                                     \\ \cline{2-3} 
                    & \multirow{2}{*}{Kinematic}   & \begin{tiny} {\tt make\_pipeline( FastICA(tol=0.45), StackingEstimator(estimator=RidgeCV()), RandomForestRegressor(bootstrap=True, max\_features=0.4, min\_samples\_leaf=16, } \end{tiny}               \\
                    &                              & \begin{tiny} {\tt min\_samples\_split=19, n\_estimators=100)) } \end{tiny}                                                                                                                              \\ \cline{2-3} 
                    & \multirow{2}{*}{Joint}       & \begin{tiny} {\tt make\_pipeline(PCA(iterated\_power=6, svd\_solver="randomized"), StackingEstimator(estimator=LassoLarsCV(normalize=False)), RandomForestRegressor(bootstrap=True, } \end{tiny}        \\
                    &                              & \begin{tiny} {\tt max\_features=0.4, min\_samples\_leaf=7, min\_samples\_split=14, n\_estimators=100)) } \end{tiny}                                                                                     \\ \hline
\multirow{8}{*}{T3} & \multirow{2}{*}{Photometric} & \begin{tiny} {\tt make\_pipeline(PCA(iterated\_power=5, svd\_solver="randomized"),RandomForestRegressor(bootstrap=True, max\_features=0.4, min\_samples\_leaf=1, min\_samples\_split=14, } \end{tiny}   \\
                    &                              & \begin{tiny} {\tt n\_estimators=100)) } \end{tiny}                                                                                                                                                      \\ \cline{2-3} 
                    & \multirow{3}{*}{Structural}  & \begin{tiny} {\tt make\_pipeline(PCA(iterated\_power=2, svd\_solver="randomized"), StackingEstimator(estimator=ExtraTreesRegressor(bootstrap=True, max\_features=0.5, } \end{tiny}                      \\
                    &                              & \begin{tiny} {\tt min\_samples\_leaf=16, min\_samples\_split=5, n\_estimators=100)), RandomForestRegressor(bootstrap=True, max\_features=0.75, min\_samples\_leaf=11, } \end{tiny}                      \\
                    &                              & \begin{tiny} {\tt min\_samples\_split=6, n\_estimators=100)) } \end{tiny}                                                                                                                               \\ \cline{2-3} 
                    & \multirow{2}{*}{Kinematic}   & \begin{tiny} {\tt make\_pipeline(PCA(iterated\_power=5, svd\_solver="randomized"), RandomForestRegressor(bootstrap=True, max\_features=0.75, min\_samples\_leaf=11,min\_samples\_split=9, } \end{tiny}  \\
                    &                              & \begin{tiny} {\tt n\_estimators=100)) } \end{tiny}                                                                                                                                                      \\ \cline{2-3} 
                    & Joint                        & \begin{tiny} {\tt RandomForestRegressor(bootstrap=False, max\_features=0.5, min\_samples\_leaf=3, min\_samples\_split=9, n\_estimators=100) } \end{tiny}                                                \\ \hline
\multirow{7}{*}{T4} & \multirow{2}{*}{Photometric} & \begin{tiny} {\tt make\_pipeline(PCA(iterated\_power=8, svd\_solver="randomized"), StackingEstimator(estimator=LassoLarsCV(normalize=False)), RandomForestRegressor(bootstrap=True, } \end{tiny}        \\
                    &                              & \begin{tiny} {\tt max\_features=0.4, min\_samples\_leaf=5, min\_samples\_split=14, n\_estimators=100)) } \end{tiny}                                                                                     \\ \cline{2-3} 
                    & \multirow{2}{*}{Structural}  & \begin{tiny} {\tt make\_pipeline(PCA(iterated\_power=6, svd\_solver="randomized"), RobustScaler(), RandomForestRegressor(bootstrap=True, max\_features=0.75, min\_samples\_leaf=5, } \end{tiny}         \\
                    &                              & \begin{tiny} {\tt min\_samples\_split=9, n\_estimators=100)) } \end{tiny}                                                                                                                               \\ \cline{2-3} 
                    & Kinematic                    & \begin{tiny} {\tt make\_pipeline(RBFSampler(gamma=0.3),RandomForestRegressor(bootstrap=True, max\_features=0.05, min\_samples\_leaf=12, min\_samples\_split=15, n\_estimators=100)) } \end{tiny}        \\ \cline{2-3} 
                    & \multirow{2}{*}{Joint}       & \begin{tiny} {\tt make\_pipeline(PCA(iterated\_power=6, svd\_solver="randomized"), StackingEstimator(estimator=ElasticNetCV(l1\_ratio=0.75, tol=0.001)), } \end{tiny}                                   \\
                    &                              & \begin{tiny} {\tt RandomForestRegressor(bootstrap=True, max\_features=0.4, min\_samples\_leaf=4, min\_samples\_split=14, n\_estimators=100)) } \end{tiny}                                               \\ \hline\hline
\end{tabular}
\label{tab:pipelines}
\end{table*}
%

\bsp	
\label{lastpage}
\end{document}